\documentclass[12pt]{article}
\usepackage{amsmath,amssymb,amsfonts,bm}
\usepackage{mathrsfs,array,graphicx,float}
\usepackage{enumerate,natbib,caption,xcolor}
\usepackage{url,subcaption}
\usepackage[toc,page]{appendix}

\usepackage{authblk}
\begin{document}
  \title{\bf Bayesian Emulation of Geotechnical Deterioration Curves Using Quadratic and B-Spline Hierarchical Models}
 \author{Jordan L. Oakley and Aleksandra Svalova\thanks{The authors are grateful for the financial support of the Engineering and Physical Sciences Research Council (EPSRC) through the programme Grant ACHILLES (EP/R034575/1).}\\
    School of Mathematics, Statistics and Physics, Newcastle University, Newcastle upon Tyne, NE1 7RU, UK\\ 
    Peter Helm\thanks{Dr. Helm’s contribution to the manuscript preparation was supported by the Research Hub for Decarbonised Adaptable and Resilient Transport Infrastructures (DARe) funded by the UK Department for Transport and EPSRC (EP/Y024257/1).}, Mohamed Rouainia, and Stephanie Glendinning\\
    School of Engineering, Newcastle University, Newcastle upon Tyne, NE1 7RU, UK\\
    Dennis Prangle\\
    School of Mathematics, University of Bristol, Bristol, BS8 1TW, UK\\
    and Darren Wilkinson,\\
    Department of Mathematical Sciences, Durham University, Durham, DH1 3LE, UK
}
  \maketitle

\begin{abstract}
The stability of geotechnical infrastructure assets, such as cuttings and embankments, is crucial to the safe and efficient delivery of transport services. The successful emulation of geotechnical models of deterioration of infrastructure slopes has the potential to inform slope design, maintenance and remediation by introducing the time dependency of deterioration into geotechnical asset management. We have performed computer experiments of deterioration, measured by the factor of safety (FoS), for a set of cutting slope geometries and soil properties that are common in the southern UK. Whilst computer experiments are an extremely useful and cost-effective method of better understanding deterioration mechanisms, it would not be practical to run enough experiments to understand relations between high-dimensional inputs and outputs. Therefore, we trained a fully-Bayesian Gaussian process emulator using an ensemble of 75 computer experiments to predict the FoS. We construct two different emulator models, one approximating the FoS temporal evolution with a quadratic model and one approximating the temporal evolution with a B-spline model; and we emulated their parameters. We also compare the ability of our models to predict failure time. The developed models could be used to inform infrastructure cutting slope design and management, and extend serviceable life.
\end{abstract}
\maketitle
\newpage
\section{Introduction}
\noindent Engineered earthworks are a key part of national infrastructure in the United Kingdom. Modern highways earthworks are designed so that their actual stability exceeds the required minimum (typically by reducing the design slope angle). The ratio between excess capacity and the minimum required to maintain stability is known as factor of safety (FoS) against ultimate limit state failure (i.e., states associated with 
collapse, see for example~\citet{burland2023} and~\citet{bsi_bs_2023}). Factor of safety can be defined in a number of ways; here the specific definition of FoS relating to slope stability is as described by~\citet{duncan1996}, whereby FoS is the ratio of available shear strength (the ability of a material to resist the component that acts parallel to a plane) of the soil to the shear strength required to barely maintain equilibrium/stability, where the slope is stable if FoS is greater than 1.\\
\noindent As the strength of slopes deteriorates due to short duration weather events, seasonal cycles, longer term climate change and other cyclic loading (for example, vehicular traffic), the FoS will reduce. This reduction in FoS due to deterioration poses an additional problem for UK 
railway earthworks, significant proportions of which are greater than 150 years old~\citep{nr2018}. These earthworks tend to be overly steep and hence have a lower FoS at construction~\citep{perry2003} compared to the more modern highways earthworks, and together with their longer exposure to processes that cause deterioration~\citep{briggs2023} mean they may be closer to ULS failure than more recently constructed engineered slopes. \\
\noindent In addition to the ultimate limit state (ULS) as mentioned above, there are additional limit state constraints that infrastructure assets tend to operate within, these are known as serviceability limit states (SLS). While ULS is concerned with the final collapse/failure of an asset and has significant implications for safety if it occurs, SLS defines the minimum performance required by an asset to satisfactorily fulfil its intended function. In the context of slope stability, this could relate to deformations driven by gradual deterioration. These deformations can affect rail track alignment, where they begin to affect ride quality and in more extreme cases, if they exceed a specified (SLS) threshold, would necessitate the imposition of reduced speed limits for safety until remediation works could be undertaken. These same drivers of deterioration can also, over long periods, lead to ULS failure. An early identification of the assets likely to be affected by these serviceability related issues in turn reduces maintenance, delay and intervention costs and can reduce the risk of ULS failures occurring. \\
\noindent While the prediction of the time at which ULS may occur has been addressed by~\citet{svalova2021} who modelled time to failure of transport infrastructure cuttings as a function of geometry and soil strength, that work did not address the rate of asset deterioration that can lead to exceedance of SLS prior to ULS failure. \\
\noindent Prior geotechnical modelling work~\citep{postill2021} has shown that the change in FoS with time is not linear and so explicitly modelling the FoS change as opposed to the time to failure allows an estimate to be made of the rate of asset deterioration as a function of material strength and slope geometry. Furthermore the magnitude of relative deterioration that has occurred when interventions are undertaken has been shown to influence their effectiveness~\citep{armstrong2024}. This shows that a model of changing FoS with time prior to failure helps identify and prioritise the selection of assets for monitoring and remediation and allows more effective management by giving a clearer picture of the proportion of the asset portfolio which may be affected at a given time. \\
\noindent As such, in this work, we address the deterioration towards shear failure of cut slopes, where shear failure can be defined as the collapse of the slope by sliding along a shear surface in the soil mass. This is performed using computer experiments of deterioration and Gaussian process emulation (GPE). Computer experiments can be used to simulate FoS evolution over time for a wide range of geotechnical properties \citep[see for example,][]{helm2024emulating} and weather scenarios \citep{rouainia2020,postill2021}. The GPE-computer experiment symbiosis has been a frequent occurrence in the statistical machine learning literature over the last thirty years~\citep[see][and references therein]{santner2018,gramacy2020}. \\
\noindent A key body of related Bayesian literature has been developed by O'Hagan and colleagues, who define the ``emulator" to be a Gaussian process conditioned on observations (i.e., training data from a computer simulator) that are assumed to be normally distributed and whose parameters are inferred using a Bayesian framework~\citep{ohagan2006,bastos2009}. This work has been extended to multiple-output problems~\citep{fricker2013}, dynamical problems~\citep{conti2009}, and non-deterministic (stochastic) simulators~\citep{oyebamiji2019}. These and other works~\citep{kennedy2001bayesian,oakley2017calibration} also outline how to perform Bayesian calibration, whilst Saltelli-style sensitivity indices~\citep{homma1996} have been developed into a fully-Bayesian sensitivity analysis and applied to GPEs to better understand input-output relations~\citep{farah2014}. Regarding its (very numerous) applications, GPE has been used in modelling mitochondrial DNA deletions~\citep{henderson2009}, influenza epidemic models~\citep{farah2014a}, and microbial communities~\citep{oyebamiji2019} to name just a few.\\
\noindent Dynamical GPE may be used to model time-dependent outputs of computer experiments~\citep{conti2009,farah2014,mohammadi2019,oyebamiji2019}. Such methods exploit the availability of the conditional form of the multivariate normal distribution to emulate the temporal structure of the data. This requires updating the values of outputs (state vector) at time $t_i$ using corresponding time-varying forcing inputs and the value of the state vector at $t_{i-1}$, as well as static parameters/initial conditions if any~\citep{conti2009}. We only have access to the initial conditions (ICs) of the experiments and are unable to use fully-dynamical emulation methods.\\
\noindent Instead, we approximate the FoS curves with a (1) single quadratic polynomial and (2) two piecewise quadratic polynomials, and emulate the resulting model/polynomial parameters by assigning Gaussian process prior distributions to them. This gives rise to a multi-output hierarchical GP model whereby we relate the parameters determining FoS curves with the ICs of the computer experiments. Therefore, the emulator that we developed uses the ICs of the slopes to predict the FoS behaviour over a time horizon. \\
\noindent We perform Bayesian inference on the unknown parameters using Markov chain Monte Carlo (MCMC). The resulting posterior distributions allow us to make FoS predictions and accurately quantify their uncertainty. This work builds on ~\citet{svalova2021} who modelled time to slope failure using the slope geometry and soil strength using Gaussian process emulation (GPE).  Bayesian emulation of computer experiments is highly cost-effective and allows full uncertainty analysis~\citep{ohagan2006}, which motivates us to use it for FoS modelling. Predictions for earthworks with specific geometries and materials can be obtained on-line and used in design, maintenance, and management of infrastructure. \\
\noindent The structure of this paper is as follows. Section~\ref{simulator_background} describes the computer experiments and the simulator or geotechnical model (GM), that simulates deterioration processes in cut slopes. In Section~\ref{Hierarchical Bayesian modelling of FoS time series} we describe hierarchical Bayesian modelling of FoS time series and introduce the quadratic model and B-spline model of FoS. In Section~\ref{Gaussian process emulator for the B-spline model} we propose the Gaussian process emulator for the quadratic and B-spline models, followed by outlining how we elicited the corresponding prior distributions. In Section~\ref{Results} we use the quadratic and B-spline emulators to model FoS. We obtain within-sample and out-of-sample posterior distributions of FoS and compare the performance of both models using the mean squared error $(MSE)$ and the continuous ranked probability score $(crps)$. We also show how the models can be used to predict time-to-failure of cut slopes. Section~\ref{Conclusions} concludes the paper and our findings and proposes areas for future research. Supplementary Material provides additional derivations and plots not included in the main manuscript.
\section{Computer experiments of cut slope deterioration} \label{simulator_background}
In this section we describe the computer experiments and the GM (our simulator) that simulates deterioration processes in cut slopes.
\subsection{The geotechnical model}
This subsection summarises the mechanisms that characterise the slope deterioration behaviour in the GM (our simulator). The GM is implemented within FLAC-TP (Fast Lagrangian Analysis of Continua with Two-Phase Flow ~\citep{FLAC}), which treats the soil as a porous medium with variable saturation and depth-dependent saturated permeability ~\citep{postill2021}. The pore fluids are separated into air and water phases and treated as immiscible fluids with differing density and viscosity. Water and airflow velocity is a function of the respective pore fluid pressures and the relative permeability of the soil. The latter is derived using the van Genuchten-Maulem relation~\citep{vangenuchten1980}.\\
\noindent The soil mechanical constitutive model uses a non-linear pressure dependent stiffness to control the bulk and shear moduli. A nonlocal Mohr-Coulomb strain softening model is used to describe the change in shear strength as a function of plastic shear strains, whereby the initial peak strength (effective peak cohesion $c'_p$ and friction angle $\phi'_p$) of the material reduces towards a minimum or residual value with increasing plastic strain.\\
\noindent The hydrological and mechanical properties were adopted from prior modelling studies \citep{potts1997,jurecic2013,tsiampousi2017, summersgill2018} and from field and laboratory data ~\citep{bromhead1986}. The effect of weather is included by application of a flux boundary condition where water is added and removed from the model upper boundary to replicate the surface water balance driven by precipitation and evapotranspiration. More details about the GM, including its validation and quality, can be found in~\citet{postill2020, rouainia2020,postill2021}, and~\citet{helm2024}.
\subsection{Initial conditions}
\noindent While over 40 model ICs can be varied~\citep{FLAC,rouainia2020, postill2021}, in this paper we only vary five of them. Namely, the slope height, $h_s$, slope angle $\nu$, derived from light detection and ranging (LiDAR) survey data provided by project stakeholders (Mott MacDonald and Network Rail), the peak shear strength parameters (effective peak cohesion $c'_p$ and friction angle $\phi'_p$) before the material has undergone any strength reduction, derived from previous laboratory data and modelling studies~\citep{apted1977, ellis2007}, and the reference coefficient of permeability of the soil at 1 metre depth with respect to water, $k_1^w$, derived from field measurements~\citep{dixon2019}. This simulation study specifically investigated high plasticity overconsolidated clay, as these were identified by relevant infrastructure asset owners as being amongst the most vulnerable to slope failures~\citep{GSRA}. The five ICs/variables were chosen as previous work has demonstrated their importance in assessing the deterioration and stability of infrastructure earthworks~\citep{potts1997,ellis2007,GSRA,rouainia2020,postill2023}. This allows us to limit the experimental time while obtaining an informative training data set. Additional parameters would increase computation time, which was around three months for 75 computer runs using 10 machines. To ensure an optimal coverage of the parameter space, a Latin hypercube experimental design is used to obtain 75 IC vectors (height, angle, peak cohesion, friction angle, and permeability). For more information on the adopted GM and the values adopted for the remaining ICs, see \citet{FLAC}, \citet{rouainia2020} and \citet{postill2021}.
\noindent Table \ref{Input_ranges} summarises ranges of the ICs used in the computer simulations which were selected based on previous studies ~\citep{rouainia2020, postill2021} and expert opinion from partners in the ACHILLES project. During emulation, permeability was scaled by $10^8$ to put the explanatory variables on a common scale.
\begin{table}[!htb]
\centering
\begin{tabular}{cccccc}
\hline
Property  & $h_s$ (m) & $\nu$ (degrees) & $c'_p$ (kPa) & $\phi'_p$ (degrees) & $\kappa^w_1$ (m/s) \\
  \hline
Range & $[4,20]$ & $[7.6,63.4]$ & $[3,10]$ & $[18.5,25]$& $[0.145, 2.5] \times 10^{-8}$\\
  \hline
\end{tabular}
\caption{Material and geometry ICs used in the computer experiments. The ranges were selected based on previous studies and expert opinion from our project partners.}
\label{Input_ranges}
\end{table}
\begin{figure}[!h]
	\centering
	\includegraphics[scale=0.6]{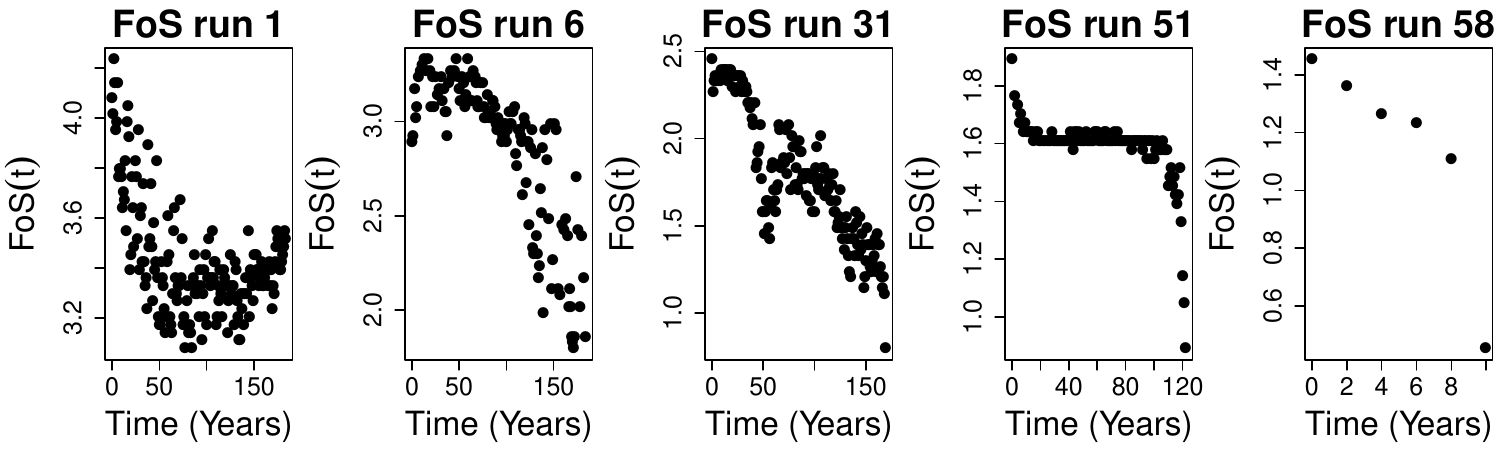}
	\caption{Examples of FoS time series illustrating different deterioration patterns.}
	\label{FoS}
\end{figure}
\subsection{Model run time}
The duration of the geotechnical modelling runs varied from approximately 75 minutes to 10 days. The latter depends on a number of factors, including model geometry, permeability, and the adopted strength parameters. The simulator runs were implemented until failure or 184 years of model time. Therefore, some computer runs provide  time series of FoS from slope initiation until slope failure and others are censored at 184 years.
\subsection{Computer experiments of FoS}
\noindent Figure~\ref{FoS} illustrates five computer runs of winter FoS which can be used to understand an earthwork's proximity to failure (a FoS of unity). Steeply-declining curves indicate rapid deterioration, whereas curves which have a plateau over a large number of years are indicative of earthworks undergoing minimal deterioration. The different behaviours are mainly driven by slope geometry, for example runs 1, 6, and 31 constitute short (in height) slopes with low angles, whereas the others are higher and steeper. The geology also influences deterioration behaviour, whereby cut slopes excavated within weaker materials will deteriorate more quickly and may also have a lower starting value of FoS. Finally, the dominant failure mechanism can change over time. For example, an initially deep seated rotational mechanism (base failure) may switch to a translational mechanism at a shallower depth within the slope (landslide) due to near-surface weathering and resultant near-surface strength reduction~\citep{briggs2023}. This would lead to a change in the FoS gradient and shape.
\noindent The time step in our geotechnical model can be between 10s to 1000s of seconds (here, it is capped at less than or equal to 3600 seconds) and the weather data is daily. However, the FoS analysis procedure is a time consuming iterative process. We are emulating yearly measurements in this work, as more frequent iterations would not be practical in terms of model run time. Instead, we use an estimate of the annual wettest and driest days, which broadly correspond to the highest and lowest values of annual FoS. In general, within a given year, the slopes have the lowest FoS and so are most vulnerable at the wettest time so the winter results are presented.\\
\noindent Despite the runs in Figure~\ref{FoS} appearing to have a stochastic component, the computer simulator is deterministic. The FoS behaviour is a function of the variability of the weather and the effects of antecedence and of the current state of deterioration. A rainfall or drying event will have a different effect on slope stability; this effect depends on the initial state of the slope prior to the event and the adopted permeability. Although each model has identical boundary conditions, due to the differences in permeability, geometry, and adopted strength properties, their responses will vary. Once models undergo significant strength reduction due to deterioration, their FoS is more strongly affected by seasonal cycles of wetting and drying, which increases the variability of annual FoS for slopes nearing ULS failure.\\
\noindent In what follows, FoS curves for a set of 75 computer experiment runs of earthwork (cutting) deterioration~\citep{svalova2021,postill2021,helm2024emulating} will be modelled as a function of slope height and angle, soil strength properties (peak cohesion, peak friction angle), and permeability. This work builds on~\citet{svalova2021} who modelled time to slope failure using Gaussian process emulation (GPE). 

\section{Hierarchical Bayesian modelling of FoS time series} \label{Hierarchical Bayesian modelling of FoS time series}
\noindent Hierarchical (multilevel) Bayesian modelling has been referred to as one of the most important statistical ideas over the past 50 years~\citep{gelman2021}, with the key idea of combining/pooling information about different categories of parameters for better inference. Here, hierarchical Bayesian modelling can be used to relate the temporal evolution of FoS to the static GM ICs. We will assume that the FoS time series can be approximated by a function $g$, the parameters of which will be emulated/related to simulator ICs.\\
\noindent To simplify some of the methodology, we will be modelling FoS measurements shifted by a unit, $Y=\,\,$FoS$\,-1$, such that failure is defined by reaching a (shifted) FoS of zero. For $i\in\{1,\ldots,N\}$ and $j\in\{1,\ldots,N_i\}$, where $N$ is the number of computer runs and $N_i$ is the number of FoS measurements in run $i$, our hierarchical Bayesian FoS model is:
\begin{equation}
\begin{cases}
\text{FoS}_{i,j}=1+g(t_{i,j}; \bm{\theta}_{g,i})+\varepsilon_{i,j},\quad\varepsilon_{i,j}\sim\text{N}(0,\sigma^2_i),\\
\text{Y}_{i,j}=g(t_{i,j}; \bm{\theta}_{g,i})+\varepsilon_{i,j},\\
\bm{\theta}_{g,i}\sim\text{GP}(m(\bm{x};\bm{\phi}_1),V(\bm{x},\bm{x}'; \bm{\phi}_2)),\\
\bm{\phi}_1\sim\Phi_1,\quad\bm{\phi}_2\sim\Phi_2.
\end{cases}
\label{hierarchical_model_g}
\end{equation}
\noindent In the above, the top level $g+\varepsilon_{i,j}$ is the model for FoS temporal evolution, and $t_{i,j}$ denotes time in years. We will compare two $g$ parametrisations, a single and a piecewise quadratic polynomial in the form of a B-spline. The parameters of $g$, $\bm{\theta}_{g}=(\bm{\theta}_{g,1},\ldots,\bm{\theta}_{g,N})^T$, are assigned GP prior distributions defined by mean and variance functions $m$ and $V$. The mean function $m$ is a function of the ICs $\bm{x}$ and hyperparameters $\bm{\phi}_1$; and the covariance function is a function of a pair of ICs $(\bm{x}, \bm{x}')$, such that $V_{i,j} = \text{Cov}(\bm{x}_i, \bm{x}_j)$, and hyperparameters $\bm{\phi}_2$. Finally, the hyperparameters (vectors) $\bm{\phi}_1$ and $\bm{\phi}_2$ will be assigned some appropriate prior distributions $\Phi_1$ and $\Phi_2$. The rest of this Section details the parameterisation of FoS time series using a quadratic polynomial curve and a quadratic B-spline with two pieces.\\
\noindent We note that both models are trained on FoS time series with at least 4 measurements, otherwise the B-spline model is not identifiable. This limits the use of the developed emulator for predicting very sudden failures; however infrastructure earthworks are typically built to last at least several decades~\citep{perry2003}, thus this should not be a problem in practice. 
\noindent Furthermore, we define three types of time-to-failure (TTF). Namely, the model TTF $(\omega)$, the predicted TTF $(\rho)$, and the true TTF. For computer run $i$, model TTF $(\omega_i)$ is defined as the minimum time such that $g(t)\leq0$, and predicted TTF $(\rho_i)$ is defined as the minimum time such that $g(t) + \varepsilon_i\leq0$. The true TTF is taken from the data. Some of our computer runs do not reach failure and, for those, the true TTF is not available.
\subsection{Quadratic model of FoS}
\noindent For a computer run $i$ at time $j$, assume that $Y_{i,j}\in\mathbb{R}$ changes with time $t_{i,j}\in\mathbb{R}^+$ through a quadratic relation
\begin{equation}
Y_{i,j}=\alpha_{0,i}+\alpha_{1,i}t_{i,j}+\alpha_{2,i}t_{i,j}^2+\varepsilon_{i,j}, \quad \varepsilon_{i,j}\sim\text{N}(0,\sigma_i^2),\label{basic}
\end{equation}
\noindent where $a_{0,i}+1$ is the FoS at time zero and $a_{1,i}$ is the FoS gradient at time zero. Equation~\eqref{basic} can be re-parametrised to include the \emph{model} TTF for computer run $i$, $\omega_i$:
\begin{equation}
Y_{i,j}=\alpha_{0,i}+\alpha_{1,i}t_{i,j}-\frac{\alpha_{1,i}+\alpha_{0,i}/\omega_i}{\omega_i}t_{i,j}^2+\varepsilon_{i,j}.\label{final}
\end{equation}
\noindent It is useful to rewrite Equation \eqref{final} as a quadratic B-spline~\citep{deboor1978,knott2000} with no interior knots. In this form, the quadratic model can be directly compared to the B-spline FoS model described in Section~\ref{B-spline model of FoS}. Therefore, 
\begin{equation}
\begin{aligned}
Y_{i,j} &= g_1(t_{i,j}; \bm{\theta}_{g_1,i}) + \varepsilon_{i,j}, \\
g_1(t_{i,j}; \bm{\theta}_{g_1,i})  &= \bigg\{\gamma_{0,i} + t_{i,j}\bigg(\frac{2 \gamma_{1,i}}{\omega_i}-\frac{2 \gamma_{0,i}}{\omega_i}\bigg)+ t_{i,j}^2\bigg(\frac{\gamma_{0,i}}{\omega_i^2}-\frac{2 \gamma_{1,i}}{\omega_i^2}\bigg)\bigg\}\text{I}_{t_{i,j}}(0;\omega_i),
\label{final2}
\end{aligned}
\end{equation}
\noindent where $\bm{\theta}_{h_1,i} = (\gamma_{0,i}, \gamma_{1,i}, \omega_i)$, 
\begin{equation}
\text{I}_{t_{i,j}}(0;\omega_i)=
\begin{cases}\nonumber
1, \quad 0 \leq t_{i,j} < \omega_i,\\
0, \quad \text{otherwise},
\end{cases}
\end{equation}
\noindent and $\gamma_{0,i} = \alpha_{0,i}$ and $\gamma_{1,i} = (\alpha_{1,i}\omega_i + 2 \alpha_{0,i})/2$. Here, $\gamma_{0,i}$ is the intercept of $Y_{i,j}$ at $t_{i,j}=0$, and we require $\gamma_{0,i}\geq 0$ to ensure non-negative $Y_{i,j}$ at $t_{i,j} = 0$. Thus, we can write $\gamma_{0,i}=\,$exp(A$_{0,i})$. Similarly, we set the failure time parameter $\omega_i=\,$exp$(\Omega_i)$ and $\sigma_i=\,$exp$(\Sigma_i)$. 
\subsubsection{Constraining the quadratic model}
\begin{figure}[!h]
	\centering
	\includegraphics[scale=0.4]{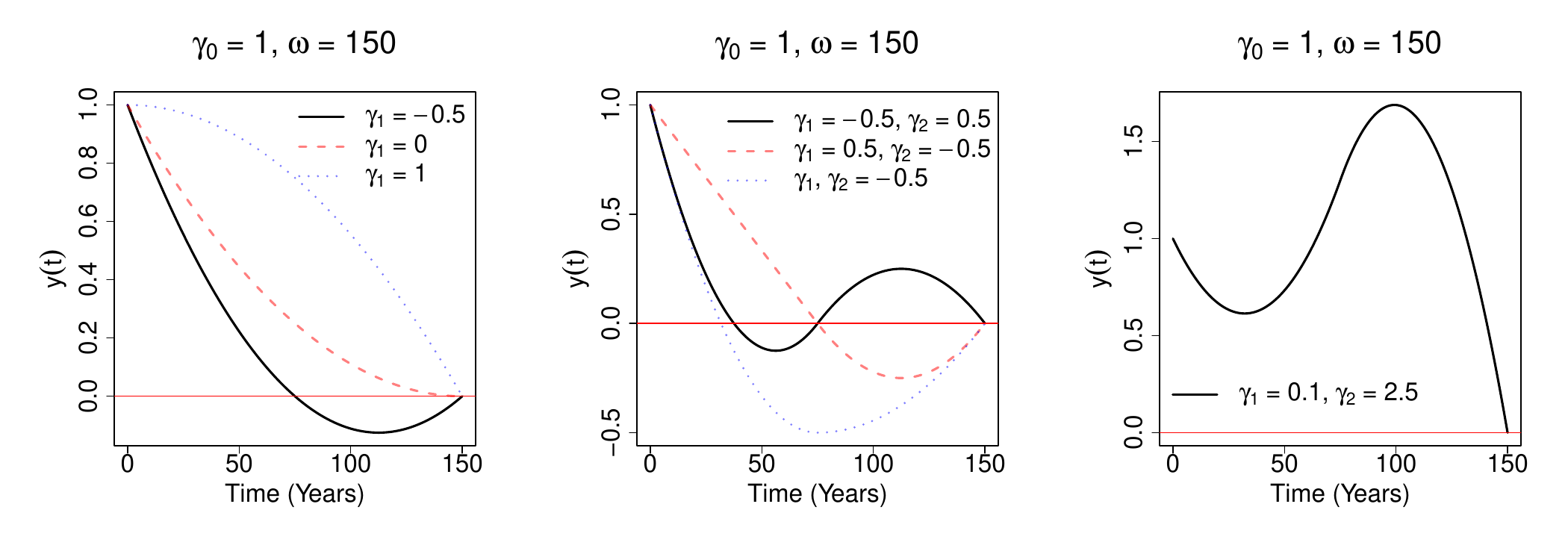}
	\caption{Constraining the quadratic and B-spline models. Plot (left) illustrates deterministic (noise-free) FoS curves when $\gamma_{1,i}<0$. Plot (middle) presents B-splines with $\gamma_{1,i}<0, \gamma_{2,i}<0$ and plot (right) illustrates an unrealistic B-spline with $\gamma_{1,i} \geq 0, \gamma_{2,i} \geq 0$.}
	\label{constraints}
\end{figure}
\noindent The model TTF, $\omega_i$, is the smallest positive time that solves the deterministic part of Equation \eqref{final2}. This can be achieved by constraining $\gamma_{1,i}>0$. Figure \ref{constraints} (left) illustrates deterministic (noise-free) FoS curves when $\gamma_{1,i}$ is unconstrained. From Figure \ref{constraints} (left) we can see that the FoS curves can cross the failure-threshold prior to $\omega_i$ when $\gamma_{1,i}<0$. Thus, we write $\gamma_{1,i}=\,$exp($\text{A}_{1,i})$. The parameter set, to be emulated, of the quadratic model defined by Equation \eqref{final2}, is $\bm{\theta}_{g_1,i}= (\text{A}_{0,i}, \text{A}_{1,i}, \Omega_i, \Sigma_i)$.

\subsection{B-spline model of FoS} \label{B-spline model of FoS}
\noindent The geometry and location of the critical failure mechanism in a slope may change over time as deterioration progresses~\citep{postill2021}. This may lead to a sudden change in the gradient of the FoS deterioration curve. Thus, a quadratic curve might not be sufficient in explaining such variation, especially as the earthwork failure time increases. A piecewise polynomial could be more advantageous over a simple polynomial, whereby local curves are fitted within non-overlapping regions. One such model is the B-spline~\citep{deboor1978,knott2000}, which is a piecewise polynomial defined within a set of knots/breakpoints that is forced to be smooth and continuous at these knots. B-splines are commonly used in shape-optimisation methods~\citep{talebitooti2015}, and can have a desired polynomial degree, which has to be the same for all piecewise curves (and greater than zero). A quadratic B-spline with no inner knots is equivalent to a simple quadratic (Equation~\eqref{final2}).\\
\noindent We use a B-spline to fit two piecewise quadratic polynomials to the FoS time series with one interior knot at half-failure time, $\omega_i/2$. This is a natural extension to the quadratic model in~\eqref{final2} as we are fitting two polynomials corresponding to a potential change in the slope failure mechanism, instead of one. What follows provides a brief definition of the B-spline, the full analytical definition can be found in Supplementary Material A.\\
\noindent The B-spline approximation of $Y_{i,j}$ is a linear combination of B-spline basis functions,
\begin{equation}
Y_{i,j}=\sum_{l=0}^{m+o-1}\gamma_{l,i}\phi_{l,o-1}(t_{i,j})+\varepsilon_{i,j}=\sum_{l=0}^{3}\gamma_{l,i}\phi_{l,2}(t_{i,j})+\varepsilon_{i,j}, \quad\varepsilon_{i,j}\sim \text{N}(0,\sigma^2_i),
\label{bspline}
\end{equation}
where $\phi_{l,2}(t_{i,j})$ are basis functions and $\gamma_{l,i}$ are linear coefficients.
Following De Boor recursive relationships~\citep[][Supplementary Material A]{deboor1978}, the spline model can be rewritten as a sum of two quadratic functions over non-overlapping intervals,
\begin{equation}
\begin{aligned}
Y_{i,j}&= g_2(t_{i,j}; \bm{\theta}_{g_2,i}) + \varepsilon_{i,j}, \\
& g_2(t_{i,j}; \bm{\theta}_{g_2,i})= \bigg\{\gamma_{0,i} + t_{i,j}\bigg(\frac{4 \gamma_{1,i}}{\omega_i}-\frac{4 \gamma_{0,i}}{\omega_i}\bigg)+ t_{i,j}^2\bigg(\frac{4 \gamma_{0,i}}{\omega_i^2}-\frac{6 \gamma_{1,i}}{\omega_i^2}+\frac{2 \gamma_{2,i}}{\omega_i^2}\bigg) \bigg\} \text{I}_{t_{i,j}}(0;\omega_i/2)+\\
&\bigg\{2(\gamma_{1,i} -\gamma_{2,i}+\gamma_{3,i})+t_{i,j}\bigg(\frac{8 \gamma_{2,i}}{\omega_i}-\frac{4 \gamma_{1,i}}{\omega_i}-\frac{4 \gamma_{3,i}}{\omega_i}\bigg) +
t_{i,j}^2\bigg(\frac{2 \gamma_{1,i}}{\omega_i^2}-\frac{6 \gamma_{2,i}}{\omega_i^2}+\frac{4 \gamma_{3,i}}{\omega_i^2}\bigg)\bigg\} \text{I}_{t_{i,j}}(\omega_i/2 ;\omega_i)\\
\label{B-spline_expanded_full}
\end{aligned}
\end{equation}
\noindent where $\bm{\theta}_{g_2,i} = (\gamma_{0,i}, \gamma_{1,i}, \gamma_{2,i}, \gamma_{3,i}, \omega_i)$, 
\begin{equation}
\text{I}_{t_{i,j} }(a;b)=
\begin{cases}\nonumber
1,\quad\text{if }a \leq t_{i,j} < b,\\
0,\quad\text{otherwise},
\end{cases}
\end{equation}
\noindent and $\varepsilon_{i,j}\sim \text{N}(0,\sigma^2_i)$. Removing $\gamma_{3,i}$ from Equation \eqref{B-spline_expanded_full} forces the B-spline curve to cross zero at the last knot, which is fitting for modelling the FoS. Then, our B-spline model becomes
\begin{equation}
\begin{aligned}
Y_{i,j} &= g_2(t_{i,j}; \bm{\theta}_{g_2,i}) + \varepsilon_{i,j}, \\
& g_2(t_{i,j}; \bm{\theta}_{g_2,i}) = \bigg\{\gamma_{0,i} + t_{i,j}\bigg(\frac{4 \gamma_{1,i}}{\omega_i}-\frac{4 \gamma_{0,i}}{\omega_i}\bigg)+ t_{i,j}^2\bigg(\frac{4 \gamma_{0,i}}{\omega_i^2}-\frac{6 \gamma_{1,i}}{\omega_i^2}+\frac{2 \gamma_{2,i}}{\omega_i^2}\bigg) \bigg\} \text{I}_{t_{i,j}}(0;\omega_i/2)+\\
&\bigg\{2(\gamma_{1,i} -\gamma_{2,i})+t_{i,j}\bigg(\frac{8 \gamma_{2,i}}{\omega_i}-\frac{4 \gamma_{1,i}}{\omega_i}\bigg) +t_{i,j}^2\bigg(\frac{2 \gamma_{1,i}}{\omega_i^2}-\frac{6 \gamma_{2,i}}{\omega_i^2}\bigg)\bigg\} \text{I}_{t_{i,j}}(\omega_i/2 ;\omega_i).
\end{aligned}
\label{B-spline_expanded}
\end{equation}
\noindent where $\bm{\theta}_{g_2,i} = (\gamma_{0,i}, \gamma_{1,i}, \gamma_{2,i},  \omega_i)$. As before, $\gamma_{0,i}$ is the intercept of $Y_{i,j}$ at $t_{i,j}=0$ and $\gamma_{0,i}\geq 0$ to ensure non-negative $Y_{i,j}$ at $t_{i,j} = 0$. We define $\gamma_{0,i}=\,$exp(A$_{0,i})$ as with the quadratic model. Similarly, $\omega_i=\,$exp$(\Omega_i)$ and $\sigma_i=\,$exp$(\Sigma_i)$.\\
\noindent The B-spline model has an additional coefficient $\gamma_{2,i}=\,$exp(A$_{2,i}$), that arises due to the second polynomial piece. Its interpretation is difficult, but we can get an intuition into its effect on the B-spline graphically (Section~\ref{sec:BsplineConstraint}). 
\subsubsection{Constraining the B-spline model}\label{sec:BsplineConstraint}
\noindent The model TTF $\omega_i$ is the smallest positive time that solves the deterministic part of \eqref{B-spline_expanded}. Figure \ref{constraints} (middle) illustrates deterministic (noise-free) FoS curves when $\gamma_{1,i}$ and $\gamma_{2,i}$ are unconstrained. From Figure \ref{constraints} (middle) we can see that the FoS curves can cross zero (failure threshold) prior to the model TTF. This behaviour can be avoided by constraining $\gamma_{1,i}, \gamma_{2,i}>0$, thus we set $\gamma_{1,i}=\,$exp($\text{A}_{1,i})$ and $\gamma_{2,i}=\,$exp(A$_{2,i})$.

\noindent Figure \ref{constraints} (right), however, is not plausible for modelling FoS as the peak of the second polynomial is higher than the initial FoS, $\gamma_{0,i}$. Indeed, Figure \ref{constraints} (right) would imply a very dramatic increase in soil strength after an initial decrease. This would typically only occur if there were significant changes made to the slope/soil such as geotechnical intervention measures, which may take the form of the installation of soil nails, piles or retaining structures or replacement of the soil with a higher strength material  \cite[see][]{nowak2012}. In the absence of such interventions this large increase in FoS should not occur. Additionally, allowing this behaviour can cause difficulties in estimating the the B-spline coefficients and using the resulting model for prediction. Therefore, we impose an additional constraint, that the FoS time series is nonincreasing after the internal knot, i.e. $Y_{i,j}(t_2) \leq Y_{i,j}(t_1)$ for $t_1\leq\omega/2 \leq t_2$. This constraint eliminates FoS curves that are not plausible while still allowing us to adequately model the FoS computer experiment data. The parameter set, to be emulated, of the B-spline model, defined by Equation \eqref{B-spline_expanded}, is $\bm{\theta}_{g_2,i} = (\text{A}_{0,i}, \text{A}_{1,i}, \text{A}_{2,i}, \Omega_i, \Sigma_i)$.
\section{Gaussian process emulator} \label{Gaussian process emulator for the B-spline model}
\noindent In this section, we define a Bayesian GPE for the B-spline model parameters, $\bm{\theta}_{g_2,i} = (\text{A}_{0,i}, \text{A}_{1,i}, \text{A}_{2,i}, \Omega_i, \Sigma_i)$, (see Equation~\eqref{B-spline_expanded}). Let $\textbf{A}_{0}=(\text{A}_{0,1},\ldots,\text{A}_{0,N})$, $\textbf{A}_{1}=(\text{A}_{1,1},\ldots,\text{A}_{1,N})$, $\textbf{A}_{2}=(\text{A}_{2,1},\ldots,\text{A}_{2,N})$, $\bm{\Omega}=(\Omega_{1,1},\ldots,\Omega_{1,N})$  and $\bm{\Sigma}=(\Sigma_{1,1},\ldots,\Sigma_{1,N})$. The Gaussian process emulator for the quadratic model is defined identically to the GPE for the B-spline model defined in this section, excluding the parameter vector $\text{\textbf{A}}_2$ and the associated hyperparameters $\bm{\beta}_2$ and $\tau_2$. For a computer run $i$, $\bm{x}_i=(x_{i,1},\ldots,x_{i,5})^T$ is a set of ICs corresponding to height $x_1$ (m), angle $x_2$ (degrees), cohesion $x_3$ (kPa), friction angle $x_4$ (degrees), and permeability $x_5$ (m/s). We define $\bm{z}_i=(z_{i,1},\ldots,z_{i,5})^T$, where 
\begin{equation}
z_{i,k} = \frac{x_{i,k} -\bm{\bar{x}}_{\text{training},k}}{\sigma_{\text{training},k}} \text{ for } k\in\{1,3,4,5\},\text{ and }z_{i,2} = \frac{x_{i,2} -\bm{\bar{x}}_{\text{training},2}}{1.5\times\sigma_{\text{training},2}}, 
\end{equation}
\noindent where $\bm{x}_{\text{training},k} = (x_{1,k}, \dots x_{N,k})$, $N=75$ is the number of training samples and $\bm{\bar{x}}_{\text{training},k}$ and $\sigma_{\text{training},k}$ correspond to the mean and standard deviation of $\bm{x}_{\text{training},k}$, respectively. This standardisation puts all ICs on the same scale. For a computer run $i$, the FoS parameters are related to $\bm{z}_i$ following a normal distribution
\begin{equation}
\begin{pmatrix}
\text{A}_{0,i}\\
\text{A}_{1,i}\\
\text{A}_{2,i}\\
\Omega_i\\
\Sigma_i
\end{pmatrix}
\sim N \left(
\begin{bmatrix}
h(\bm{z}_i)^T\bm{\beta}_0\\
h(\bm{z}_i)^T\bm{\beta}_1\\
h(\bm{z}_i)^T\bm{\beta}_2\\
h(\bm{z}_i)^T\bm{\beta}_{\Omega}\\
h(\bm{z}_i)^T\bm{\beta}_{\Sigma}\\
\end{bmatrix},\begin{bmatrix}
\tau_0&0&0&0&0\\
0&\tau_1&0&0&0\\
0&0&\tau_2&0&0\\
0&0&0&\tau_{\Omega}&0\\
0&0&0&0&\tau_{\Sigma}\\
\end{bmatrix}
\right).
\label{SCR}
\end{equation}
\noindent The parameter vectors $\bm{\beta}_l= (\beta_{l,0},\ldots,\beta_{l,5})^T, l\in\{0,1,2,\Omega,\Sigma\}$ are regressor coefficients and $h(\bm{z}_i)=(1,z_{i,1},\ldots,z_{i,5})^T$ is a regressor function of the ICs. The variance matrix in~\eqref{SCR} is diagonal with marginal variances $\tau_k$ and zero prior correlation between $\text{A}_{0,i}, \text{A}_{1,i}, \text{A}_{2,i}, \Omega_i$, and $\Sigma_i$, which therefore can be expressed as individual normal distributions. Transforming the data from $\bm{x}_i$ to $\bm{z}_i$ makes the B-spline parameters ($\text{A}_{0,i}, \text{A}_{1,i}, \text{A}_{2,i}, \Omega_i$ and $\Sigma_i$) easier to interpret. For example, when $\bm{z}_i = \bm{0}$, $\exp(\beta_{\Omega,0})$ can now be interpreted as the expected failure time of a slope with ``average'' characteristics. On the other hand, low or zero values for some of the (untransformed) ICs have a non-linear relationship with deterioration, for example, zero cohesion would imply a slope buillt with ``dry sand'' which becomes difficult to model and interpret. 
\noindent Jointly, \textbf{A}\textsubscript{0},  \textbf{A}\textsubscript{1}, \textbf{A}\textsubscript{2}, $\bm{\Omega}$, and $\bm{\Sigma}$ follow a matrix-normal distribution,
\begin{eqnarray}
&\text{vec}(\text{\textbf{A}}_0,\text{\textbf{A}}_1,\text{\textbf{A}}_2,\bm{\Omega},\bm{\Sigma})\sim \text{N} \left(
\text{vec}(H\bm{\beta}_0, H\bm{\beta}_1, H\bm{\beta}_2,H\bm{\beta}_{\Omega},H\bm{\beta}_{\Sigma}),\bm{V}\otimes\bm{U}\right),\\
&\bm{U} = C(\bm{x},\bm{x}',\bm{\delta},\zeta),\quad \bm{V}= \text{diag}(\tau_0, \tau_1, \tau_2, \tau_\Omega, \tau_\Sigma).
\nonumber
\end{eqnarray}
\noindent In the above, $H$ is the regressor matrix, whose $i$\textsuperscript{th} row is $h(\bm{z}_i)$, $\bm{U}$ and $\bm{V}$ are the row- and column-wise covariance matrices, respectively. We use a non-isotropic Gaussian correlation function $C(\bm{z},\bm{z}',\bm{\delta}, \zeta) =\exp\left\{-\sum_{i=1}^5\frac{(z_i-z_i')^2}{\delta_i^2}\right\} +  \zeta\mathbb{I}(\bm{z},\bm{z}')$, where $\bm{\delta}=(\delta_1,\ldots,\delta_5)^T$ is a vector of correlation lengths. The nugget~\citep{andrianakis2012} $\zeta$ is added to the diagonal of $C$ to improve numerical stability of inversion calculations, and the function $\mathbb{I}(\bm{z},\bm{z}')$ is an indicator for the event $\bm{z}=\bm{z}'$. Note that as \textbf{A}\textsubscript{0}, \textbf{A}\textsubscript{1},  \textbf{A}\textsubscript{2}, $\bm{\Omega}$ and $\bm{\Sigma}$ are uncorrelated a priori, their joint prior distribution can be written as a product of the following marginal distributions
\begin{eqnarray}
&\text{\textbf{A}}_0\sim \text{N}\left(H\bm{\beta}_0,\tau_0\bm{U}\right),\quad\text{\textbf{A}}_1\sim \text{N}\left(H\bm{\beta}_1,\tau_1\bm{U}\right), \quad \text{\textbf{A}}_2\sim \text{N}\left(H\bm{\beta}_2,\tau_2\bm{U}\right),\label{secondlevel}\\
&\bm{\Omega}\sim \text{N}\left(H\bm{\beta}_{\Omega},\tau_{\Omega}\bm{U}\right), \quad \bm{\Sigma}\sim \text{N}\left(H\bm{\beta}_{\Sigma},\tau_{\Sigma}\bm{U}\right).\nonumber
\end{eqnarray}
\noindent Concerning the hierarchical model given by Equation \eqref{hierarchical_model_g}, we have defined two models for FoS temporal evolution; the quadratic model, defined by Equation \eqref{final2}, and the B-spline model, defined by Equation \eqref{B-spline_expanded}. Furthermore, Equation~\eqref{secondlevel} completes the second level of the hierarchical model, assigning GP priors to the temporal model parameters. Removing $\text{\textbf{A}}_2$ from Equation~\eqref{secondlevel}  provides the GPE for the quadratic model. In Sections \ref{Priors_B_spline} and \ref{Priors_quadratic} we complete the hierarchical model defined by Equation \eqref{hierarchical_model_g} by assigning prior distributions to the hyperparameters of the B-spline and quadratic model parameters, respectively.

\subsection{Prior distributions for the B-spline model} \label{Priors_B_spline}
\noindent This section describes how we elicited prior distributions for the  B-spline models, using engineering judgement where applicable. The IC $\bm{z} = \bm{0}$ corresponds to a slope with $x_1=9.77$ (height), $x_2=20$ (angle), $x_3=6.81$ (cohesion), $x_4=21.7$ (friction) and $x_5=0.77$ (permeability). Using Equation \eqref{SCR}, $\bm{z} = \bm{0}$ corresponds to a slope with ``average'' characteristics of our training data with the following prior distributions
\begin{eqnarray}
&\text{A}_{0}(\bm{0}) \sim \text{N}(\beta_{0,0},\tau_0),\quad
\text{A}_{1}(\bm{0}) \sim \text{N}(\beta_{1,0},\tau_1), \quad 
\text{A}_{2}(\bm{0}) \sim \text{N}(\beta_{2,0},\tau_2), \label{eq_z0}\\ 
&\Omega(\bm{0})  \sim \text{N}(\beta_{\omega,0},\tau_\omega), \quad
\Sigma(\bm{0}) \sim \text{N}(\beta_{\Sigma,0},\tau_\Sigma). \nonumber
\end{eqnarray}
\noindent The means of distributions in Equation~\eqref{eq_z0} reduce to intercept-only instead of the full regressor function form. These $\beta_{l,0}$, $l\in\{0,1,2,\Omega,\Sigma\}$, represent our uncertainty about the initial FoS averaged across $\bm{x}$. By generating synthetic data and ensuring reasonable relations between the ICs and the resulting FoS behaviour in the prior predictive distributions, we elicited the following prior distributions 
\begin{eqnarray}
&\beta_{0,0} \sim \text{N}(\text{log}(1), 0.5), \quad
\beta_{1,0} \sim \text{N}(\text{log}(0.6), 0.4), \quad
\beta_{2,0} \sim \text{N}(-0.5, 2.5), \label{beta_0}\\
&\beta_{\omega,0} \sim \text{N}(5.25, 1), \quad
\beta_{\Sigma,0} \sim \text{N}(\text{log}(0.1),0.5).\nonumber
\end{eqnarray}
In addition,
\begin{equation}
\beta_{0,i} \sim \text{N}(0, 0.5), \quad
\beta_{1,i} \sim \text{N}(0, 0.5), \quad
\beta_{2,i} \sim \text{N}(0, 1), \quad
\beta_{\omega,i} \sim \text{N}(0, 1), \quad
\beta_{\Sigma_i} \sim \text{N}(0,0.5),
\label{beta_i}
\end{equation}
\noindent for $i = 1,\dots,5$. Table \ref{Expectations_prior} presents prior mean and central $95\%$ credible intervals for the expected initial FoS, model failure time and noise using the priors given by Equation \eqref{beta_0}. 
The shape parameters $\text{A}_1$ and $\text{A}_2$ have no intuitive interpretation and are not included in Table \ref{Expectations_prior}. From Table \ref{Expectations_prior}, we see that a priori we expect the initial FoS for a slope with $\bm{z} = \bm{0}$ to be between 1.38 and 3.66, $95\%$ of the time with a mean of 2; $\omega_i$ is expected to be between 26.8 days and 1350 days, $95\%$ of the time with a mean of 191 days; and we expect the noise to be between 0.0375 and 0.266, $95\%$ of the time with a mean of 0.1. These priors are vaguely informative and represent our uncertainty about the parameters of the B-spline model for a slope with $\bm{z} = 0$ and put prior weighting in regions of the parameter space that produce plausible synthetic data.

\begin{table}[!htb]
\centering
\begin{tabular}{rrr}
\hline
Prior & Mean & Central $95\%$ Credible Interval\\
  \hline
$\text{E}[\gamma_0] + 1$&2.00& $(1.38, 3.66)$\\
$\text{E}[\omega]$& 191.0  & $(26.8, 1350.0)$\\
$\text{E}[\sigma]$&  0.1000 & $(0.0375, 0.2660)$\\
  \hline
\end{tabular}
\caption{Prior mean and central $95\%$ credible intervals for $\text{E}[\gamma_0] + 1$, $\text{E}[\omega]$ and $\text{E}[\sigma]$.}
\label{Expectations_prior}
\end{table}

\begin{figure}[H]
	\centering
	\includegraphics[scale=0.5]{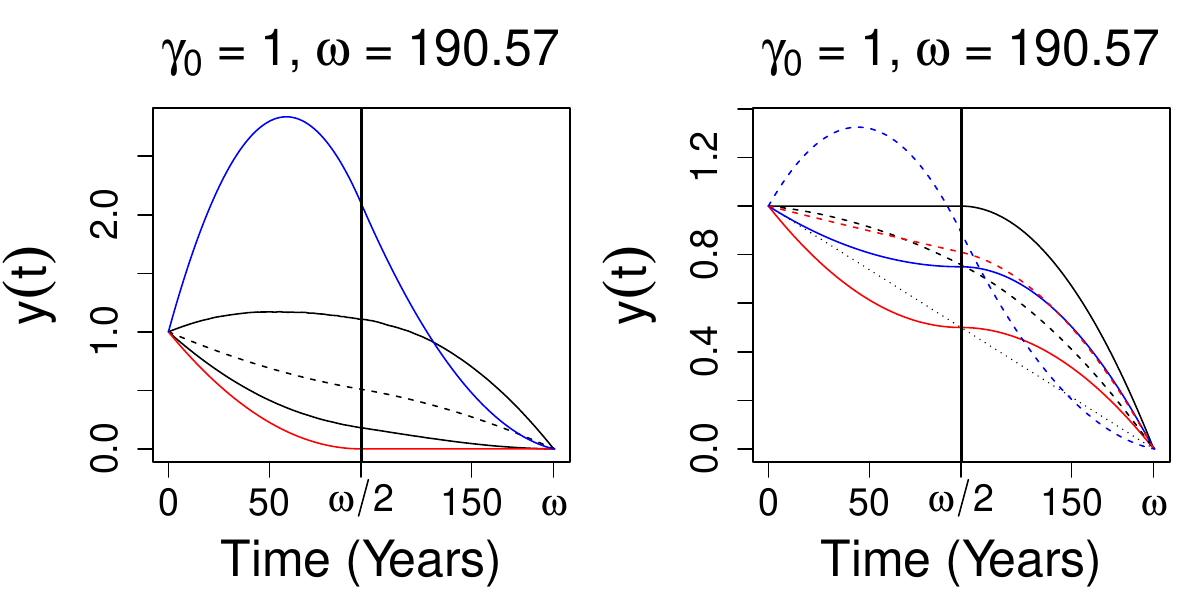}
	\caption{Some example prior deterioration curves with $\gamma_0 = 1,\Omega=5.25$ sampling from $\beta_{1,0}$ and $\beta_{2,0}$ in Equation \eqref{beta_0}.}
	\label{noise_free_prior_shapes}
\end{figure}

\noindent Similarly to ~\citet{svalova2021}, we selected vague prior distributions for the covariance function parameters. Specifically, 
\begin{equation}
\tau_l \sim \text{IGa}(3,0.5), \quad
\delta_k \sim \text{Exp}(0.2),
\end{equation}
\noindent for $l \in\{0,1,2,\Omega,\Sigma\}$ and $k=1,\dots,5$. Figure \ref{noise_free_prior_shapes} illustrates some example prior deterministic deterioration curves with $\gamma_0 = 1$ (the prior mean for a slope with $\bm{z}=\bm{0}$) and $\Omega=5.25$ (the prior mean for a slope with $\bm{z}=\bm{0}$). Figure \ref{noise_free_prior_shapes} (left) illustrates the mean deterministic deterioration curve (dashed line) and the central $95\%$ prediction interval (solid lines) when sampling from $\beta_{1,0}$ and $\beta_{2,0}$ in Equation \eqref{beta_0}; the red line corresponds to $\gamma_1, \gamma_2 = 0$; and the evolution depicted by the blue line is a B-spline curve we believe is not plausible. From Figure \ref{noise_free_prior_shapes} (left) we can see that the priors specified on $\beta_{1,0}$ and $\beta_{2,0}$ are vague. These priors represent our uncertainty in the evolution of the FoS of the slope with $\bm{z} = \bm{0}$. It is difficult to specify how the FoS of a slope with $\bm{z} = \bm{0}$ will evolve over time. However, there is little weighting between the 0.025 quantile and the red line, as we do not expect curves to deteriorate rapidly (up to the knot at $\omega/2$, shown by the vertical black line) and then asymptotically deteriorate to zero. Through simulation of synthetic data we were able to elicit priors that put little weight on implausible curves, but there is no published data on how the FoS of a slope with $\bm{z} = \bm{0}$ will evolve over time.\\ 
\noindent Figure \ref{noise_free_prior_shapes} (right) presents some example deterministic FoS time series sampled from $\beta_{1,0}$ and $\beta_{2,0}$ in Equation \eqref{beta_0}. The priors specified on $\beta_{1,0}$ and $\beta_{2,0}$ allow for a variety of time series of slope deterioration that we see in reality. The priors for $\beta_{1,0}$ and $\beta_{2,0}$ govern the evolution of deterioration for a slope with $\bm{z}=\bm{0}$; and these priors put little weighting on deterioration curves that we do not expect to see in reality (such as the red and blue curves in Figure \ref{noise_free_prior_shapes} (left)), but it is difficult to be more specific than that. 

\noindent The remaining parameters, i.e., $\beta_{l,i}$, for $l \in\{0,1,2,\Omega,\Sigma\}$ and $i=1,\dots,5$, were given vaguely informative priors centered around zero, as can be seen by Equation \eqref{beta_i}. 

\noindent The posterior distributions for the GPE parameters are intractable and will be sampled using Markov chain methods.
\subsection{Prior distributions for the quadratic model} \label{Priors_quadratic}

\begin{figure}[!h]
	\centering
	\includegraphics[scale=0.5]{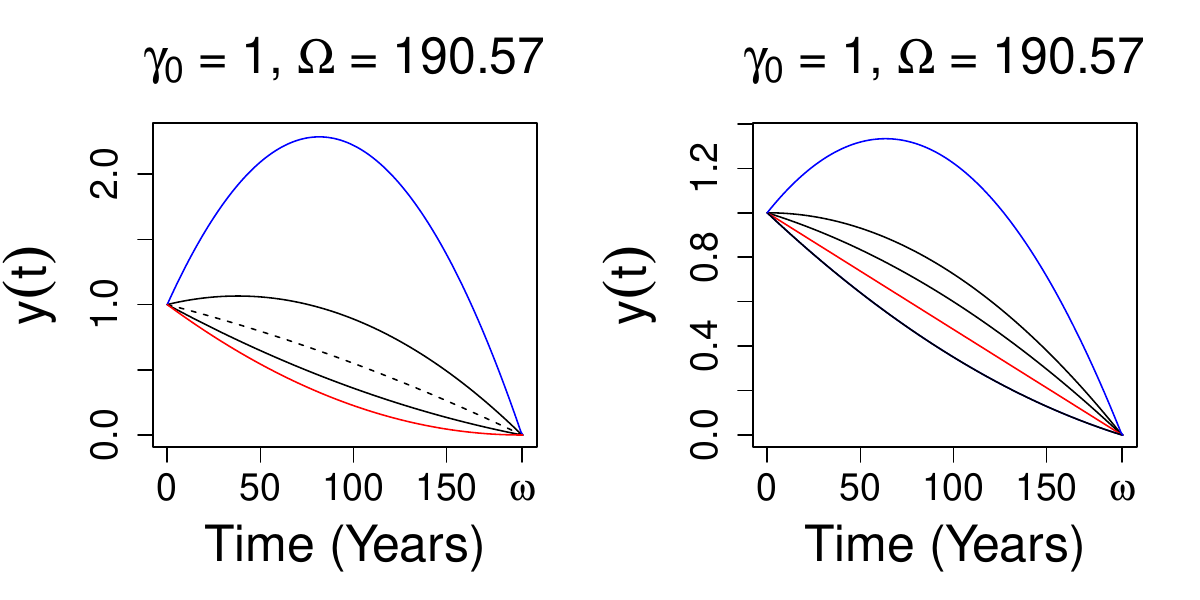}
	\caption{Example prior deterioration curves with $\gamma_0 = 1,\Omega=5.25$ sampling from $\beta_{1,0}$ in Equation \eqref{beta_0}.}
	\label{noise_free_prior_shapes_quadratic}
\end{figure}

The prior distributions for the Gaussian process emulator for the quadratic model are identical to those used under the B-spline model. Once again, the posterior distributions for the GPE parameters are intractable and will be sampled using Markov chain methods. \\
\noindent Figure \ref{noise_free_prior_shapes_quadratic} presents some example prior deterministic deterioration curves with $\gamma_0 = 1$ (the prior mean for a slope with $\bm{z}=\bm{0}$) and $\Omega=5.25$ (the prior mean for a slope with $\bm{z}=\bm{0}$) . Figure \ref{noise_free_prior_shapes_quadratic} (left) presents the mean deterministic deterioration curve (dashed line) and the central $95\%$ prediction interval (solid lines) when sampling from $\beta_{1,0}$ in Equation \eqref{beta_0}; the red line corresponds to $\gamma_1 = 0$; and the evolution depicted by the blue line is a quadratic curve we believe is not plausible. From Figure \ref{noise_free_prior_shapes_quadratic} (left) we can see that the prior specified on $\beta_{1,0}$ is vague. This prior represents our uncertainty in the evolution of the FoS of the slope with $\bm{z} = \bm{0}$. Similarly to Section \ref{Priors_B_spline}, through simulation of synthetic data, we were able to elicit a prior that put little weight on implausible curves, but there is no published data on how the FoS of a slope with $\bm{z} = \bm{0}$ will evolve over time. Figure \ref{noise_free_prior_shapes_quadratic} (right) presents some example deterministic FoS time series sampled from $\beta_{1,0}$ in Equation \eqref{beta_0}. The prior specified on $\beta_{1,0}$ allows for a variety of time series of slope deterioration that we see in reality. Similarly to Section \ref{Priors_B_spline}, the prior for $\beta_{1,0}$ governs the evolution of deterioration for a slope with $\bm{z}=\bm{0}$; and this prior puts little weighting on deterioration curves that we do not expect to see in reality (such as the red and blue curves in Figure \ref{noise_free_prior_shapes_quadratic} (left)), but it is difficult to be more specific than that. 
\section{Results} \label{Results}
\noindent We obtained posterior distributions of the two hierarchical models using RStan, the R interface to Stan \citep{carpenter2017stan, STAN}, with 4 chains. The quadratic and B-spline emulators took approximately 5 and 33 hours, respectively, to run using 4 cores with Intel CPU (Xeon, E5-2699 v4, base frequency 2.2GHz) through the Slurm workload manager \citet{Slurm} on Newcastle University's high performance computing service, Rocket.\\
\noindent Figure B1-B3 in the Supplementary Material B present some example trace plots, for the upper-level parameters, under the quadratic and B-spline models, respectively. In addition, Figures B5 and B6 present trace plots for the emulator parameters under the quadratic and B-spline models, respectively. We can see that all four chains have converged to the same values for the emulator parameters under the quadratic and B-spline models, respectively. Similar conclusions can be drawn from Figures B1 and B2, although FoS run 33 in Figure C5 may hint at multi-modality under the quadratic model or that the chains have not converged to the stationary distribution. Some trace plots show multi-modality under the B-spline model (see Figure B3). Figure B4 presents three within-sample computer runs with (at least one) multimodal (upper level) posterior distribution; we overlay a sample posterior FoS curve sampled from each mode. The chains in FoS run 34 switch between modes and the chains in FoS runs 39 and 68 do not switch between modes. The three computer runs in Figure B4 all follow a similar time series, with a relatively stable initial period followed by a rapid decline in FoS. Both modes fit reasonably well to the data. It may be possible to eliminate one mode by penalising FoS curves that are non-decreasing (by penalising values of $\gamma_1 > \gamma_0$).
\subsection{Posterior distributions of FoS}
\begin{figure}[!h]
	\centering
	\includegraphics[scale=0.5]{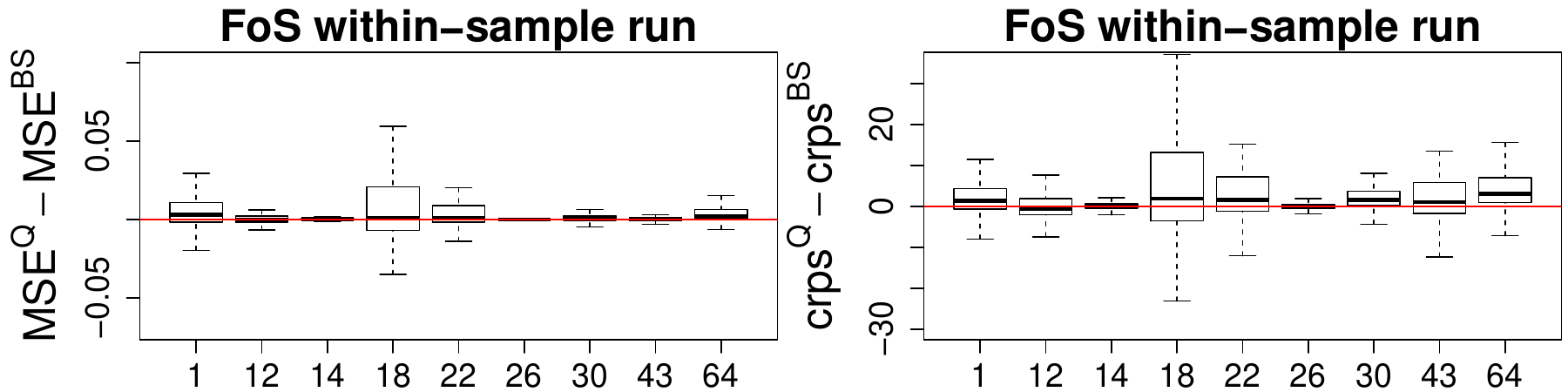}
	\caption{ Difference between the quadratic model and the B-spline model for nine within-sample computer runs in the in the $MSE$ (left) and $crps$ (right).}
	\label{within_sample1}
\end{figure}

\noindent Figure \ref{Plots1}, and Figures C1 - C8 in the Supplementary Material C, present posterior distributions of FoS obtained under the quadratic model (blue) and the B-spline model (red), for examples of computer experiment data. From Figure \ref{Plots1}, and Figures C1 - C8, we can see that the B-spline model captures the FoS trends well. The quadratic model and the B-spline model perform approximately the same for computer runs 12, 14 and 26, within the range of the data (\ref{Plots1}). However, the quadratic model is unable to capture the trends of computer runs 1, 18, 22, 30, and 64. Similar conclusions can be drawn from Figures C1 - C8 in the Supplementary Material C, e.g. run 43 in Figure C1. This is expected as some computer runs have trends that are not adequately modelled by a quadratic curve.\\
\begin{figure}[!h]
	\centering
	\includegraphics[scale=0.54]{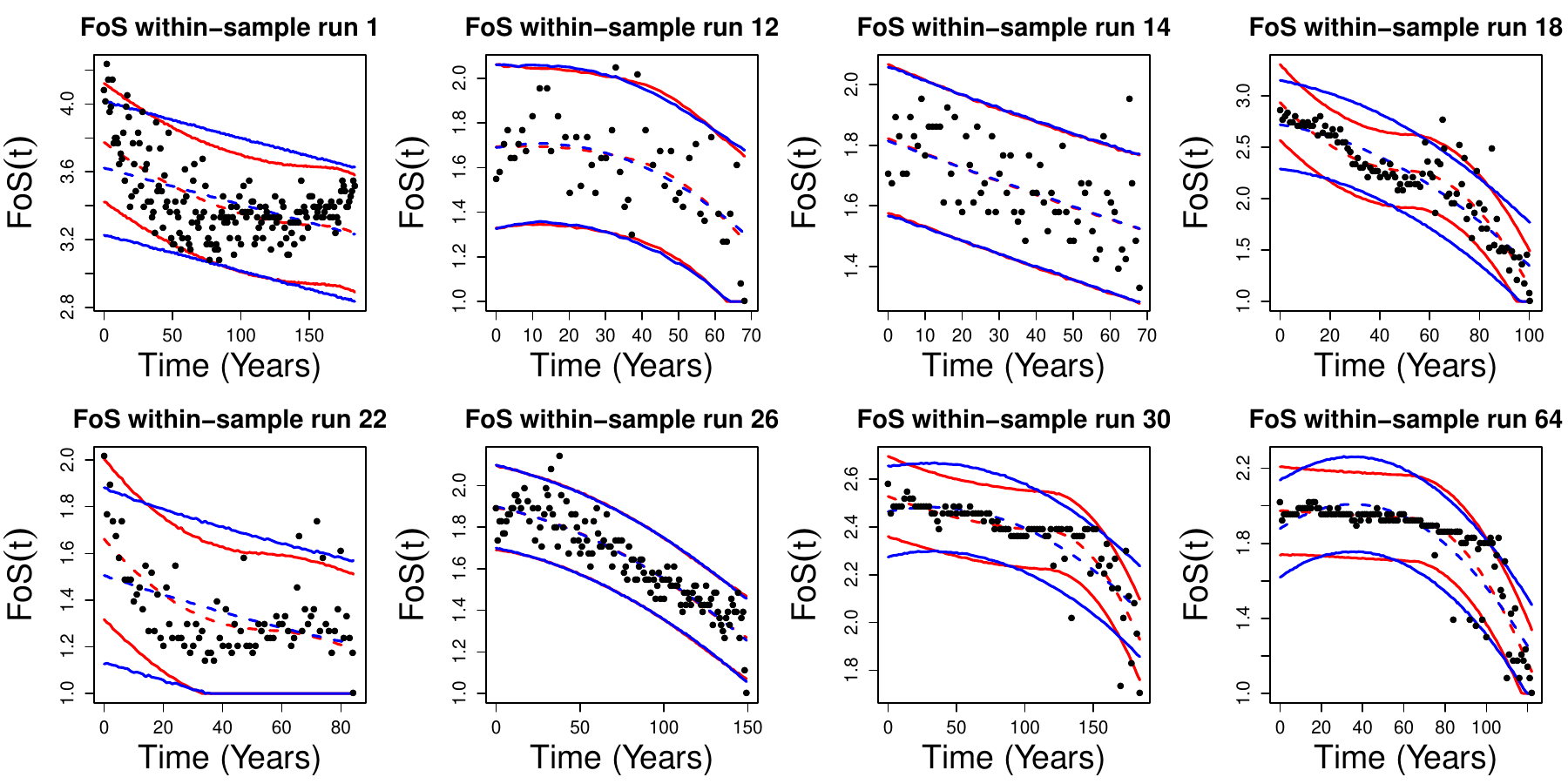}
	\caption{Examples of posterior distributions of FoS using the quadratic model (blue) and the B-spline model (red). The dashed lines represent the posterior means and the solid lines represent the central posterior 95\% prediction intervals.}
	\label{Plots1}
\end{figure}
\noindent Figure~\ref{within_sample1} presents the difference in the mean squared error $(MSE)$ and the continuous ranked probability score $(crps)$~\citep{gneiting2007probabilistic} between the quadratic model ($MSE^Q$ and $crps^Q$ respectively) and the B-spline model ($MSE^{BS}$ and $crps^{BS}$ respectively) for the nine within-sample computer runs presented in Figure \ref{Plots1}. The $crps$ is calculated at each time point for each computer run and is defined in terms of the predictive CDF, $F_t$, and an observation, $x_t$, 
\begin{equation}
crps(F_t, x_t) = \int_{-\infty}^\infty \{F_t(y) - \textbf{I}(y \geq x_t)\}^2 dy,
\end{equation}
\noindent where $F_t$ is the CDF of FoS at time $t$ and $x_t$ is the observed FoS at time $t$. The $crps$ is a proper scoring rule that evaluates the performance of a full predictive density.\\
\noindent As expected for computer runs 12, 14 and 26 the quadratic model and the B-spline model perform approximately the same, as their distributions are narrowly concentrated around zero. On the other hand, the $MSE$ and the $crps$ are larger for the quadratic model for computer runs 1, 18, 22, 30, 43 and 64 (c.f. Figure \ref{Plots1}). For the latter runs, the boxplots are centred in the positive range and have a longer tail in the same direction. Figure D1 in Supplementary Material D provides the difference in the $MSE$ and the $crps$ between the quadratic model and the B-spline model for the remaining sixty-six within-sample computer runs (Figures C1 - C8). 
\subsection{Posterior distributions of TTF}
\begin{figure}[!h]
	\centering
	\includegraphics[scale=0.55]{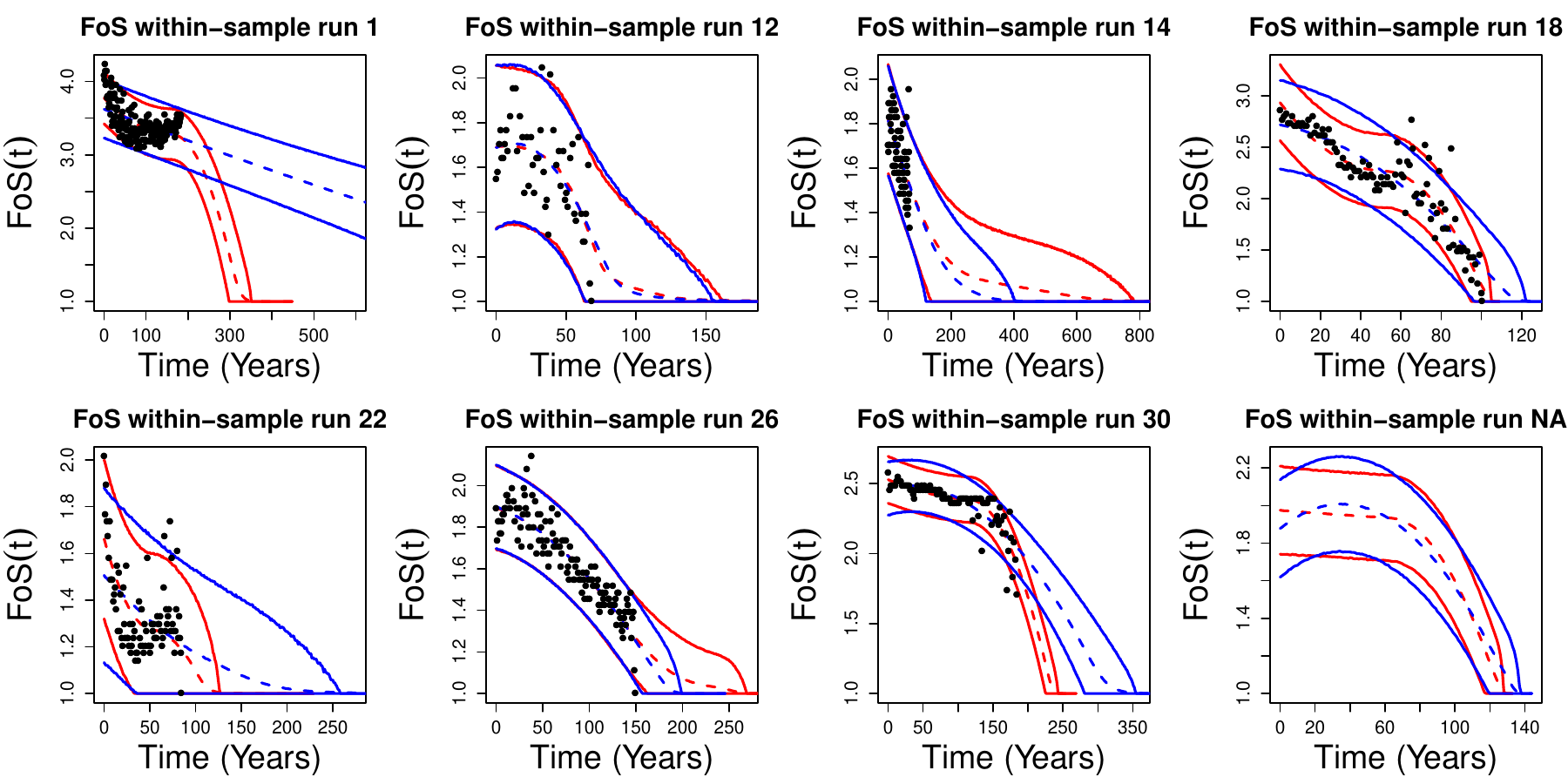}
	\caption{Examples of posterior distributions of FoS using the quadratic model (blue) and the B-spline model (red). The dashed lines represent the posterior means and the solid lines represent the central posterior 95\% prediction intervals.}
	\label{Plots_long_time_scale}
\end{figure}

\noindent We can obtain posterior distributions of the \emph{predicted} TTF, $\rho_i$, for $i=1,\dots,N$, by simulating FoS time series for every posterior draw of the model parameters and obtaining a distribution of the first time when FoS reaches failure, in most cases due to the error term. The distributions of predicted TTF, $\rho_i$, are presented in Figure~\ref{Plots_TTF1} and Figures E1 - E8 in the Supplementary Material E. The solid red lines indicate true TTF for computer runs that reached failure.
\begin{figure}[!h]
	\centering
	\includegraphics[scale=0.55]{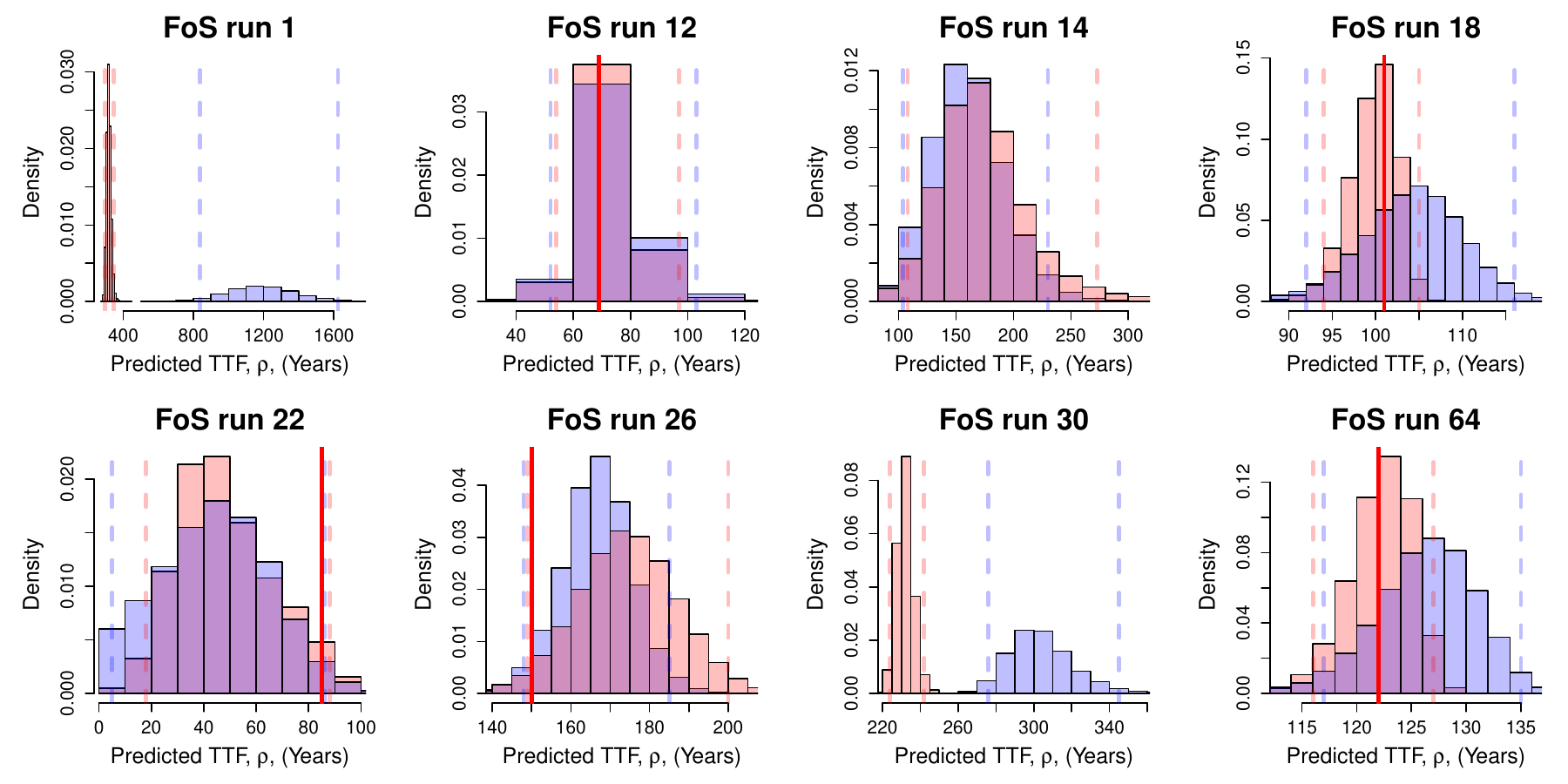}
	\caption{Examples of posterior distributions of predicted TTF, $\rho_i$, using the quadratic model (blue) and the B-spline model (red). The dashed lines represent the central posterior 95\% prediction intervals under each model. The solid red lines indicate true TTF for computer runs that reached failure.}
	\label{Plots_TTF1}
\end{figure}
\noindent From Figure \ref{Plots_TTF1}, and Figures E1 - E8, we can see that the TTF predictions are noticeably different under the quadratic model compared to the B-spline model for some computer runs. In general, the B-spline model leads to a posterior uncertainty decrease in the predicted TTF, and in some cases, a dramatic one. Figure~\ref{Plots_long_time_scale} presents examples of posterior distributions of FoS using the quadratic model (blue) and the B-spline model (red) for the nine computer runs shown in Figure \ref{Plots_TTF1}, extrapolating beyond the range of the data. The quadratic model is unable to capture some of the trends in the FoS data, such as computer runs 1, 18, 22, 30, and 64 in Figure \ref{Plots_long_time_scale} or run 43 in Figure E1. For computer run 1, we can see there is a substantial difference in the TTF prediction between the two models. Computer run 1 decays for around 50-75 years and then reaches a stable plateau. The quantiles of the quadratic model fail to capture this behaviour, whereas the B-spline model captures the decay and stability periods well. The B-spline model then predicts the slope will begin to deteriorate to failure. This is a consequence of having the internal knot fixed at $k = \omega/2$.\\
 \noindent Computer run 1 could remain stable for a long time after year 184, as illustrated in Figure E9 (Supplementary Material E). Emulating the internal knot, $k$, as well as the model failure time, $\omega$, would allow for time series such as those shown in Figure E9. At present, we avoid modelling the internal knot as our training data is of moderate size, but this should be considered in future work. 
\subsection{Model validation}
\noindent We also validated both models using 20 held-out computer experiment runs. As GPs are closed under conditioning, it is possible to derive an analytical expression for the GPE conditioned on a set of computer training runs. Assume a (finite) collection of $n$ ``observed'' experimental outputs, $\bm{y} = \big(\text{A}_0(\bm{z}_1), \text{A}_0(\bm{z}_2), \dots, \text{A}_0(\bm{z}_n)\big)$ performed on the ICs $\bm{z}_1, \bm{z}_2, \dots, \bm{z}_n$, where $n=75$, which comprise the training data. We assume no repeated inputs, so $\bm{z}_i = \bm{z}_j$ if and only if $i=j$. The $n-$vector $\bm{y}$ follows a multivariate normal distribution,
\begin{equation}
\bm{y} \mid \bm{\beta}_0, \tau_0, \bm{\delta}, \zeta \sim \text{N}(H_z\bm{\beta}_0, \tau_0\Sigma_z), 
\end{equation}
\noindent where $H_z$ is a matrix of regressors whose $i$th row is $h(\bm{z}_i)$, $\Sigma_{z}(i,j) = C(\bm{z}_i,\bm{z}_j)  + \zeta \mathbb{I}(\bm{z},\bm{z}')$, $\zeta$ is the nugget and the function $\mathbb{I}(\bm{z},\bm{z}')$ is an indicator for the event $\bm{z}=\bm{z}'$. Using standard rules for conditioning on a subset of observations~\citep{gramacy2020}, we find 
\begin{eqnarray}
&\text{A}_0(\cdot)|\bm{y},\bm{\beta}_0, \tau_0, \bm{\delta}, \zeta \sim \text{GP}(m^*(\cdot), V^*(\cdot,\cdot)),\label{A_0}\\
&m^*(\bm{z}) = h(\bm{z})^\top\bm{\beta}_0 + t(\bm{z})^\top\Sigma^{-1}_z(\bm{y} - H_z\bm{\beta}_0),\quad
V^*(\bm{z},\bm{z}') = \tau_0\big(C(\bm{z},\bm{z}',\bm{\delta}) - t(\bm{z})^\top\Sigma^{-1}_zt(\bm{z}')\big),\nonumber
\end{eqnarray}
\noindent where $t(\bm{z}) = \big(C(\bm{z},\bm{z}_1,\bm{\delta}),C(\bm{z},\bm{z}_2,\bm{\delta}),\dots,C(\bm{z},\bm{z}_n,\bm{\delta})\big)$ is a column vector of correlations between the (generic) emulator IC $\bm{z}$ and training ICs $\bm{z}_1, \bm{z}_2, \dots, \bm{z}_n$. Similar expressions can be derived for the conditional distributions of $\text{A}_1(\cdot), \text{A}_2(\cdot), \Omega(\cdot)$ and $\Sigma(\cdot)$. \\
\noindent Using Equation \eqref{A_0} for a held-out computer run with $\bm{z}$, provides us with a distribution for $\text{A}_0(\bm{z})|\bm{y},\bm{\beta}_0, \tau_0, \bm{\delta}, \zeta$, and this procedure can be repeated for $\text{A}_1(\bm{z}), \text{A}_2(\bm{z}), \Omega(\bm{z})$ and $\Sigma(\bm{z})$. Using these distributions, we can sample from Equation \eqref{B-spline_expanded}, and obtain FoS time series for a computer run with some IC $\bm{z}$. \\
\noindent Figure \ref{FoS1} and Figures F1 - F2 in Supplementary Material F present posterior distributions of FoS under the quadratic (blue) and B-spline (red) models for 20 held-out computer runs. For both models, the posterior means capture the FoS behaviour well. Validation runs 1 (Figure F1) and 2 (Figure~\ref{FoS1}) have the widest 95\% highest posterior density intervals (HPDI) under both models. These computer runs are quite unusual, as they both begin with high initial FoS measurements which then increase before plateauing at values larger than their initial strength. Both models do not reach failure at the end of simulation and at 184 years (the simulation end time) both computer experiements have higher FoS measurements than their starting FoS. This behaviour leads to a large uncertainty in the model and predicted TTF and in the posterior predictions of FoS. Despite this uncertainty, the mean predicted FoS is extremely close to the validation data. For the remaining runs in Figure~\ref{FoS1} and Figures F1 - F2 in Supplementary Material F, the 95\% HPDIs are much narrower (under both models), and the mean posterior behaviour is close to the true values for most out-of-sample computer runs.\\
\begin{figure}[!h]
	\centering
	\includegraphics[scale=0.55]{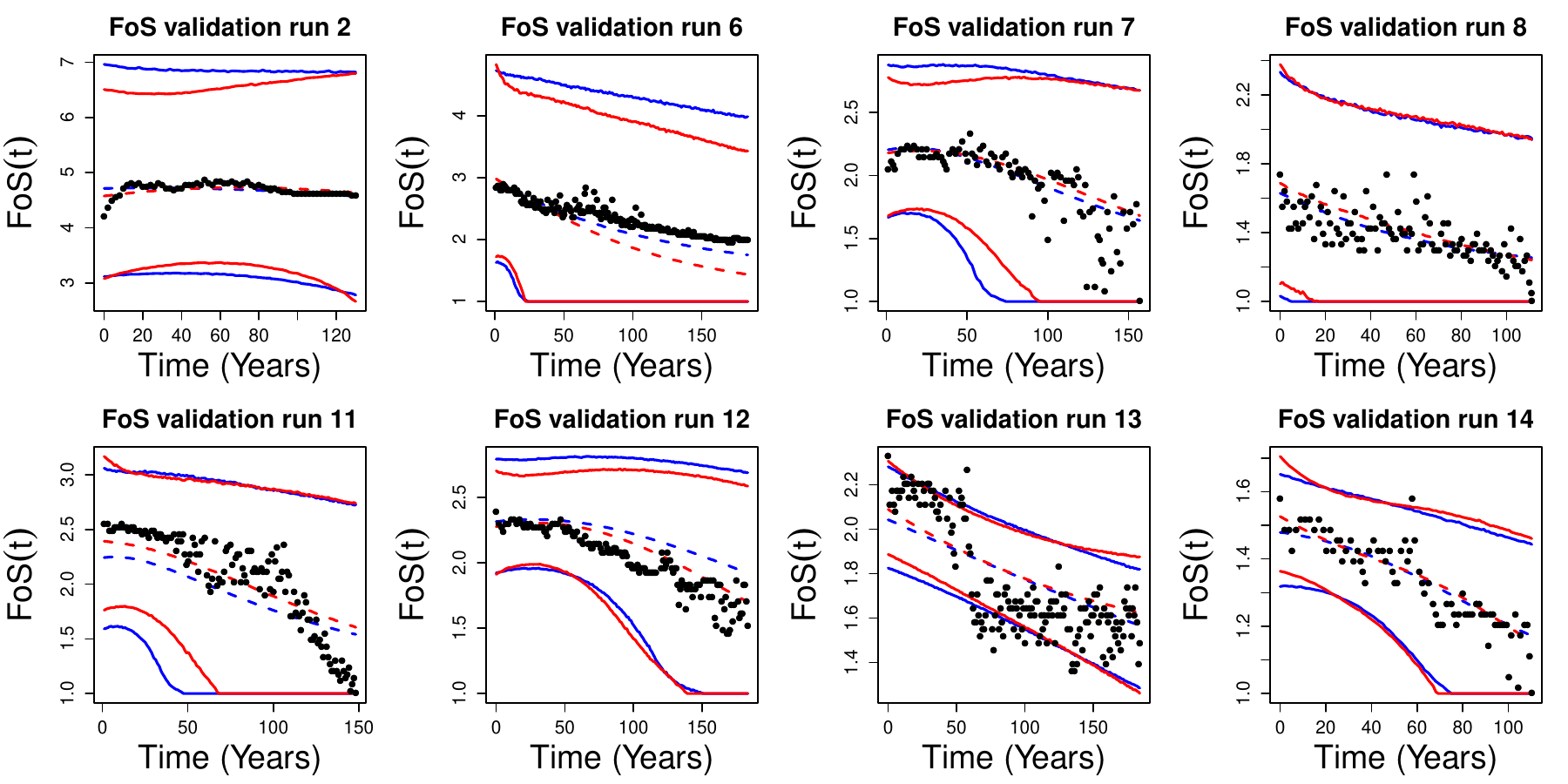}
	\caption{Examples of posterior distributions of FoS using the quadratic model (blue) and the B-spline model (red) for out-of sample computer runs. The dashed lines represent the posterior means and the solid lines represent the central posterior 95\% prediction intervals.}
	\label{FoS1}
\end{figure}
\begin{figure}[!h]
	\centering
	\includegraphics[scale=0.45]{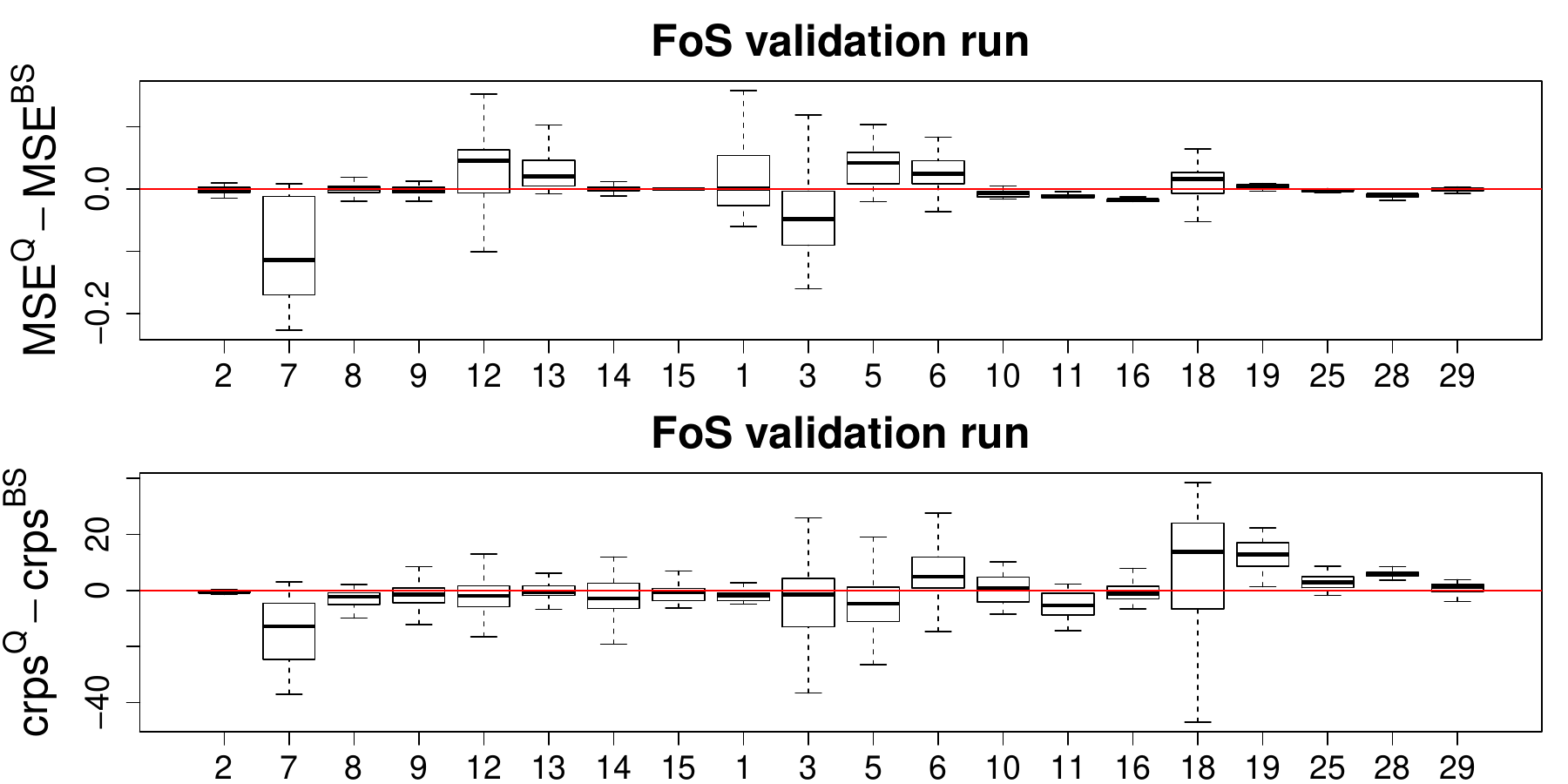}
	\caption{Plots presenting the difference in the $MSE$ and $crps$ between the quadratic model and the B-spline model for 20 out-of-sample computer runs.}
	\label{MSE_crps_out_of_sample}
\end{figure}
\noindent Figure \ref{MSE_crps_out_of_sample} presents the difference in the $MSE$ and the $crps$ between the quadratic model and the B-spline model for the 20 out-of-sample computer runs presented in Figure~\ref{FoS1}, and Figures F1 - F2 in Supplementary Material F. The differences in the $MSE$ are small for most out-of-sample runs. For computer runs e.g. 5 (Figure F1), 11, and 12, the mean behaviour of the B-spline model more closely captures the FoS behaviour compared to the quadratic model. The opposite can be observed for computer runs 3 (Figure F1) and 6, although for run 3 the mean FoS at failure time is closer to the true FoS than that of the quadratic model. For runs with unoberved failure times (e.g. 7, 12 (Figure~\ref{FoS1}), and 16 (Figure F1)), the B-spline model tends to provide a more conservative estimate of the FoS at the last time point. An exception is run 14 where the mean FoS is marginally larger.\\
\noindent The $95\%$ prediction intervals tend to be narrower under the B-spline model, except for failure times of 10 years or less. Computer run 16 (Figure F1) illustrates behaviour that is difficult to model, as it begins with a stable period followed by a sudden, rapid deterioration before an increase in FoS. Most of its measurements are outside of the quadratic model prediction intervals from initiation until around 100 years. The corresponding $95\%$ interval is wider under the B-spline model, however it captures all of the FOS measurements. 
\section{Conclusions} \label{Conclusions}
\noindent Factor of safety is an important diagnostic tool to predict and monitor the safety of transport infrastructure. For a given slope, the change in FoS can be estimated using computer experiments of deterioration, however they are very time-consuming to run. Here, we built a GP emulator of FoS for cuttings in high-plasticity soils using Bayesian hierarchical modelling to approximate FoS measurements faster for any IC combination. We approximated the FoS curves with a single quadratic model and a B-spline model of two quadratic polynomials with one interior knot; and we emulated the model parameters by assigning Gaussian process prior distributions to them. This produced a multi-output hierarchical GP model that relates the parameters determining FoS curves with the computer experiment ICs.\\
\noindent We obtained within-sample posterior distributions of FoS and demonstrated how the models could be used to obtain posterior distributions of predicted TTF. The quadratic model was able to emulate some computer experiments well, as several FoS time series deteriorated with quadratic trends. However, in many cases it did not have the flexibility required to emulate FoS curves that were driven by more than one failure mechanism. For example, computer run 1 (Figure~\ref{Plots1}), decays until around year 40, and then stabilises for approximately 20 years, before deteriorating to failure; and computer run 64 remains stable for approximately 75-100 years, before rapidly declining. These time series are much better captured under the B-spline model. We compared model performance using the $MSE$ and the $crps$ and found that the B-spline model performed as well as or outperformed the quadratic model for most computer runs. Furthermore, we illustrated how to use the models for out-of-sample prediction by conditioning on the ``observed'' upper level model parameters. The successful emulation of geotechnical models of deterioration of infrastructure slopes, has the potential to inform slope design, maintenance and remediation by introducing the time dependency of deterioration into geotechnical asset management.\\
\noindent Future developments in our work include learning the interior knot of the B-spline from the data rather than fixing it at half-model TTF, as setting $k=\omega/2$ limits the B-spline's predictive capability outside of the range of the data. For example, for computer run 1, the B-spline model predicts the FoS time series will deteriorate rapidly to failure and predicts the expected failure time of computer run 1 to be 318 years with a central 95\% prediction interval of $(295, 347)$ years. Computer run 1 could remain stable for a relatively long time after year 184, as illustrated in Figure E9. Other developments would include emulating the FoS using linear dynamical models and Gaussian processes, for example, as in~\citet{oyebamiji2017}, as well as including wider ranges of soil properties and earthwork types in the computer model. 
\bibliographystyle{Chicago}
\bibliography{Oakley_Svalova_arXiv_Submission.bbl}

\begin{thebibliography}{}

\bibitem[\protect\citeauthoryear{Abbott}{Abbott}{2018}]{nr2018}
Abbott, S. (2018, June).
\newblock Earthworks technical strategy.
\newblock Technical report, Network Rail.

\bibitem[\protect\citeauthoryear{Andrianakis and Challenor}{Andrianakis and
  Challenor}{2012}]{andrianakis2012}
Andrianakis, I. and P.~G. Challenor (2012).
\newblock The effect of the nugget on {G}aussian process emulators of computer
  models.
\newblock {\em Computational Statistics \& Data Analysis\/}~{\em 56},
  4215--4228.

\bibitem[\protect\citeauthoryear{Apted}{Apted}{1977}]{apted1977}
Apted, J.~P. (1977).
\newblock {\em {Effects of weathering on some geotechnical properties of London
  clay}}.
\newblock {PhD} thesis, Imperial College (University of London).

\bibitem[\protect\citeauthoryear{Armstrong, Helm, Preston, and
  Loveridge}{Armstrong et~al.}{2024}]{armstrong2024}
Armstrong, J., P.~Helm, J.~Preston, and F.~Loveridge (2024).
\newblock Economics of geotechnical asset deterioration, maintenance and
  renewal.
\newblock {\em Transportation Geotechnics\/}, 101185.

\bibitem[\protect\citeauthoryear{Bastos and O'Hagan}{Bastos and
  O'Hagan}{2009}]{bastos2009}
Bastos, L.~S. and A.~O'Hagan (2009).
\newblock Diagnostics for {G}aussian process emulators.
\newblock {\em Technometrics\/}~{\em 51}, 425--438.

\bibitem[\protect\citeauthoryear{Briggs, Helm, Smethurst, Smith, Stirling,
  Svalova, {Trinidad González}, Loveridge, and Glendinning}{Briggs
  et~al.}{2023}]{briggs2023}
Briggs, K.~M., P.~R. Helm, J.~A. Smethurst, A.~Smith, R.~Stirling, A.~Svalova,
  Y.~{Trinidad González}, F.~A. Loveridge, and S.~Glendinning (2023).
\newblock Evidence for the weather-driven deterioration of ageing
  transportation earthworks in the uk.
\newblock {\em Transportation Geotechnics\/}~{\em 43}, 101130.

\bibitem[\protect\citeauthoryear{Bromhead and Dixon}{Bromhead and
  Dixon}{1986}]{bromhead1986}
Bromhead, E.~N. and N.~Dixon (1986).
\newblock The field residual strength of {L}ondon {C}lay and its correlation
  with laboratory measurements, especially ring shear tests.
\newblock {\em G{\'{e}}otechnique\/}~{\em 36}, 449--452.

\bibitem[\protect\citeauthoryear{{BSI}}{{BSI}}{2023}]{bsi_bs_2023}
{BSI} (2023).
\newblock {BS} {EN} 1990: {Eurocode}. {Basis} of structural and geotechnical
  design.
\newblock OCLC: 1400083613.

\bibitem[\protect\citeauthoryear{Burland, Chapman, and Walsh}{Burland
  et~al.}{2023}]{burland2023}
Burland, J.~B., T.~Chapman, and M.~Walsh (2023).
\newblock Foundations and other geotechnical elements in context--their role.
\newblock In {\em ICE Manual of Geotechnical Engineering, Second edition,
  Volume I: Geotechnical engineering principles, problematic soils and site
  investigation}, pp.\  5--12. Emerald Publishing Limited.

\bibitem[\protect\citeauthoryear{Carpenter, Gelman, Hoffman, Lee, Goodrich,
  Betancourt, Brubaker, Guo, Li, and Riddell}{Carpenter
  et~al.}{2017}]{carpenter2017stan}
Carpenter, B., A.~Gelman, M.~D. Hoffman, D.~Lee, B.~Goodrich, M.~Betancourt,
  M.~A. Brubaker, J.~Guo, P.~Li, and A.~Riddell (2017).
\newblock Stan: A probabilistic programming language.
\newblock {\em Journal of statistical software\/}~{\em 76}.

\bibitem[\protect\citeauthoryear{Conti, Gosling, Oakley, and O'Hagan}{Conti
  et~al.}{2009}]{conti2009}
Conti, S., J.~P. Gosling, J.~E. Oakley, and A.~O'Hagan (2009).
\newblock Gaussian process emulation of dynamic computer codes.
\newblock {\em Biometrika\/}~{\em 3}, 663--676.

\bibitem[\protect\citeauthoryear{De~Boor}{De~Boor}{1978}]{deboor1978}
De~Boor, C. (1978).
\newblock {\em A practical guide to splines}.
\newblock New York: Springer-Verlag.

\bibitem[\protect\citeauthoryear{Dixon, Crosby, Stirling, Hughes, Smethurst,
  Briggs, Hughes, Gunn, Hobbs, Loveridge, Glendinning, Dijkstra, and
  Hudson}{Dixon et~al.}{2019}]{dixon2019}
Dixon, N., C.~J. Crosby, R.~Stirling, P.~N. Hughes, J.~Smethurst, K.~Briggs,
  D.~Hughes, D.~Gunn, P.~Hobbs, F.~Loveridge, S.~Glendinning, T.~Dijkstra, and
  A.~Hudson (2019).
\newblock In situ measurements of near-surface hydraulic conductivity in
  engineered clay slopes.
\newblock {\em Quarterly Journal of Engineering Geology and
  Hydrogeology\/}~{\em 52}, 123--135.

\bibitem[\protect\citeauthoryear{Duncan}{Duncan}{1996}]{duncan1996}
Duncan, J.~M. (1996).
\newblock State of the art: Limit equilibrium and finite-element analysis of
  slopes.
\newblock {\em Journal of Geotechnical Engineering\/}~{\em 122}, 577--596.

\bibitem[\protect\citeauthoryear{Ellis and O'Brien}{Ellis and
  O'Brien}{2007}]{ellis2007}
Ellis, E.~A. and A.~S. O'Brien (2007).
\newblock Effect of height on delayed collapse of cuttings in stiff clay.
\newblock {\em Proceedings of the Institution of Civil Engineers- Geotechnical
  Engineering\/}~{\em 160}, 73--84.

\bibitem[\protect\citeauthoryear{Farah, Birrell, Conti, and De~Angelis}{Farah
  et~al.}{2014}]{farah2014a}
Farah, M., P.~Birrell, S.~Conti, and D.~De~Angelis (2014).
\newblock Bayesian emulation and calibration of a dynamic epidemic model for
  {A/H1N1} influenza.
\newblock {\em Journal of the American Statistical Association\/}~{\em 109},
  1398--1411.

\bibitem[\protect\citeauthoryear{Farah and Kottas}{Farah and
  Kottas}{2014}]{farah2014}
Farah, M. and A.~Kottas (2014).
\newblock Bayesian inference for sensitivity analysis of computer simulations,
  with an application to radiative transfer models.
\newblock {\em Technometrics\/}~{\em 56}, 159--173.

\bibitem[\protect\citeauthoryear{Fricker, Oakley, and Urban}{Fricker
  et~al.}{2013}]{fricker2013}
Fricker, T.~E., J.~E. Oakley, and N.~M. Urban (2013).
\newblock Multivariate {G}aussian process emulators with nonseparable
  covariance structures.
\newblock {\em Technometrics\/}~{\em 55}.

\bibitem[\protect\citeauthoryear{Gelman and Vehtari}{Gelman and
  Vehtari}{2021}]{gelman2021}
Gelman, A. and A.~Vehtari (2021).
\newblock What are the most important statistical ideas of the past 50 years?
\newblock {\em Journal of the American Statistical Association\/}~{\em 116},
  2087--2097.

\bibitem[\protect\citeauthoryear{Gneiting, Balabdaoui, and Raftery}{Gneiting
  et~al.}{2007}]{gneiting2007probabilistic}
Gneiting, T., F.~Balabdaoui, and A.~E. Raftery (2007).
\newblock Probabilistic forecasts, calibration and sharpness.
\newblock {\em Journal of the Royal Statistical Society Series B: Statistical
  Methodology\/}~{\em 69\/}(2), 243--268.

\bibitem[\protect\citeauthoryear{Gramacy}{Gramacy}{2020}]{gramacy2020}
Gramacy, R. (2020).
\newblock {\em Surrogates}.
\newblock Boca Raton: CRC Press.

\bibitem[\protect\citeauthoryear{Helm, Svalova, Morsy, Rouainia, Smith,
  El-Hamalawi, Wilkinson, Postill, and Glendinning}{Helm
  et~al.}{2024a}]{helm2024emulating}
Helm, P., A.~Svalova, A.~Morsy, M.~Rouainia, A.~Smith, A.~El-Hamalawi,
  D.~Wilkinson, H.~Postill, and S.~Glendinning (2024a).
\newblock Emulating long-term weather-driven transportation earthworks
  deterioration models to support asset management.
\newblock {\em Transportation Geotechnics\/}~{\em 44}, 101155.

\bibitem[\protect\citeauthoryear{Helm, Svalova, Morsy, Rouainia, Smith,
  El-Hamalawi, Wilkinson, Postill, and Glendinning}{Helm
  et~al.}{2024b}]{helm2024}
Helm, P.~R., A.~Svalova, A.~M. Morsy, M.~Rouainia, A.~Smith, A.~El-Hamalawi,
  D.~J. Wilkinson, H.~Postill, and S.~Glendinning (2024b).
\newblock Emulating long-term weather-driven transportation earthworks
  deterioration models to support asset management.
\newblock {\em Transportation Geotechnics\/}~{\em 44}, 101155.

\bibitem[\protect\citeauthoryear{Henderson, Boys, Krishnan, Lawless, and
  Wilkinson}{Henderson et~al.}{2009}]{henderson2009}
Henderson, D.~A., R.~J. Boys, K.~J. Krishnan, C.~Lawless, and D.~J. Wilkinson
  (2009).
\newblock Bayesian emulation and calibration of a stochastic computer model of
  mitochondrial {DNA} deletions in substantia nigra neurons.
\newblock {\em Journal of the American Statistical Association\/}~{\em 104},
  76--87.

\bibitem[\protect\citeauthoryear{Homma and Saltelli}{Homma and
  Saltelli}{1996}]{homma1996}
Homma, T. and A.~Saltelli (1996).
\newblock Importance measures in global sensitivity analysis of nonlinear
  models.
\newblock {\em Reliability Engineering and System Safety\/}~{\em 52}, 1--17.

\bibitem[\protect\citeauthoryear{{Itasca}}{{Itasca}}{2016}]{FLAC}
{Itasca} (2016).
\newblock {\em {F}ast {L}agrangian {A}nalysis of {C}ontinua, {V}er. 8.0}.
\newblock Minneapolis: Itasca: Itasca Consulting Group, Inc.

\bibitem[\protect\citeauthoryear{Jette and Wickberg}{Jette and
  Wickberg}{2023}]{Slurm}
Jette, M.~A. and T.~Wickberg (2023).
\newblock Architecture of the slurm workload manager.
\newblock In D.~Klusáček, J.~Corbalán, and G.~Rodrigo (Eds.), {\em Job
  Scheduling Strategies for Parallel Processing. JSSPP 2023}, Lecture Notes in
  Computer Science, Cham. Springer.

\bibitem[\protect\citeauthoryear{Jure{\v{c}}i{\v{c}}, Zdravkovi{\'{c}}, and
  Jovi{\v{c}}i{\'{c}}}{Jure{\v{c}}i{\v{c}} et~al.}{2013}]{jurecic2013}
Jure{\v{c}}i{\v{c}}, N., L.~Zdravkovi{\'{c}}, and V.~Jovi{\v{c}}i{\'{c}}
  (2013).
\newblock {Predicting ground movements in London Clay}.
\newblock {\em Proceedings of the Institution of Civil Engineers - Geotechnical
  Engineering\/}~{\em 166}, 466--482.

\bibitem[\protect\citeauthoryear{Kennedy and O'Hagan}{Kennedy and
  O'Hagan}{2001}]{kennedy2001bayesian}
Kennedy, M.~C. and A.~O'Hagan (2001).
\newblock Bayesian calibration of computer models.
\newblock {\em Journal of the Royal Statistical Society: Series B (Statistical
  Methodology)\/}~{\em 63\/}(3), 425--464.

\bibitem[\protect\citeauthoryear{Knott}{Knott}{2000}]{knott2000}
Knott, G.~D. (2000).
\newblock {\em Interpolating cubic splines}.
\newblock Boston: Springer.

\bibitem[\protect\citeauthoryear{Mellor, Parry, and Power}{Mellor
  et~al.}{2017}]{GSRA}
Mellor, R., L.~Parry, and C.~Power (2017, August).
\newblock {CP6} earthworks asset policy developemnt. {T}ask 63 - global
  stability and resilience appraisal. {I}nterim report.
\newblock Technical report, Network Rail.

\bibitem[\protect\citeauthoryear{Mohammadi, Challenor, and
  Goodfellow}{Mohammadi et~al.}{2019}]{mohammadi2019}
Mohammadi, H., P.~Challenor, and M.~Goodfellow (2019).
\newblock Emulating dynamic non-linear simulators using {G}aussian processes.
\newblock {\em Computational Statistics and Data Analysis\/}~{\em 139},
  178--196.

\bibitem[\protect\citeauthoryear{Nowak}{Nowak}{2012}]{nowak2012}
Nowak, P. (2012).
\newblock Slope stabilisation methods.
\newblock In J.~B. Burland, T.~Chapman, H.~Skinner, and M.~Brown (Eds.), {\em
  {ICE} {Manual} of geotechnical engineering. 2: {Geotechnical} design,
  construction and verification}, Volume~2. London: ICE Publishing.

\bibitem[\protect\citeauthoryear{Oakley and Youngman}{Oakley and
  Youngman}{2017}]{oakley2017calibration}
Oakley, J.~E. and B.~D. Youngman (2017).
\newblock Calibration of stochastic computer simulators using likelihood
  emulation.
\newblock {\em Technometrics\/}~{\em 59\/}(1), 80--92.

\bibitem[\protect\citeauthoryear{O'Hagan}{O'Hagan}{2006}]{ohagan2006}
O'Hagan, A. (2006).
\newblock Bayesian analysis of computer code outputs: A tutorial.
\newblock {\em Reliability Engineering \& System Safety\/}~{\em 91},
  1290--1300.

\bibitem[\protect\citeauthoryear{Oyebamiji, Wilkinson, Jayathilake, Curtis,
  Rushton, Li, and Gupta}{Oyebamiji et~al.}{2017}]{oyebamiji2017}
Oyebamiji, O.~K., D.~J. Wilkinson, P.~G. Jayathilake, T.~P. Curtis, S.~P.
  Rushton, B.~Li, and P.~Gupta (2017).
\newblock Gaussian process emulation of an individual-based model simulation of
  microbial communities.
\newblock {\em Journal of Computational Science\/}~{\em 22}, 69--84.

\bibitem[\protect\citeauthoryear{Oyebamiji, Wilkinson, Li, Jayathilake,
  Zuliani, and P.}{Oyebamiji et~al.}{2019}]{oyebamiji2019}
Oyebamiji, O.~K., D.~J. Wilkinson, B.~Li, P.~G. Jayathilake, P.~Zuliani, and
  C.~T. P. (2019).
\newblock Bayesian emulation and calibration of an individual-based model of
  microbial communities.
\newblock {\em Journal of Computational Science\/}~{\em 30}, 194--208.

\bibitem[\protect\citeauthoryear{Perry, Pedley, and Brady}{Perry
  et~al.}{2003}]{perry2003}
Perry, J., M.~Pedley, and K.~Brady (2003).
\newblock Infrastructure cuttings - condition appraisal and remedial treatment.
\newblock Technical Report C591, {CIRIA}, London.

\bibitem[\protect\citeauthoryear{Postill, Dixon, Fowmes, El-Hamalawi, and
  Take}{Postill et~al.}{2020}]{postill2020}
Postill, H., N.~Dixon, G.~Fowmes, A.~El-Hamalawi, and W.~A. Take (2020).
\newblock Modelling seasonal ratcheting and progressive failure in clay slopes:
  a validation.
\newblock {\em Canadian Geotechnical Journal\/}~{\em 57\/}(9), 1265--1279.

\bibitem[\protect\citeauthoryear{Postill, Helm, Dixon, El-Hamalawi,
  Glendinning, and Take}{Postill et~al.}{2023}]{postill2023}
Postill, H., P.~R. Helm, N.~Dixon, A.~El-Hamalawi, S.~Glendinning, and W.~A.
  Take (2023).
\newblock Strength parameter selection framework for evaluating the design life
  of clay cut slopes.
\newblock {\em Proceedings of the Institution of Civil Engineers - Geotechnical
  Engineering\/}~{\em 176\/}(3), 254--273.

\bibitem[\protect\citeauthoryear{Postill, Helm, Dixon, Glendinning, Smethurst,
  Rouainia, Briggs, El-Hamalawi, and Blake}{Postill et~al.}{2021}]{postill2021}
Postill, H., P.~R. Helm, N.~Dixon, S.~Glendinning, J.~A. Smethurst,
  M.~Rouainia, K.~M. Briggs, A.~El-Hamalawi, and A.~P. Blake (2021).
\newblock {Forecasting the long-term deterioration of a cut slope in
  high-plasticity clay using a numerical model}.
\newblock {\em Engineering Geology\/}~{\em 280}, 1--19.

\bibitem[\protect\citeauthoryear{Potts, Kovacevic, and Vaughan}{Potts
  et~al.}{1997}]{potts1997}
Potts, D.~M., N.~Kovacevic, and P.~R. Vaughan (1997).
\newblock Delayed collapse of cut slopes in stiff clay.
\newblock {\em G{\'{e}}otechnique\/}~{\em 47}, 953--982.

\bibitem[\protect\citeauthoryear{Rouainia, Helm, Davies, and
  Glendinning}{Rouainia et~al.}{2020}]{rouainia2020}
Rouainia, M., P.~Helm, O.~Davies, and S.~Glendinning (2020).
\newblock Deterioration of an infrastructure cutting subjected to climate
  change.
\newblock {\em Acta Geotechnica\/}~{\em 15}, 2997--3016.

\bibitem[\protect\citeauthoryear{Santner}{Santner}{2018}]{santner2018}
Santner, T.~J. (2018).
\newblock {\em The Design and Analysis of Computer Experiments\/} (2 ed.).
\newblock Springer Series in Statistics.

\bibitem[\protect\citeauthoryear{{Stan~Development~Team}}{{Stan~Development~Team}}{2022}]{STAN}
{Stan~Development~Team} (2022).
\newblock {\em Stan Modeling Language Users Guide and Reference Manual\/}
  ({VERSION 2.29} ed.).
\newblock \url{https://mc-stan.org}.

\bibitem[\protect\citeauthoryear{Summersgill, Kontoe, and Potts}{Summersgill
  et~al.}{2018}]{summersgill2018}
Summersgill, F.~C., S.~Kontoe, and D.~M. Potts (2018).
\newblock Stabilisation of excavated slopes in strain-softening materials with
  piles.
\newblock {\em G{\'{e}}otechnique\/}~{\em 68}, 1--14.

\bibitem[\protect\citeauthoryear{Svalova, Helm, Prangle, Rouainia, Glendinning,
  and Wilkinson}{Svalova et~al.}{2021}]{svalova2021}
Svalova, A., P.~Helm, D.~Prangle, M.~Rouainia, S.~Glendinning, and D.~J.
  Wilkinson (2021).
\newblock Emulating computer experiments of transport infrastructure slope
  stability using {G}aussian processes and {B}ayesian inference.
\newblock {\em Data-Centric Engineering\/}~{\em 2}, e12.

\bibitem[\protect\citeauthoryear{Talebitooti, Shojaeefard, and
  Yarmohammadisatri}{Talebitooti et~al.}{2015}]{talebitooti2015}
Talebitooti, R., M.~H. Shojaeefard, and S.~Yarmohammadisatri (2015).
\newblock Shape design optimization of cylindrical tank using {B}-spline
  curves.
\newblock {\em Computers \& Fluids\/}~{\em 109}, 100--112.

\bibitem[\protect\citeauthoryear{Tsiampousi, Zdravkovic, and Potts}{Tsiampousi
  et~al.}{2017}]{tsiampousi2017}
Tsiampousi, A., L.~Zdravkovic, and D.~Potts (2017).
\newblock {Numerical study of the effect of soil–atmosphere interaction on
  the stability and serviceability of cut slopes in London Clay}.
\newblock {\em Canadian Geotechnical Journal\/}~{\em 54}, 405--418.

\bibitem[\protect\citeauthoryear{van Genuchten}{van
  Genuchten}{1980}]{vangenuchten1980}
van Genuchten, M.~T. (1980).
\newblock {A Closed-form Equation for Predicting the Hydraulic Conductivity of
  Unsaturated Soils}.
\newblock {\em Soil Science Society of America Journal\/}~{\em 44}, 892--898.

\end{thebibliography}
\newpage
{\bf\LARGE Supplementary Materials}
\begin{appendices}
\section{B-spline definition}\label{B-spline_details}
Consider a vector of knots $\bm{k}=(k_0,\ldots,k_{m+1})$ over which the B-spline is to be fitted. The \textit{boundary knots} are equivalent to zero-time ($k_0=0$) and the failure time $k_{m+1}=\omega$, and there are $m$ \textit{interior knots}. As in the quadratic model, we assume that $\omega$ is unobserved and will be estimated with the remaining parameters. To fit a B-spline of order $o$ (degree $o-1$) over $\bm{k}$, define an augmented knot sequence $\bm{k}^*$ with $m+2o$ elements
\begin{equation}
k_{-o+1}=\cdots=k_0\leq\cdots\leq k_m\leq k_{m+1}=\cdots=k_{m+o}.
\label{knots}
\end{equation}
\noindent It is noteworthy that in Equation~\eqref{knots} the first and the last $o$ knots are repeated to fix the boundary constraints. For easier computation, the subscripts of the elements of $\bm{k}^*$ can be re-labelled as $k_l\subset \bm{k}^*, l\in\{0,\ldots,m+2o-1\}$.
\noindent We assume that two piecewise polynomials could be sufficient to explain the two dominant failure modes (deep and shallow), thus $m=1$ and $o=3$. This gives the augmented knot sequence $\bm{k}^*=(k_0,k_0,k_0,k_1,k_2,k_2,k_2)=(0,0,0,k,\omega,\omega,\omega)$. Frequently, the interior knots are equidistant, and here we assume that $k=0.5\omega$.\\
\noindent The B-spline is defined using the following recursive relationships. Firstly, we define a set of basis (indicator) functions $\phi_{l,j}$ for $j\in\{0,1,\ldots,o-1\}$ and $l\in\{0,1,\ldots,m+2o-2\}$:
\begin{equation}
\phi_{l,0}(x)=
\begin{cases}
1, & \text{if } k_l\leq x < k_{l+1},\\
0, & \text{otherwise}.
\end{cases}
\label{phi}
\end{equation}
\noindent Then, for $j\in\{0,1,\ldots,o-1\}$ and $l\in\{0,1,\ldots,m+2o-2-(j+1)\}$ we define the set of functions:
\begin{equation}
\phi_{l,j+1}(x)=\alpha_{l,j+1}(x)\phi_{l,j}(x)+[1-\alpha_{l+1,j+1}(x)]\phi_{l+1,j}(x),
\label{phi2}
\end{equation} 
where
\begin{equation}
\alpha_{l,j}(x)=
\begin{cases}
\frac{x-k_l}{k_{l+j}-k_l}, &\text{if } k_{l+j}\neq k_l,\\
0, & \text{otherwise}.
\end{cases}
\label{alpha}
\end{equation}
\noindent The B-spline approximation of $Y_{i,j}=\,$FoS$_{i,j}-1$ can be written as
\begin{equation}
Y_{i,j}=\sum_{l=0}^{m+o-1}\gamma_{l,i}\phi_{l,o-1}(t_{i,j})+\varepsilon_{i,j}=\sum_{l=0}^{3}\gamma_{l,i}\phi_{l,2}(t_{i,j})+\varepsilon_{i,j}, \quad\varepsilon_{i,j}\sim \text{N}(0,\sigma^2_i).
\label{appbspline}
\end{equation}
\noindent Expanding Equation~\eqref{appbspline} using Equations~\eqref{phi},~\eqref{phi2} and~\eqref{alpha} we find that the spline model can be written as 
\begin{equation}
\begin{aligned}
Y_{i,j} = &\bigg\{\gamma_{0,i} + t_{i,j}\bigg(\frac{4 \gamma_{1,i}}{\omega_i}-\frac{4 \gamma_{0,i}}{\omega_i}\bigg)+ t_{i,j}^2\bigg(\frac{4 \gamma_{0,i}}{\omega_i^2}-\frac{6 \gamma_{1,i}}{\omega_i^2}+\frac{2 \gamma_{2,i}}{\omega_i^2}\bigg) \bigg\} I(0 \leq t_{i,j} < \omega_i/2)+\\
&\bigg\{2(\gamma_{1,i} -\gamma_{2,i}+\gamma_{3,i})+t_{i,j}\bigg(\frac{8 \gamma_{2,i}}{\omega_i}-\frac{4 \gamma_{1,i}}{\omega_i}-\frac{4 \gamma_{3,i}}{\omega_i}\bigg) +\\
&t_{i,j}^2\bigg(\frac{2 \gamma_{1,i}}{\omega_i^2}-\frac{6 \gamma_{2,i}}{\omega_i^2}+\frac{4 \gamma_{3,i}}{\omega_i^2}\bigg)\bigg\} I(\omega_i/2 \leq t_{i,j} < \omega_i) + \varepsilon_{i,j},
\end{aligned}
\end{equation}

\noindent where $\varepsilon_{i,j}\sim \text{N}(0,\sigma^2_i)$. Conveniently, removing $\gamma_{3,i}$ forces the B-spline curve to cross zero at the last knot, which is well-suited for modelling the FoS. Therefore, the B-spline that will be used in this study is

\begin{equation}
\begin{aligned}
Y_{i,j}& = \bigg\{\gamma_{0,i} + t_{i,j}\bigg(\frac{4 \gamma_{1,i}}{\omega_i}-\frac{4 \gamma_{0,i}}{\omega_i}\bigg)+ t_{i,j}^2\bigg(\frac{4 \gamma_{0,i}}{\omega_i^2}-\frac{6 \gamma_{1,i}}{\omega_i^2}+\frac{2 \gamma_{2,i}}{\omega_i^2}\bigg) \bigg\} I(0 \leq t_{i,j} < \omega_i/2)+\\
&\bigg\{2(\gamma_{1,i} -\gamma_{2,i})+t_{i,j}\bigg(\frac{8 \gamma_{2,i}}{\omega_i}-\frac{4 \gamma_{1,i}}{\omega_i}\bigg) +t_{i,j}^2\bigg(\frac{2 \gamma_{1,i}}{\omega_i^2}-\frac{6 \gamma_{2,i}}{\omega_i^2}\bigg)\bigg\} I(\omega_i/2 \leq t_{i,j} < \omega_i) + \varepsilon_{i,j}.
\end{aligned}
\end{equation}
\newpage 

\section{Posterior diagnostic plots}\label{DiagSupplement}

\begin{figure}[H]
	\centering
	\includegraphics[scale=0.5]{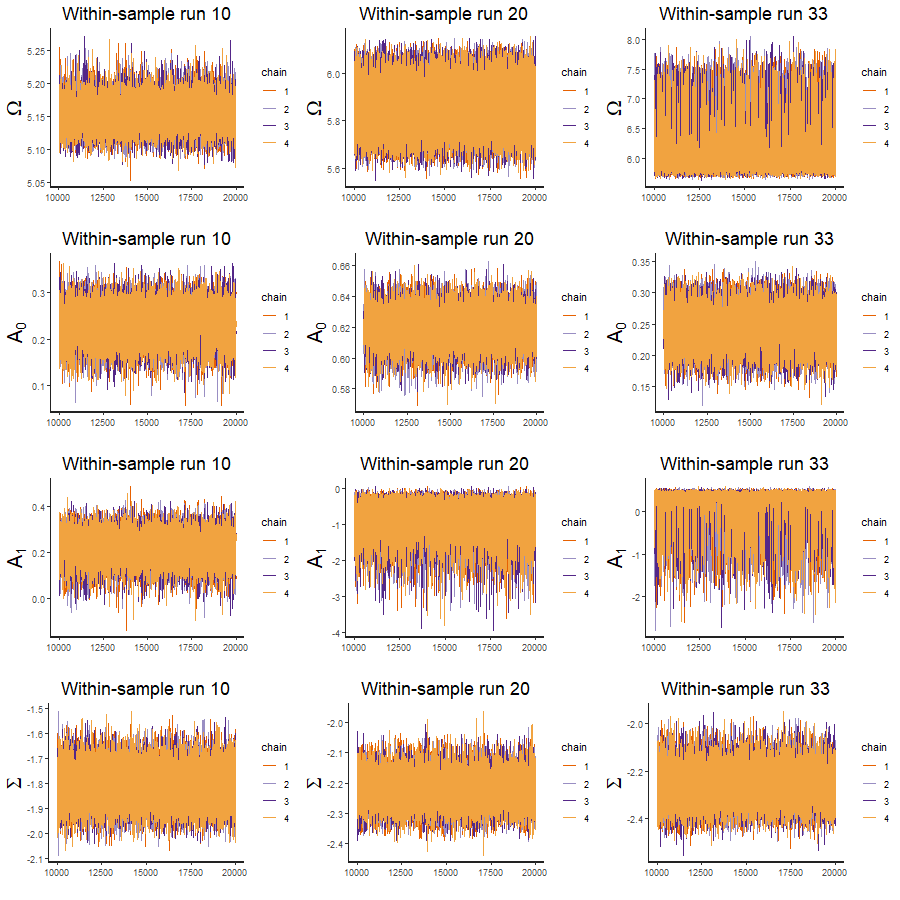}
	\caption*{Figure B1: Examples of posterior trace plots of the quadratic model parameters.}
	\label{traceplots_quadratic}
\end{figure}

\begin{figure}[H]
	\centering
	\includegraphics[scale=0.5]{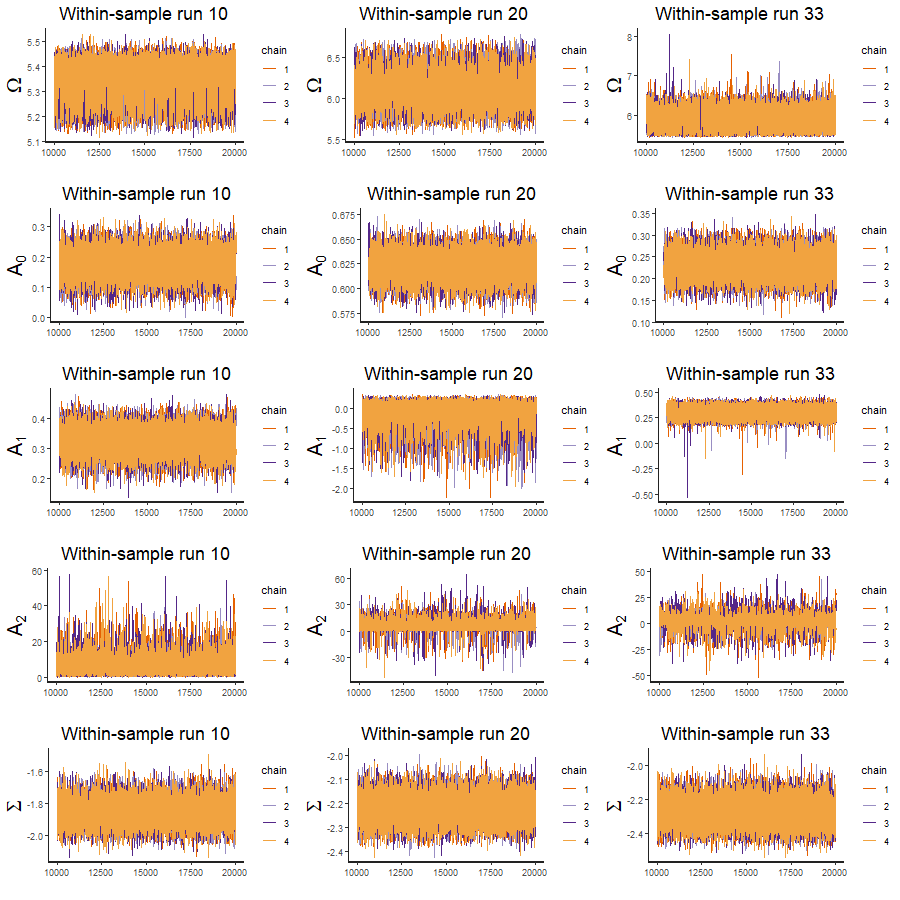}
	\caption*{Figure B2: Examples of posterior trace plots of the B-spline model parameters.}
	\label{traceplots_splines}
\end{figure}

\begin{figure}[H]
	\centering
	\includegraphics[scale=0.5]{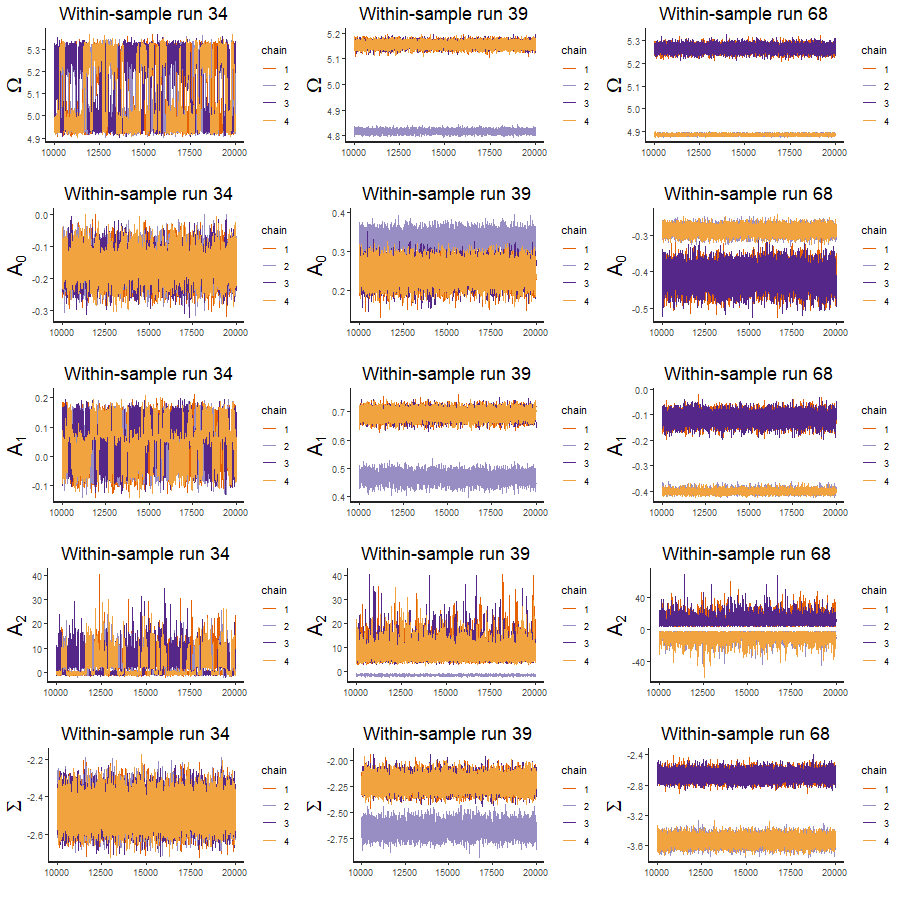}
	\caption*{Figure B3: Examples of multi-modal posterior trace plots of the B-spline model parameters corresponding to the FoS plots shown in Figure B4.}
	\label{multi_modal_traceplots}
\end{figure}

\begin{figure}[H]
	\centering
	\includegraphics[scale=0.5]{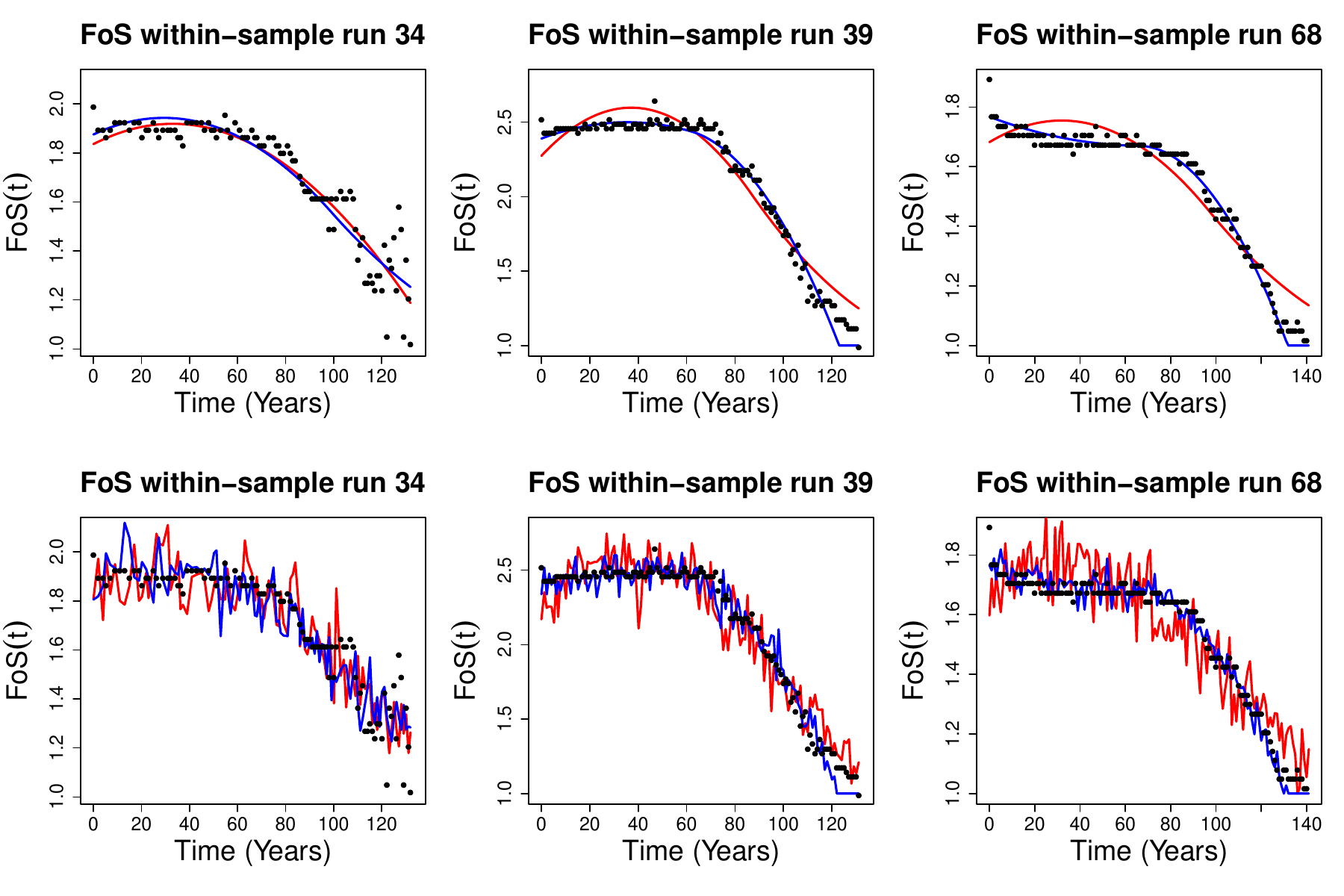}
	\caption*{Figure B4: Examples of multi-modal posterior plots of FoS.}
	\label{multi_modal_FoS}
\end{figure}

\begin{figure}[H]
	\centering
	\includegraphics[scale=0.5]{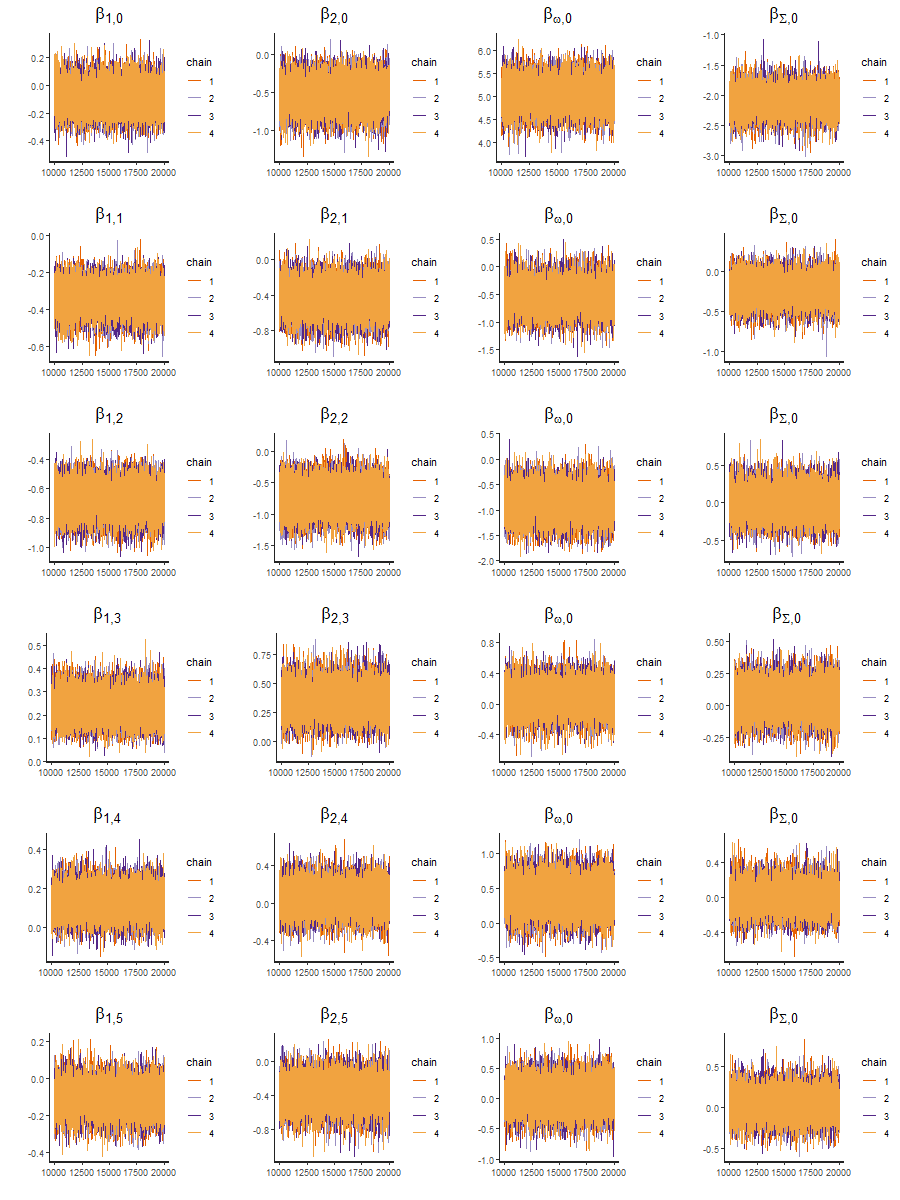}
	\caption*{Figure B5: Posterior trace plots of the emulator parameters using the quadratic model.}
	\label{traceplots_quadratic_betas}
\end{figure}

\begin{figure}[H]
	\centering
	\includegraphics[scale=0.4]{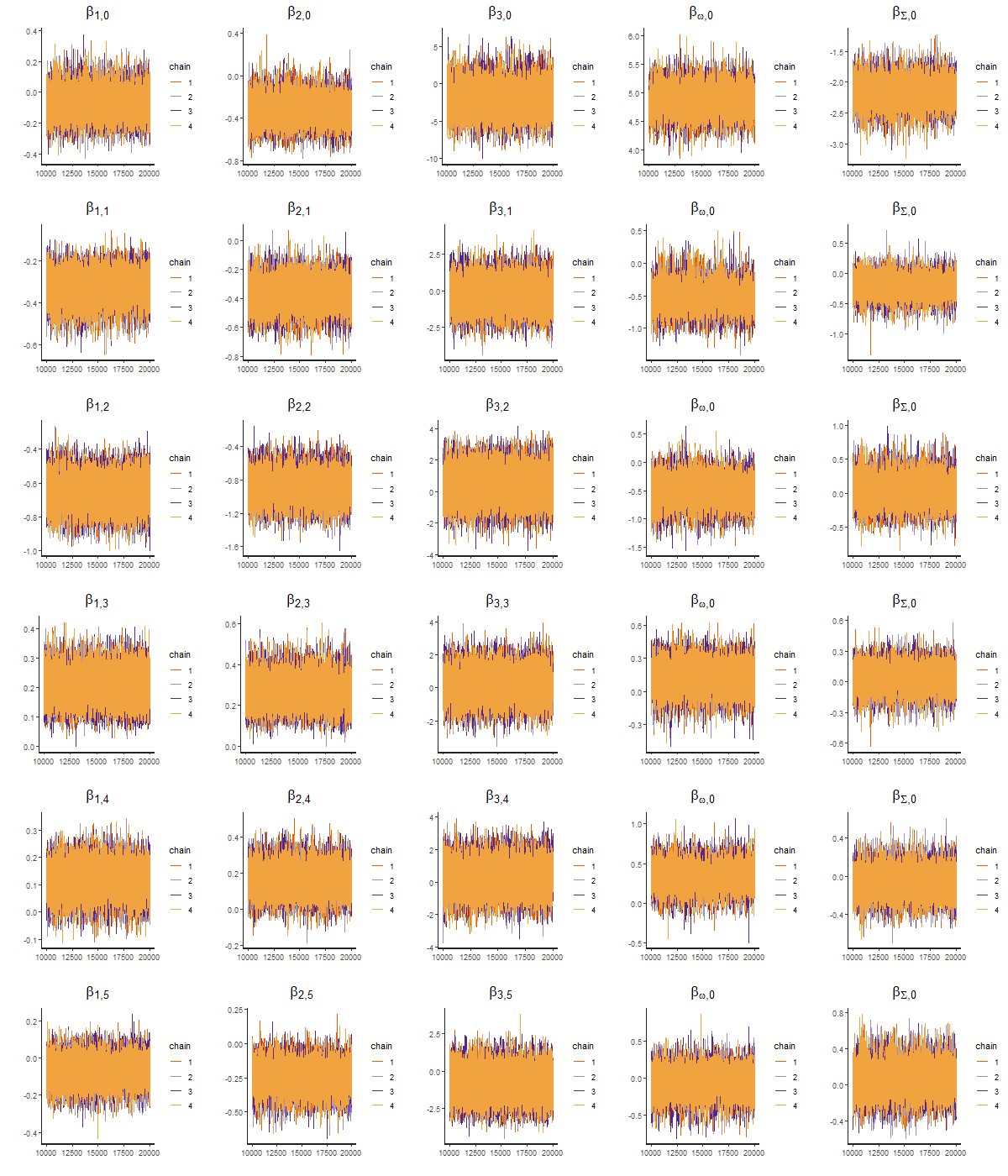}
	\caption*{Figure B6: Posterior trace plots of the emulator parameters using the B-spline model.}
	\label{traceplots_splines_betas}
\end{figure}

\section{Posterior plots of FoS}\label{PostFoS}
\begin{figure}[H]
	\centering
	\includegraphics[scale=0.7]{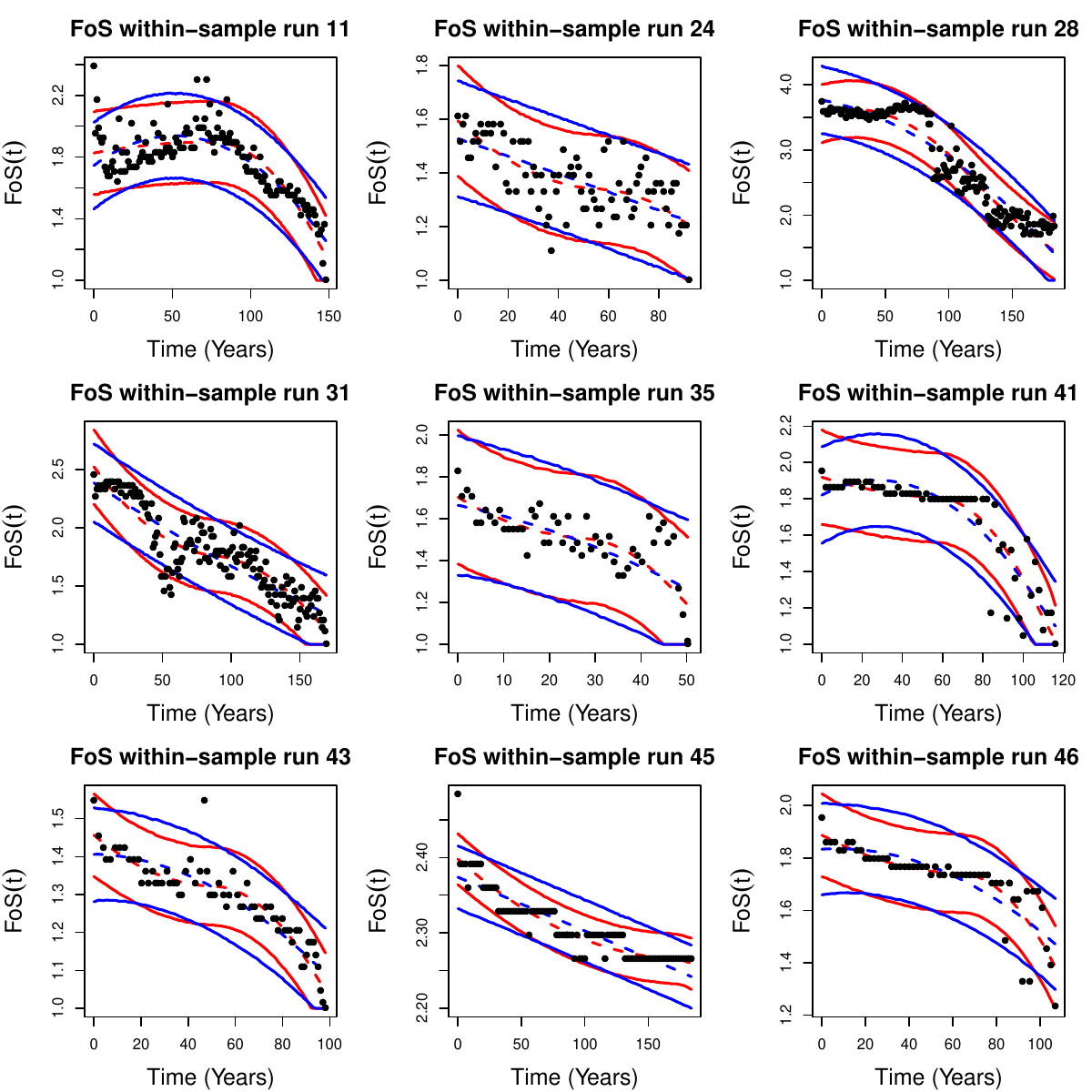}
	\caption*{Figure C1: Examples of posterior distributions of FoS using the quadratic model (blue) and the B-spline model (red). The dashed lines represent the posterior means and the solid lines represent the central posterior 95\% prediction intervals.}
	\label{Plots2}
\end{figure}

\begin{figure}[H]
	\centering
	\includegraphics[scale=0.7]{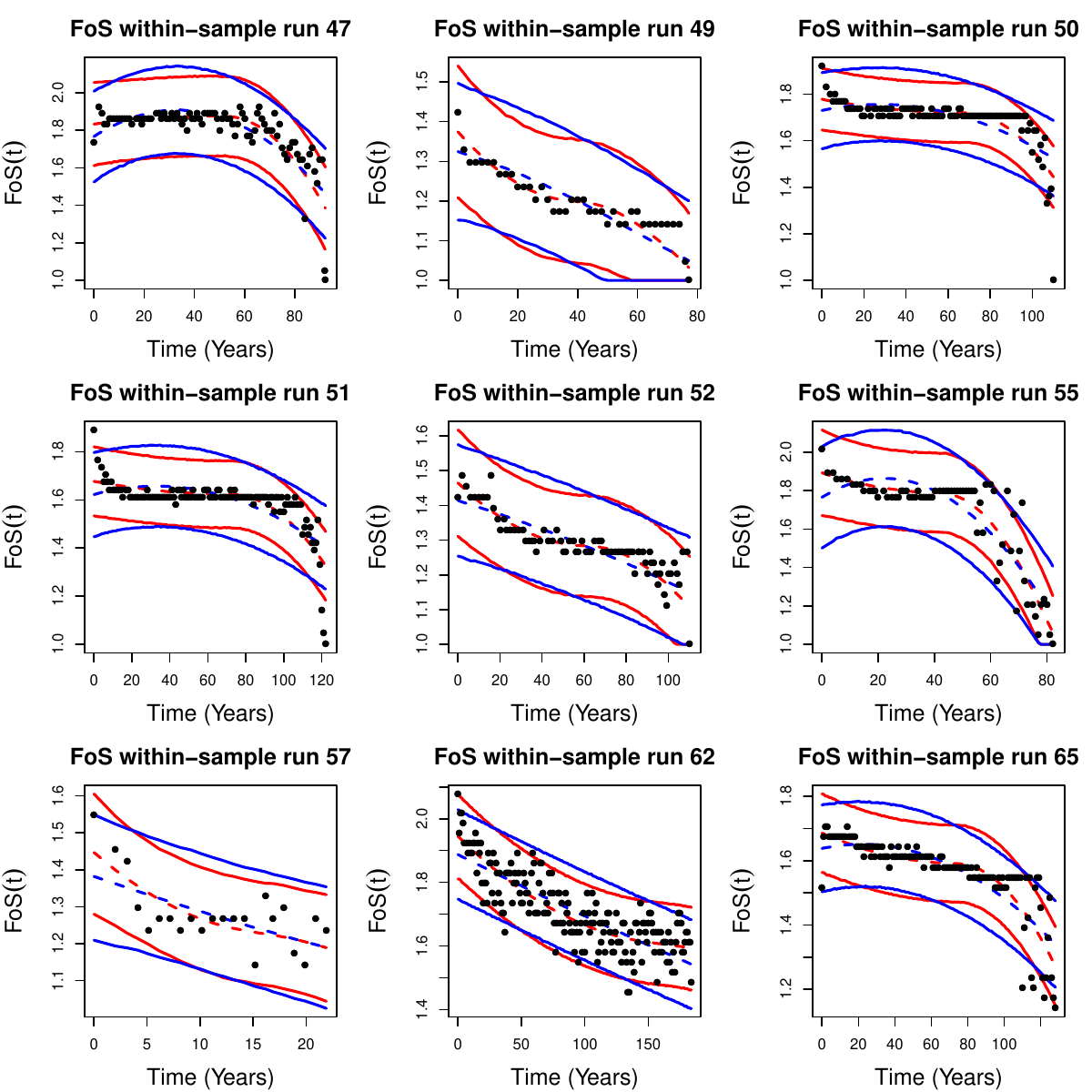}
	\caption*{Figure C2: Examples of posterior distributions of FoS using the quadratic model (blue) and the B-spline model (red). The dashed lines represent the posterior means and the solid lines represent the central posterior 95\% prediction intervals.}
	\label{Plots3}
\end{figure}

\begin{figure}[H]
	\centering
	\includegraphics[scale=0.7]{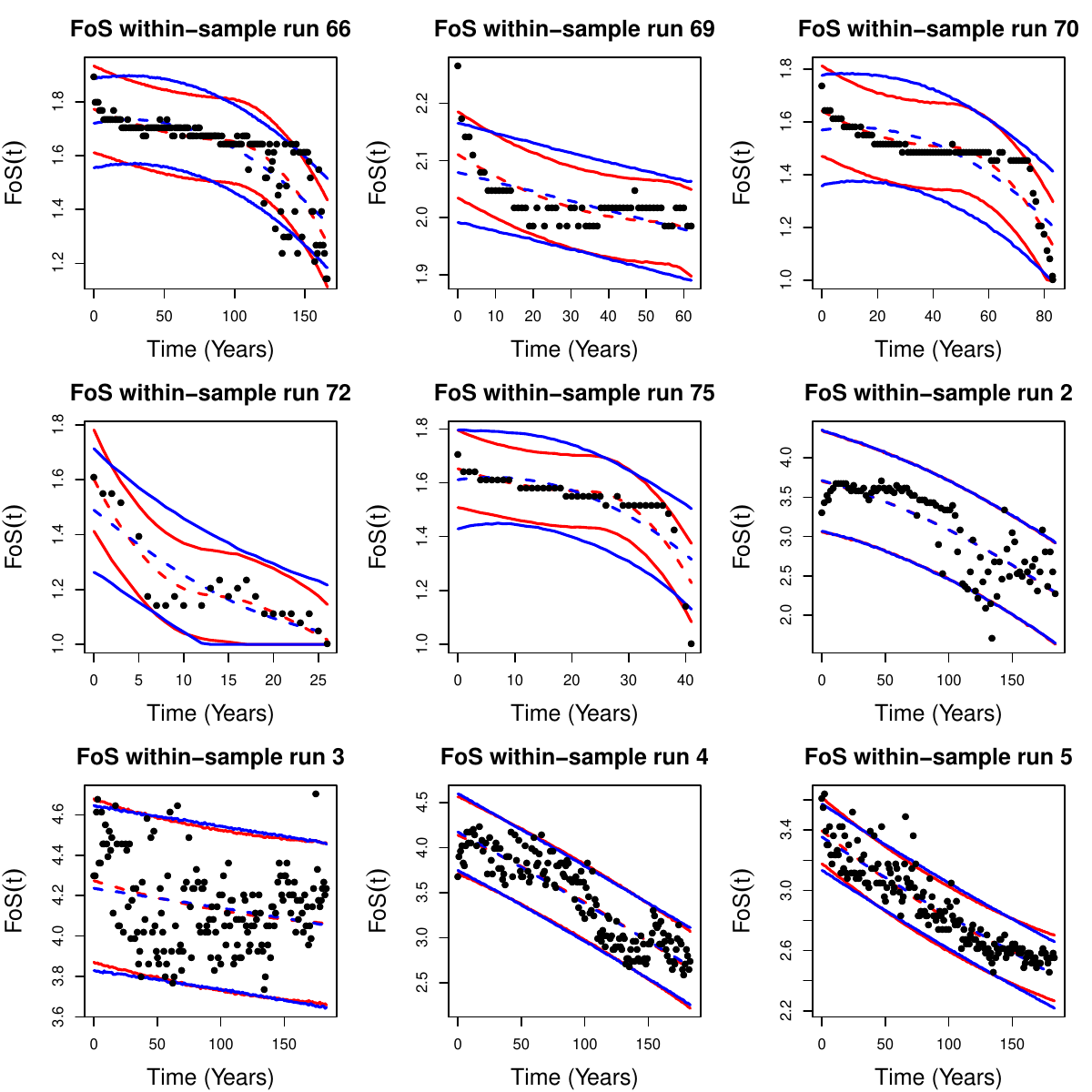}
	\caption*{Figure C3: Examples of posterior distributions of FoS using the quadratic model (blue) and the B-spline model (red). The dashed lines represent the posterior means and the solid lines represent the central posterior 95\% prediction intervals.}
	\label{Plots4}
\end{figure}

\begin{figure}[H]
	\centering
	\includegraphics[scale=0.7]{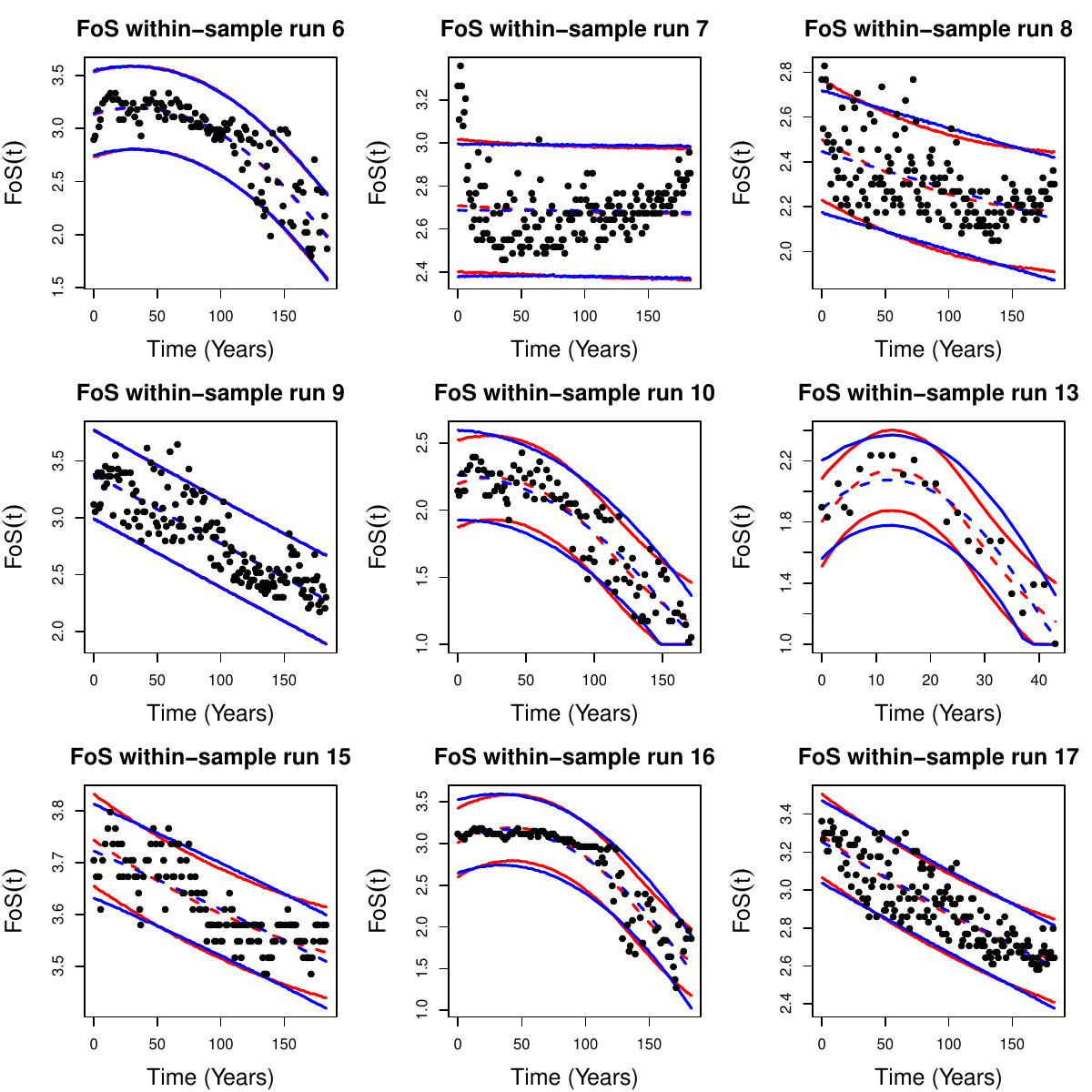}
	\caption*{Figure C4: Examples of posterior distributions of FoS using the quadratic model (blue) and the B-spline model (red). The dashed lines represent the posterior means and the solid lines represent the central posterior 95\% prediction intervals.}
	\label{Plots5}
\end{figure}

\begin{figure}[H]
	\centering
	\includegraphics[scale=0.7]{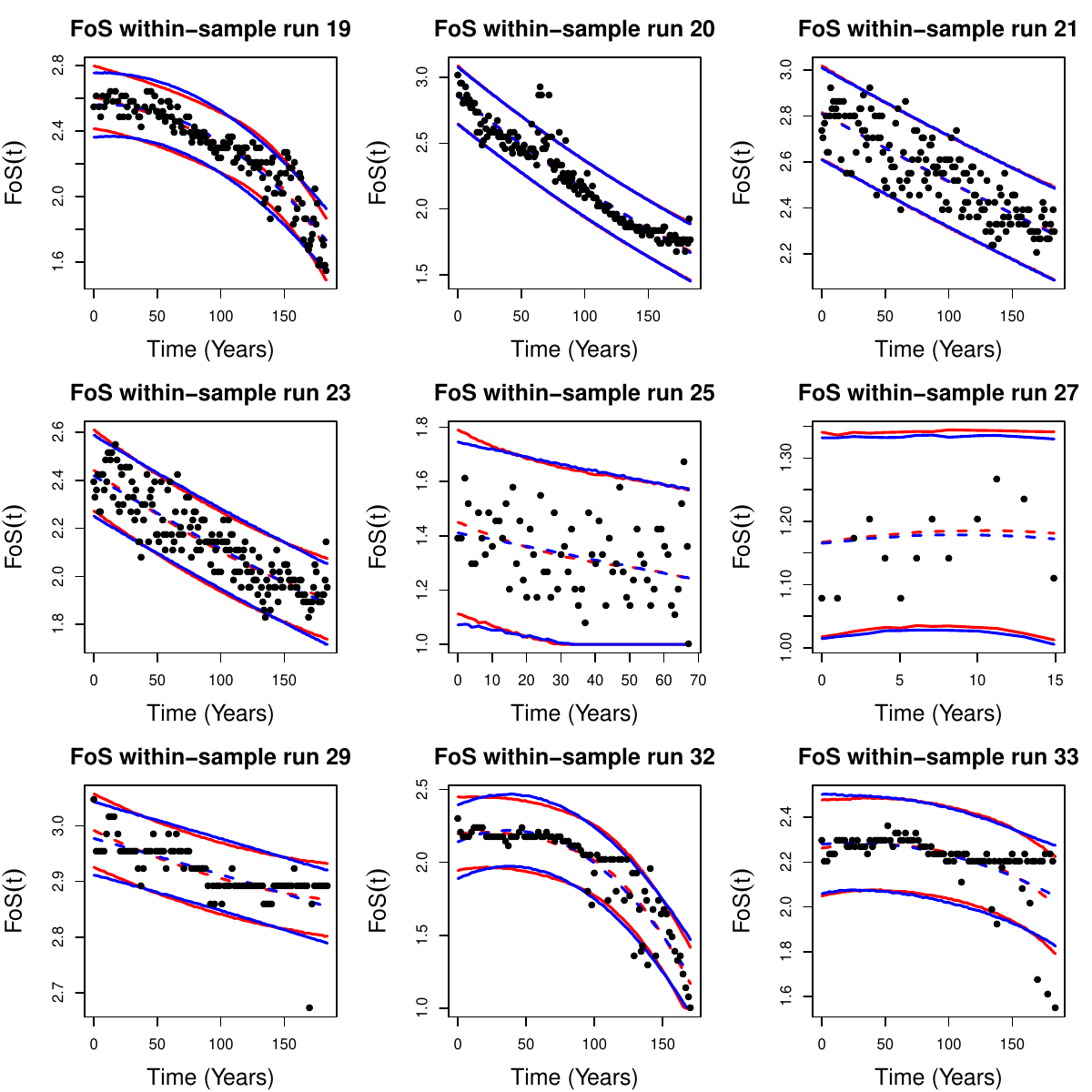}
	\caption*{Figure C5: Examples of posterior distributions of FoS using the quadratic model (blue) and the B-spline model (red). The dashed lines represent the posterior means and the solid lines represent the central posterior 95\% prediction intervals.}
	\label{Plots6}
\end{figure}
\begin{figure}[H]
	\centering
	\includegraphics[scale=0.7]{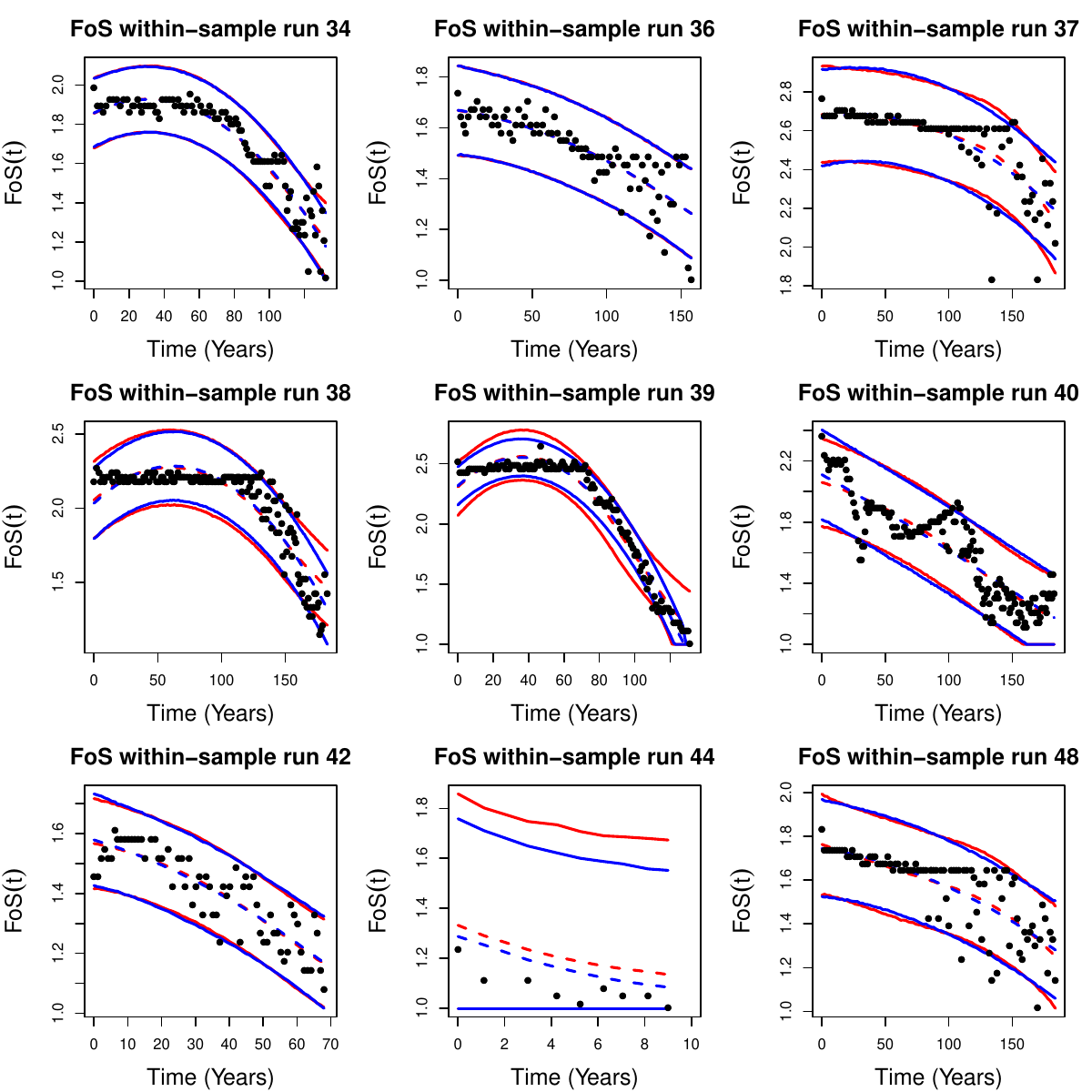}
	\caption*{Figure C6: Examples of posterior distributions of FoS using the quadratic model (blue) and the B-spline model (red). The dashed lines represent the posterior means and the solid lines represent the central posterior 95\% prediction intervals.}
	\label{Plots7}
\end{figure}

\begin{figure}[H]
	\centering
	\includegraphics[scale=0.7]{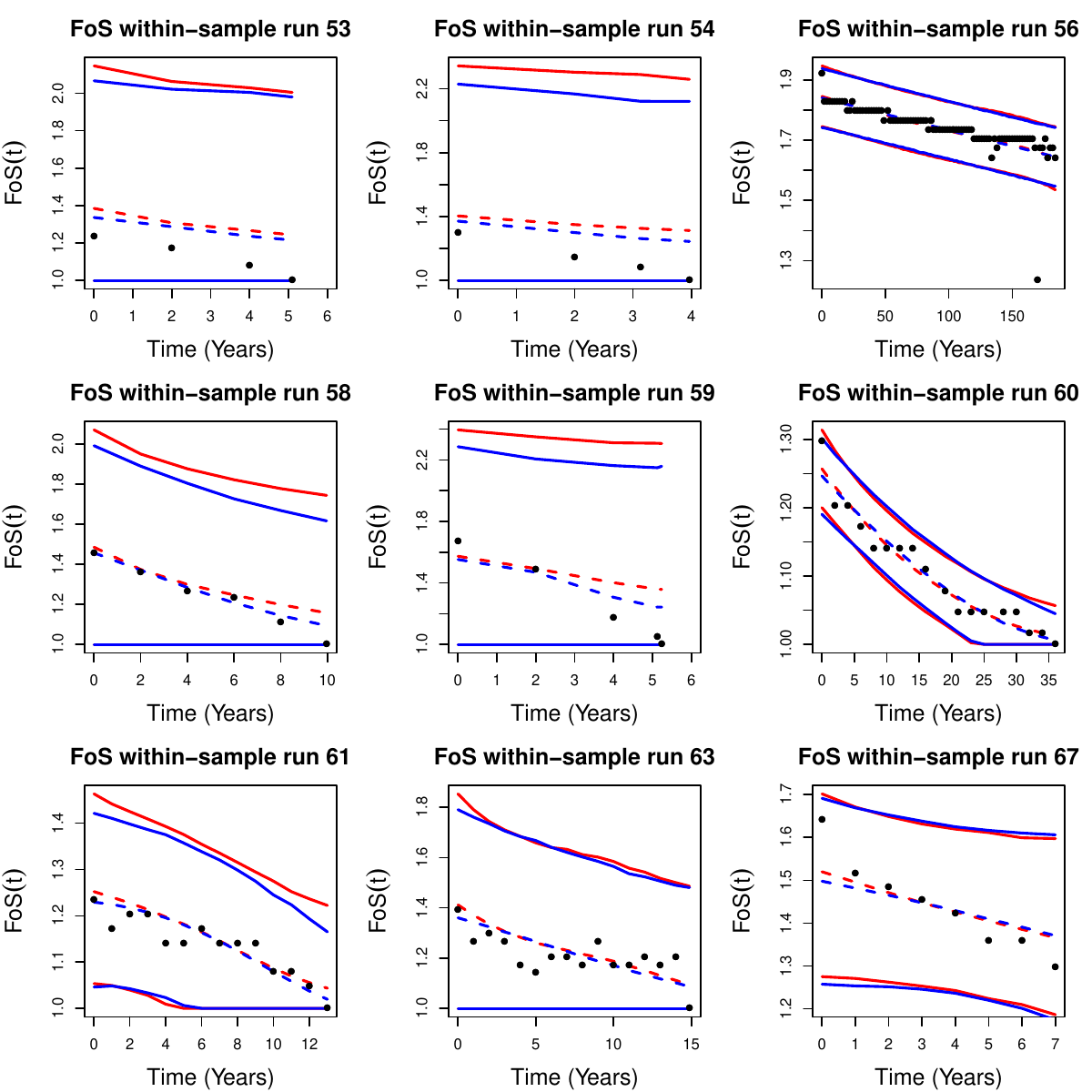}
	\caption*{Figure C7: Examples of posterior distributions of FoS using the quadratic model (blue) and the B-spline model (red). The dashed lines represent the posterior means and the solid lines represent the central posterior 95\% prediction intervals.}
	\label{Plots8}
\end{figure}

\begin{figure}[H]
	\centering
	\includegraphics[scale=0.7]{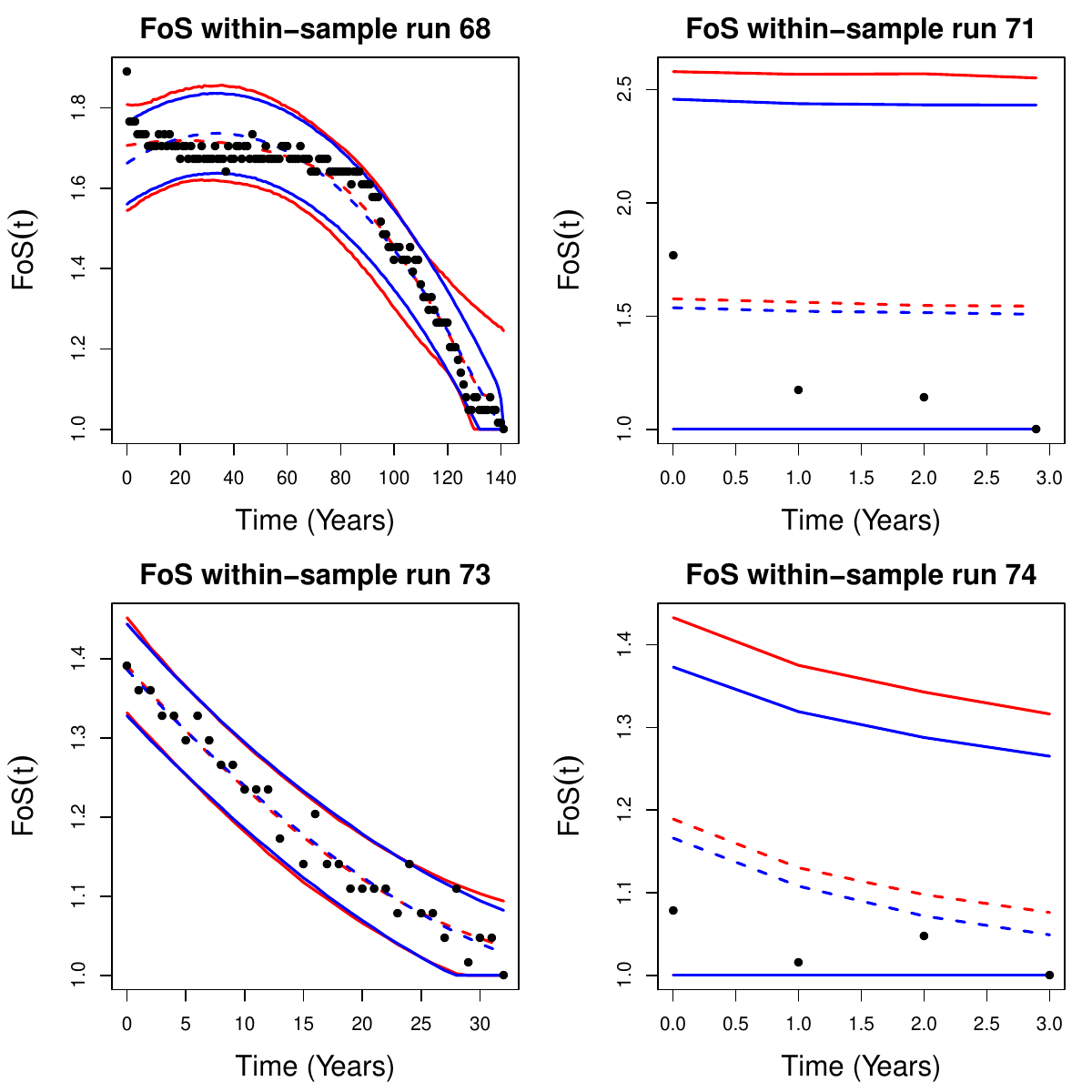}
	\caption*{Figure C8: Examples of posterior distributions of FoS using the quadratic model (blue) and the B-spline model (red). The dashed lines represent the posterior means and the solid lines represent the central posterior 95\% prediction intervals.}
	\label{Plots9}
\end{figure}
\newpage
\section{Diagnostics of posterior fit}\label{PostDiag}
\begin{figure}[H]
	\centering
	\includegraphics[scale=0.5]{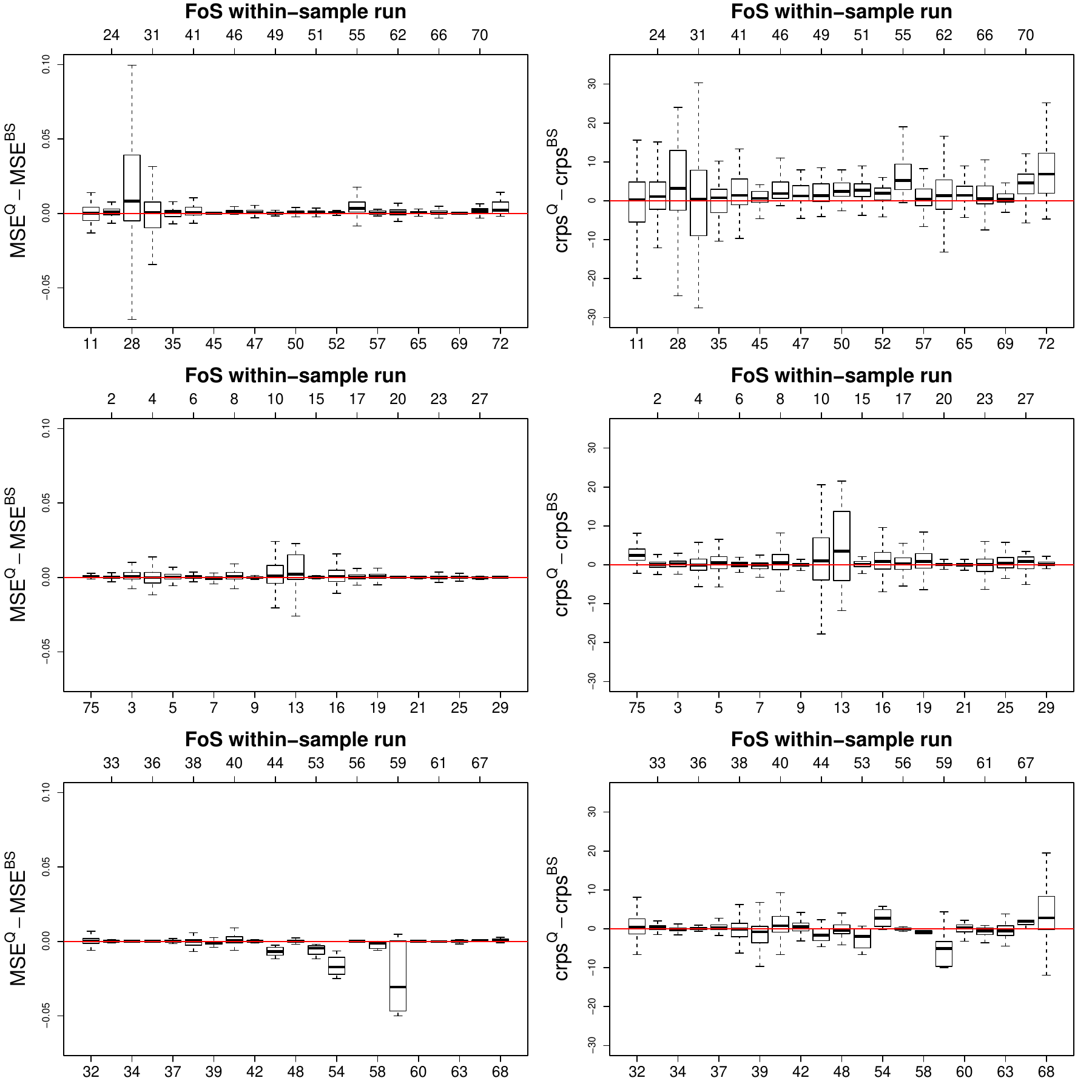}
	\caption*{Figure D1: Plots (left) presents the difference in the $MSE$ between the quadratic model and the B-spline model for sixty-six within-sample computer runs. Plot (right) presents the difference in the $crps$ between the quadratic model and the B-spline model for sixty six within-sample computer runs. The numbers on the x-axis correspond to run numbers.}
	\label{within_sample66}
\end{figure}

\newpage

\newpage
\section{Posterior distributions of predicted TTF}\label{PostTTF}
\begin{figure}[H]
	\centering
	\includegraphics[scale=0.7]{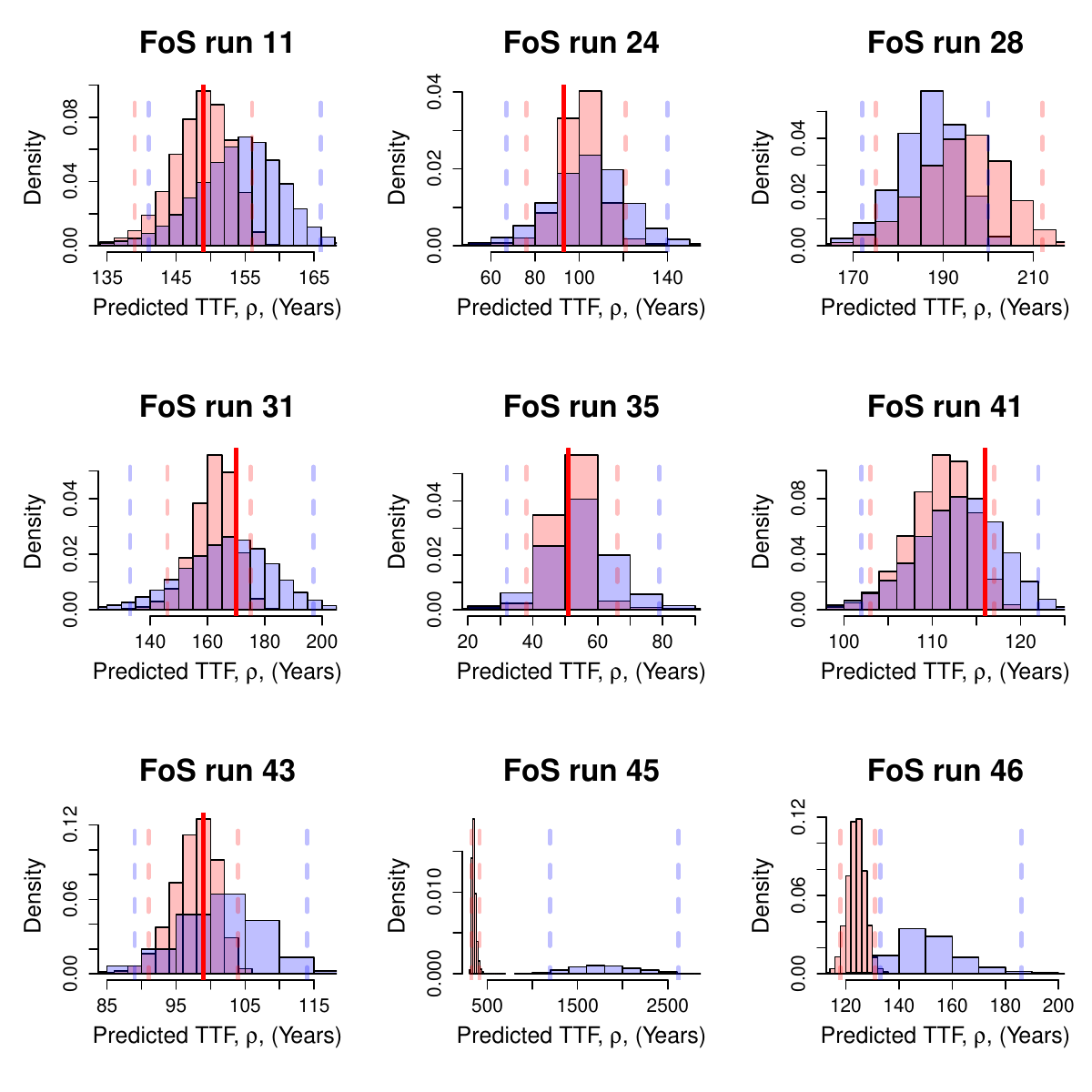}
	\caption*{Figure E1: Examples of posterior distributions of predicted TTF, $\rho_i$, using the quadratic model (blue) and the B-spline model (red). The dashed lines represent the central posterior 95\% prediction intervals under each model. The solid red lines indicate true TTF for computer runs that reached failure.}
	\label{Plots_TTF2}
\end{figure}

\begin{figure}[H]
	\centering
	\includegraphics[scale=0.7]{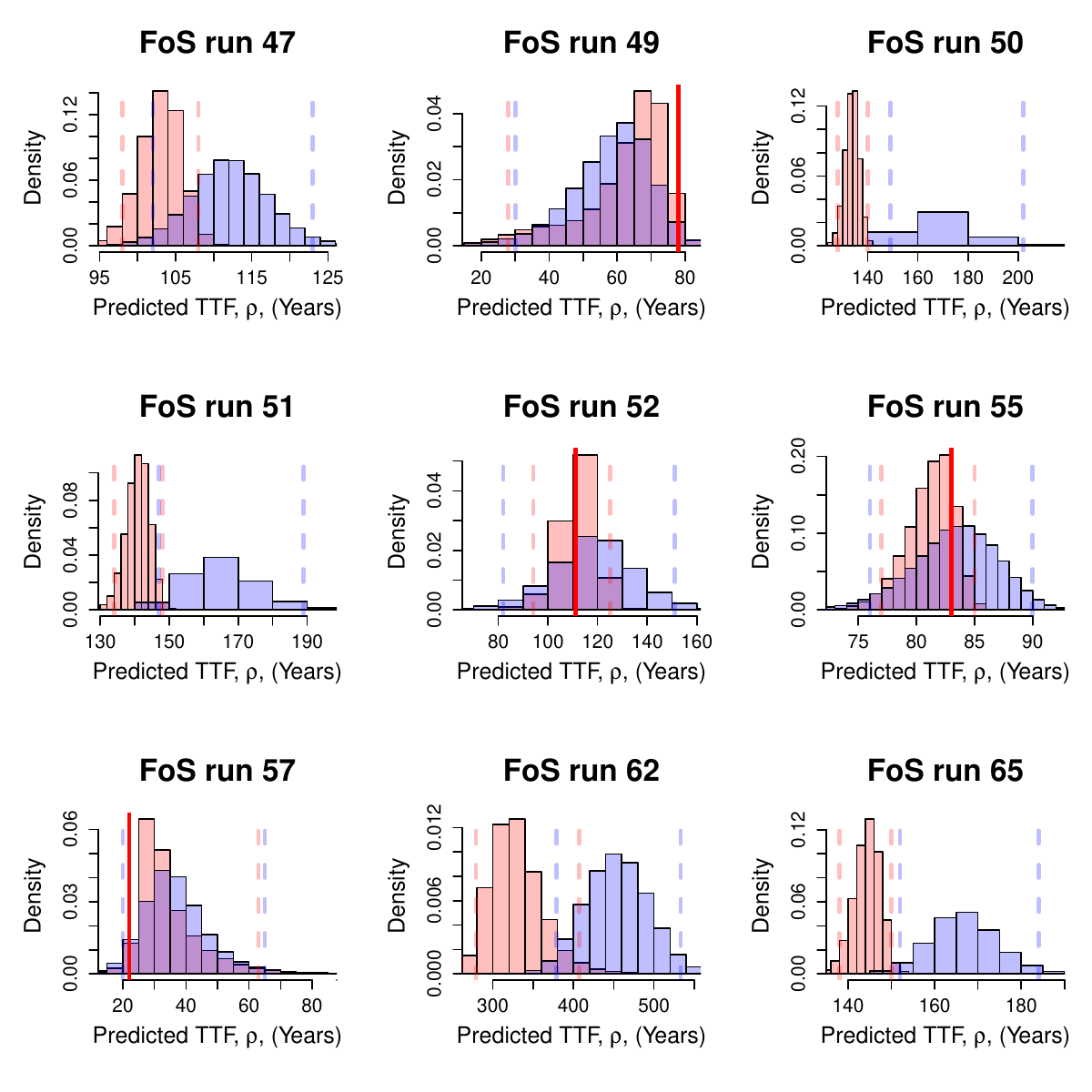}
	\caption*{Figure E2: Examples of posterior distributions of predicted TTF, $\rho_i$, using the quadratic model (blue) and the B-spline model (red). The dashed lines represent the central posterior 95\% prediction intervals under each model. The solid red lines indicate true TTF for computer runs that reached failure.}
	\label{Plots_TTF3}
\end{figure}

\begin{figure}[H]
	\centering
	\includegraphics[scale=0.7]{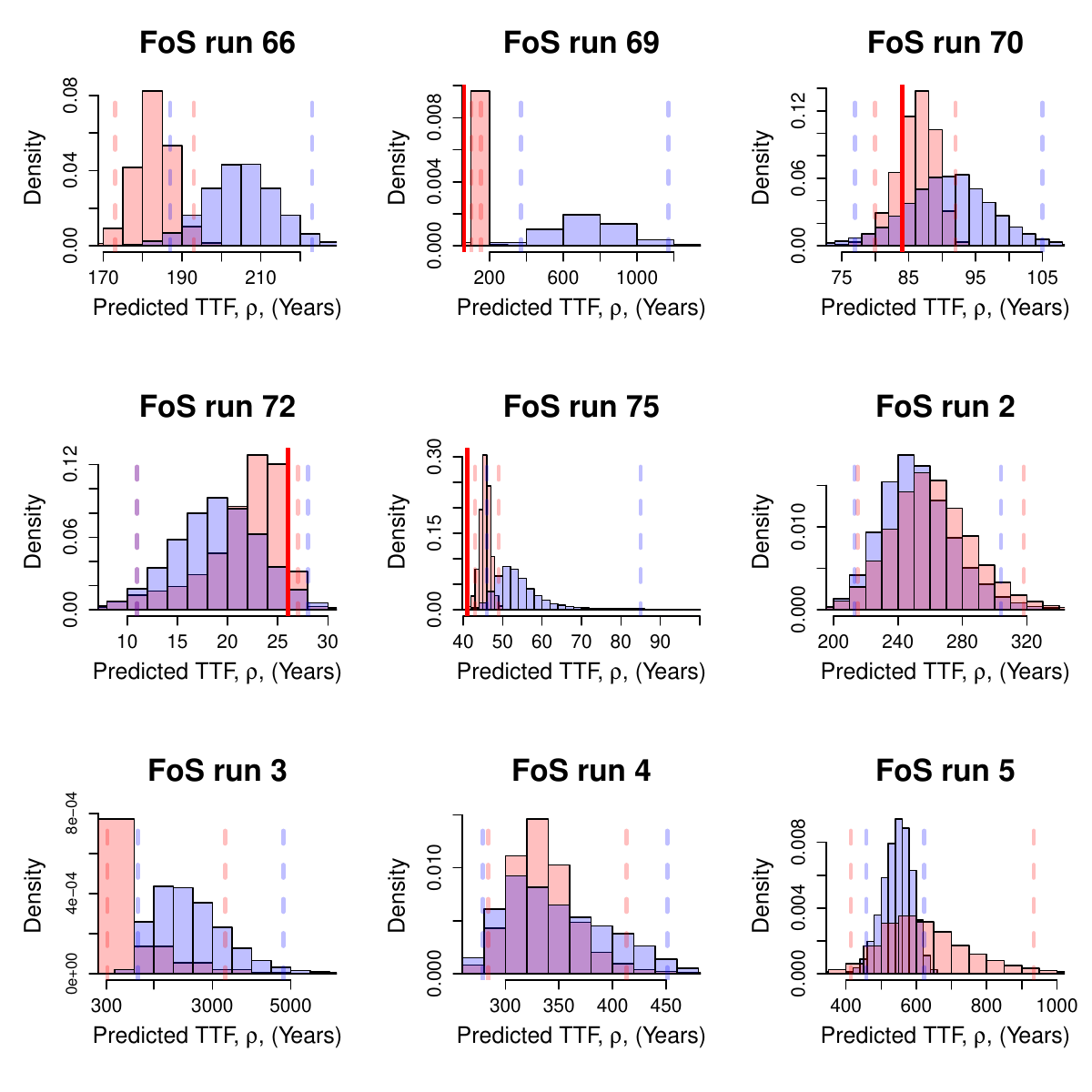}
	\caption*{Figure E3: Examples of posterior distributions of predicted TTF, $\rho_i$, using the quadratic model (blue) and the B-spline model (red). The dashed lines represent the central posterior 95\% prediction intervals under each model. The solid red lines indicate true TTF for computer runs that reached failure.}
	\label{Plots_TTF4}
\end{figure}

\begin{figure}[H]
	\centering
	\includegraphics[scale=0.7]{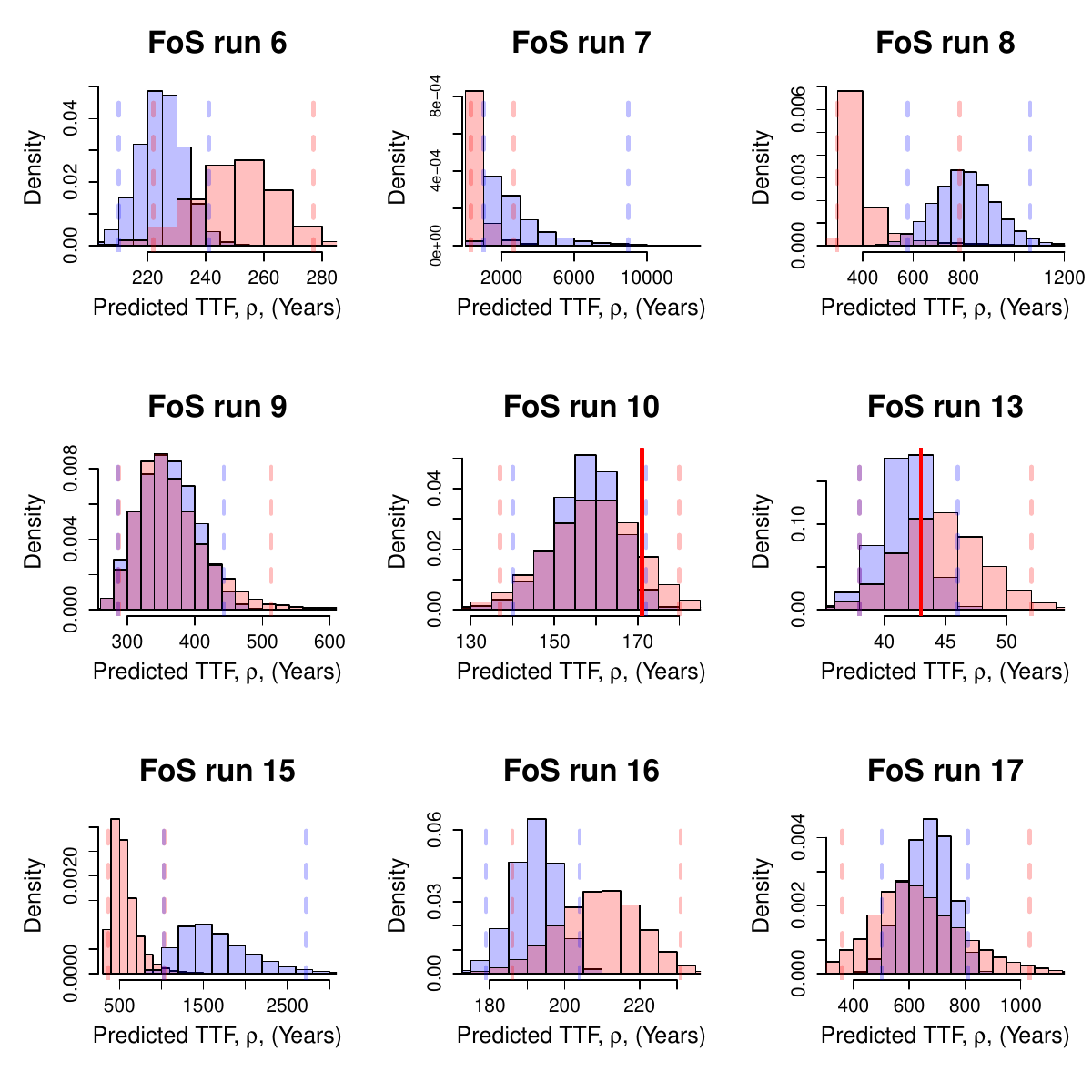}
	\caption*{Figure E4: Examples of posterior distributions of predicted TTF, $\rho_i$, using the quadratic model (blue) and the B-spline model (red). The dashed lines represent the central posterior 95\% prediction intervals under each model. The solid red lines indicate true TTF for computer runs that reached failure.}
	\label{Plots_TTF5}
\end{figure}

\begin{figure}[H]
	\centering
	\includegraphics[scale=0.7]{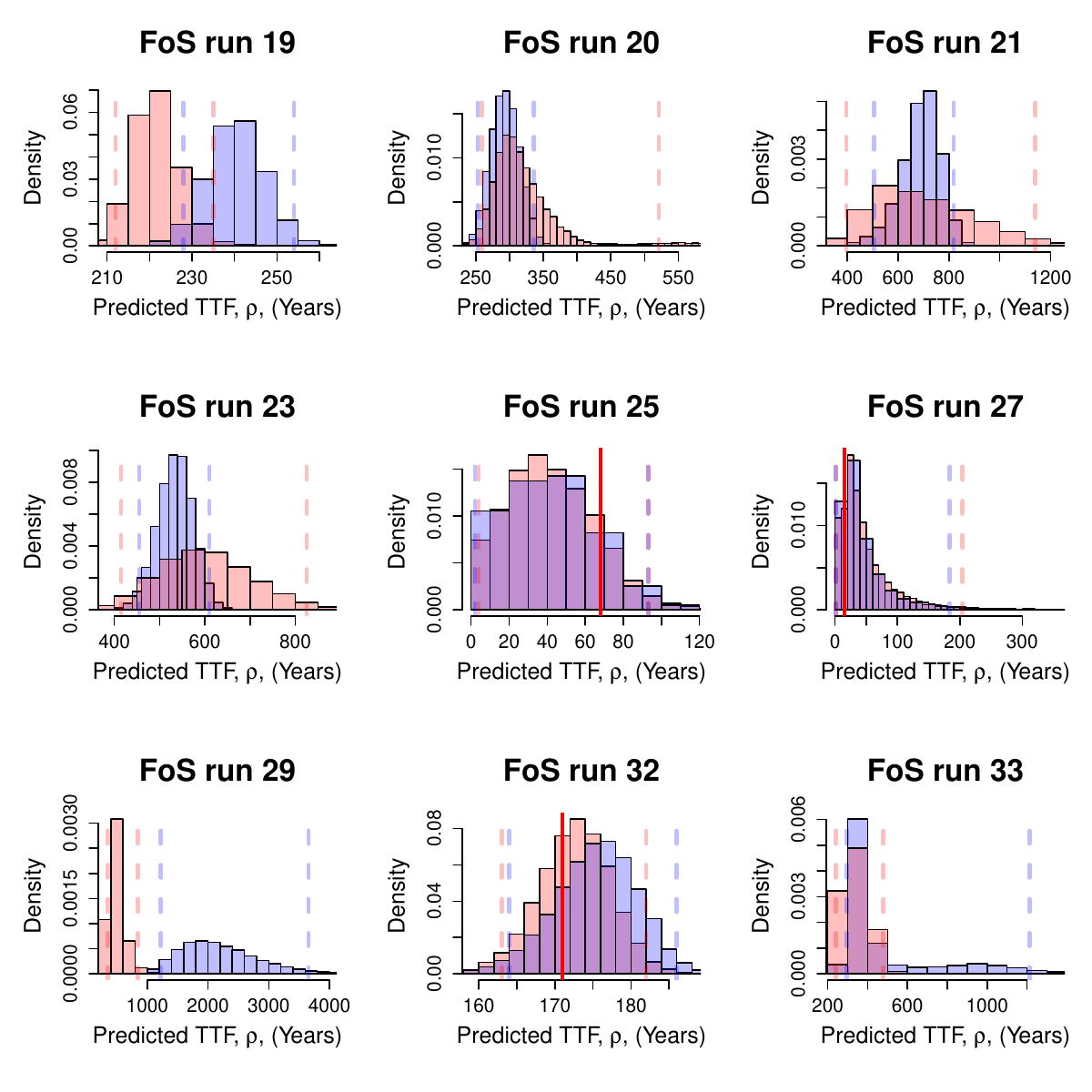}
	\caption*{Figure E5: Examples of posterior distributions of predicted TTF, $\rho_i$, using the quadratic model (blue) and the B-spline model (red). The dashed lines represent the central posterior 95\% prediction intervals under each model. The solid red lines indicate true TTF for computer runs that reached failure.}
	\label{Plots_TTF6}
\end{figure}

\begin{figure}[H]
	\centering
	\includegraphics[scale=0.7]{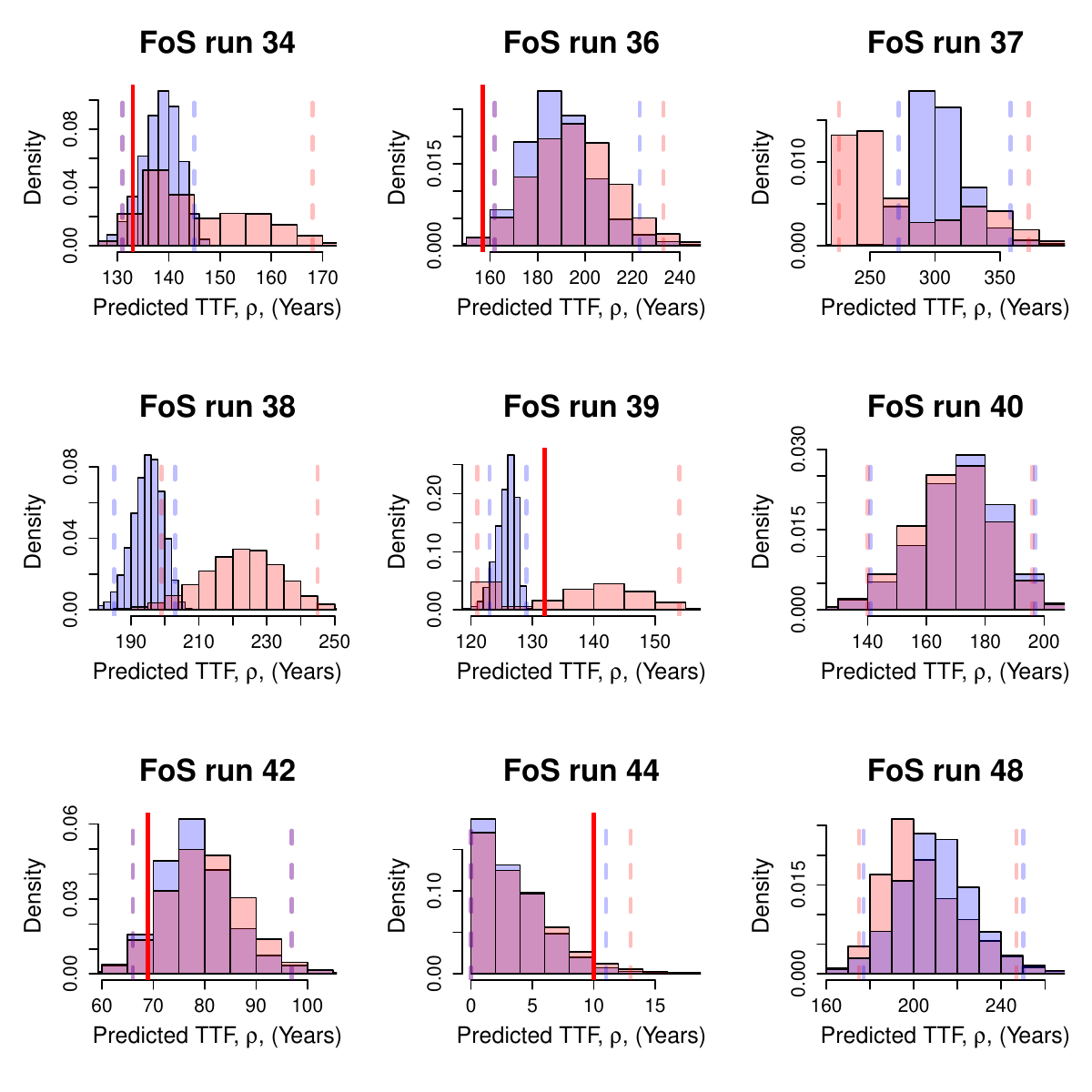}
	\caption*{Figure E6: Examples of posterior distributions of predicted TTF, $\rho_i$, using the quadratic model (blue) and the B-spline model (red). The dashed lines represent the central posterior 95\% prediction intervals under each model. The solid red lines indicate true TTF for computer runs that reached failure.}
	\label{Plots_TTF7}
\end{figure}

\begin{figure}[H]
	\centering
	\includegraphics[scale=0.7]{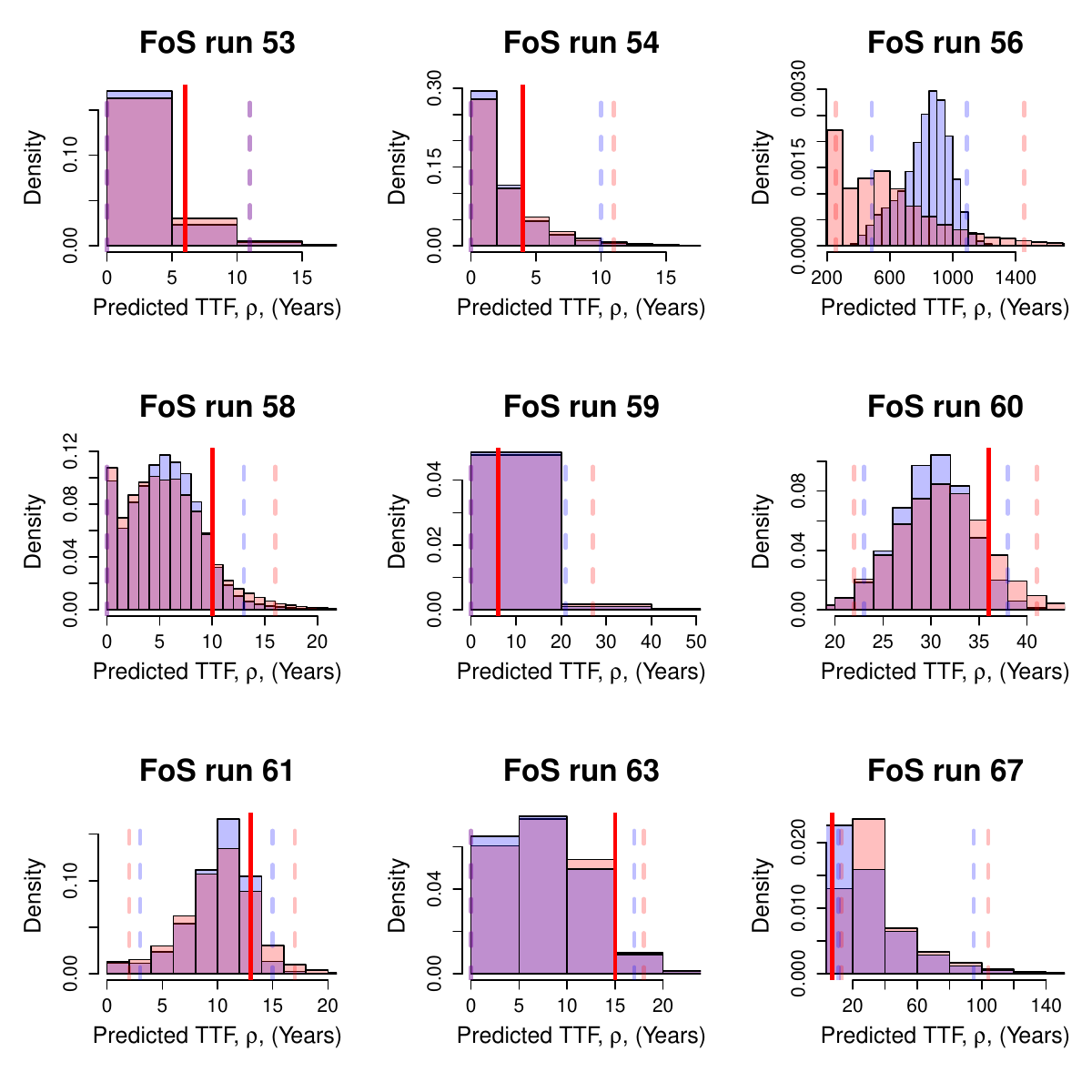}
	\caption*{Figure E7: Examples of posterior distributions of predicted TTF, $\rho_i$, using the quadratic model (blue) and the B-spline model (red). The dashed lines represent the central posterior 95\% prediction intervals under each model. The solid red lines indicate true TTF for computer runs that reached failure.}
	\label{Plots_TTF8}
\end{figure}

\begin{figure}[H]
	\centering
	\includegraphics[scale=0.6]{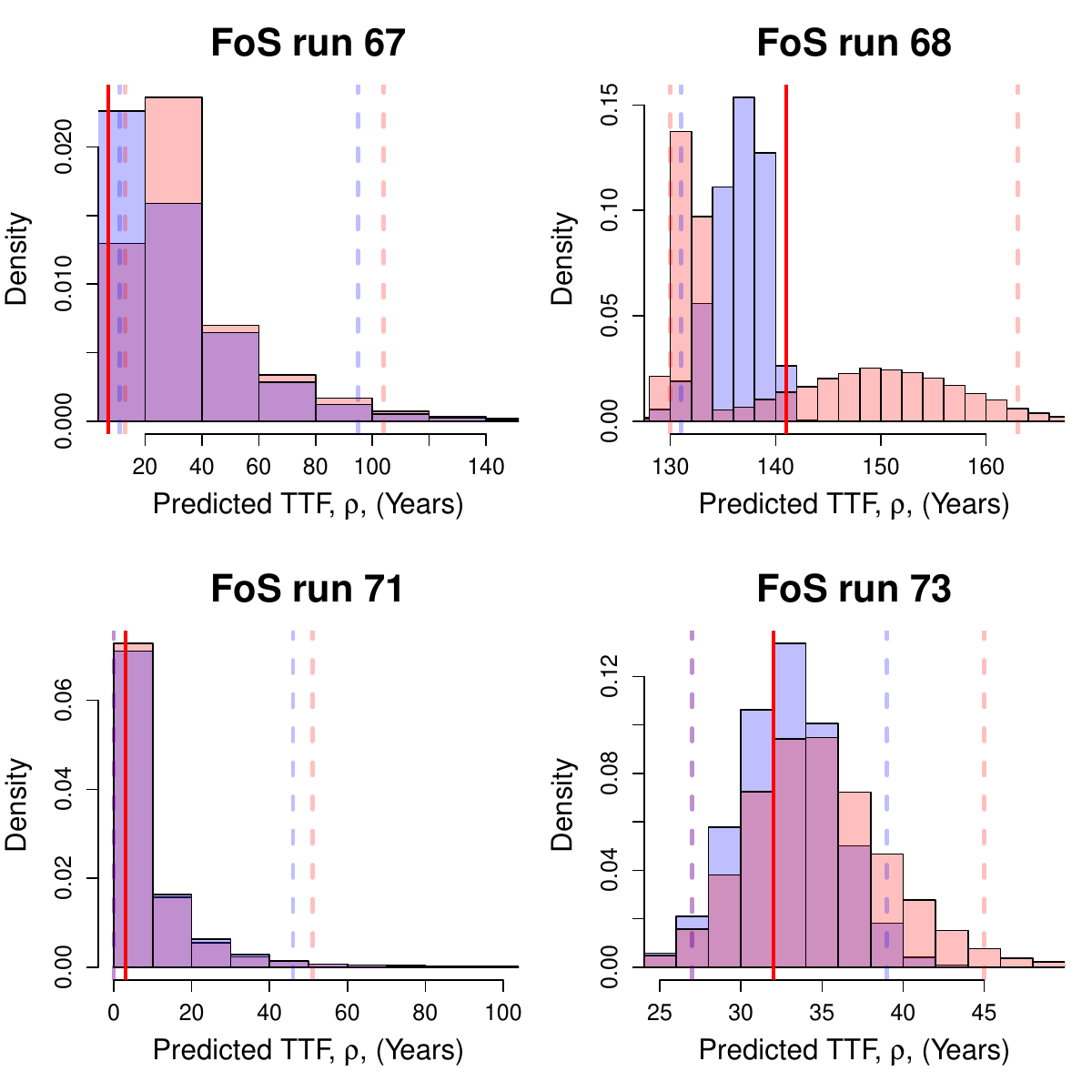}
	\caption*{Figure E8: Examples of posterior distributions of predicted TTF, $\rho_i$, using the quadratic model (blue) and the B-spline model (red). The dashed lines represent the central posterior 95\% prediction intervals under each model. The solid red lines indicate true TTF for computer runs that reached failure.}
	\label{Plots_TTF9}
\end{figure}
\begin{figure}[H]
	\centering
	\includegraphics[scale=0.5]{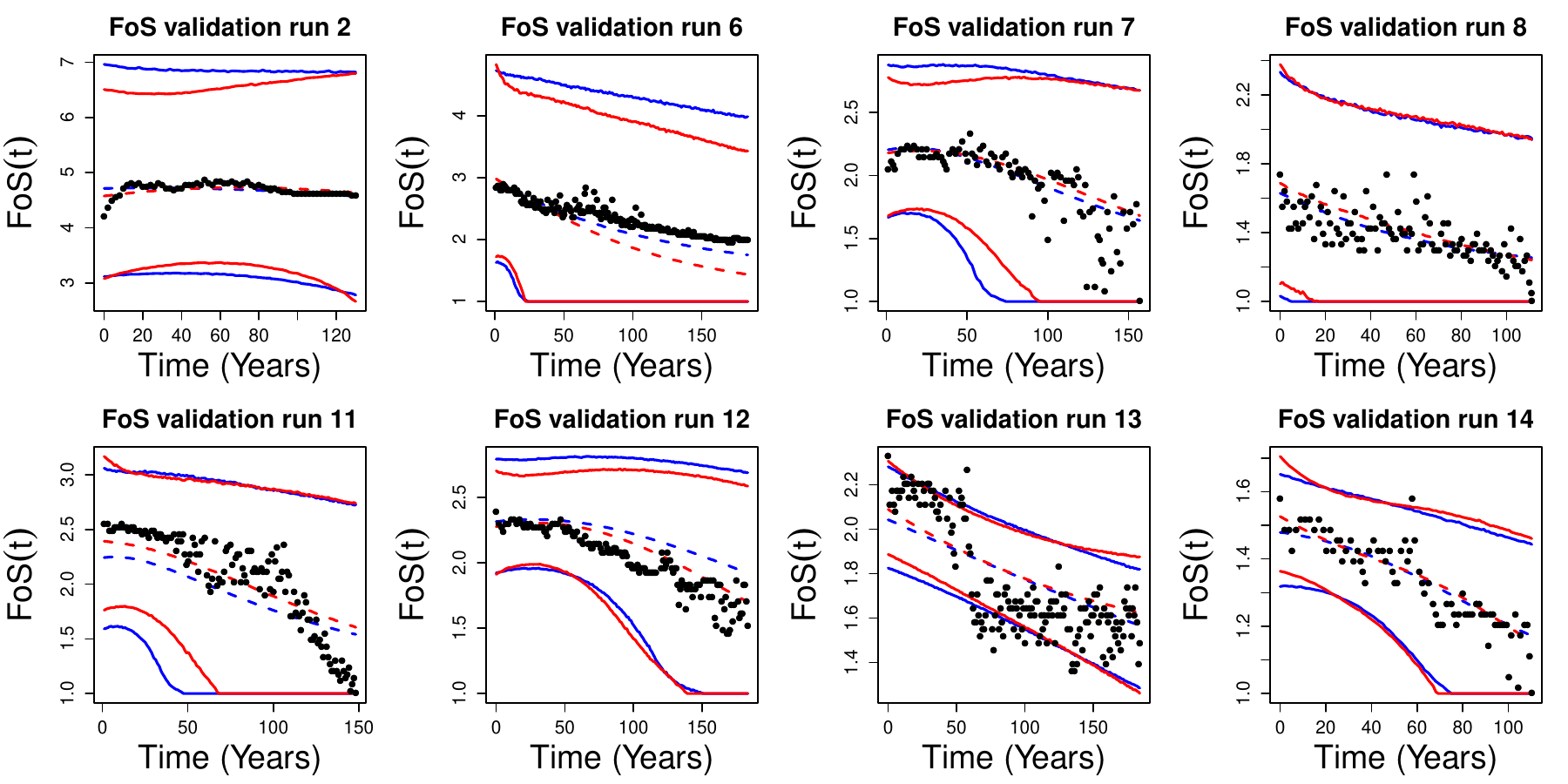}
	\caption*{Figure E9: Examples of deterministic (noise-free) FoS curves with different internal knot positions.}
	\label{different_knot}
\end{figure}
\section{Posterior distributions of FoS for the validation data sets}\label{PostValid}
\begin{figure}[H]
	\centering
	\includegraphics[scale=0.54]{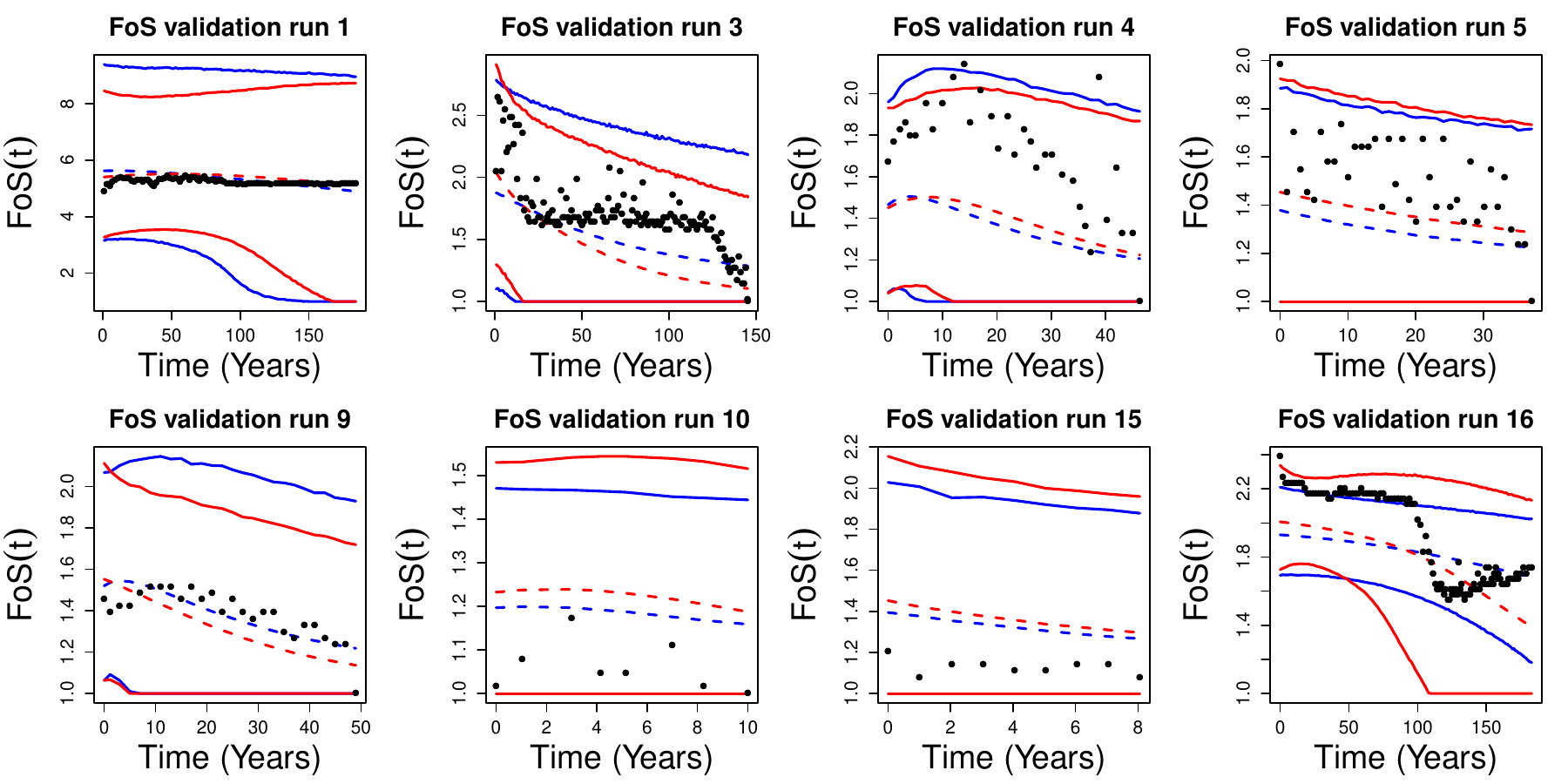}
	\caption*{Figure F1: Examples of posterior distributions of FoS using the quadratic model (blue) and the B-spline model (red) for four out-of sample computer runs. The dashed lines represent the posterior means and the solid lines represent the central posterior 95\% prediction intervals.}
	\label{FoS3}
\end{figure}

\begin{figure}[H]
	\centering
	\includegraphics[scale=0.6]{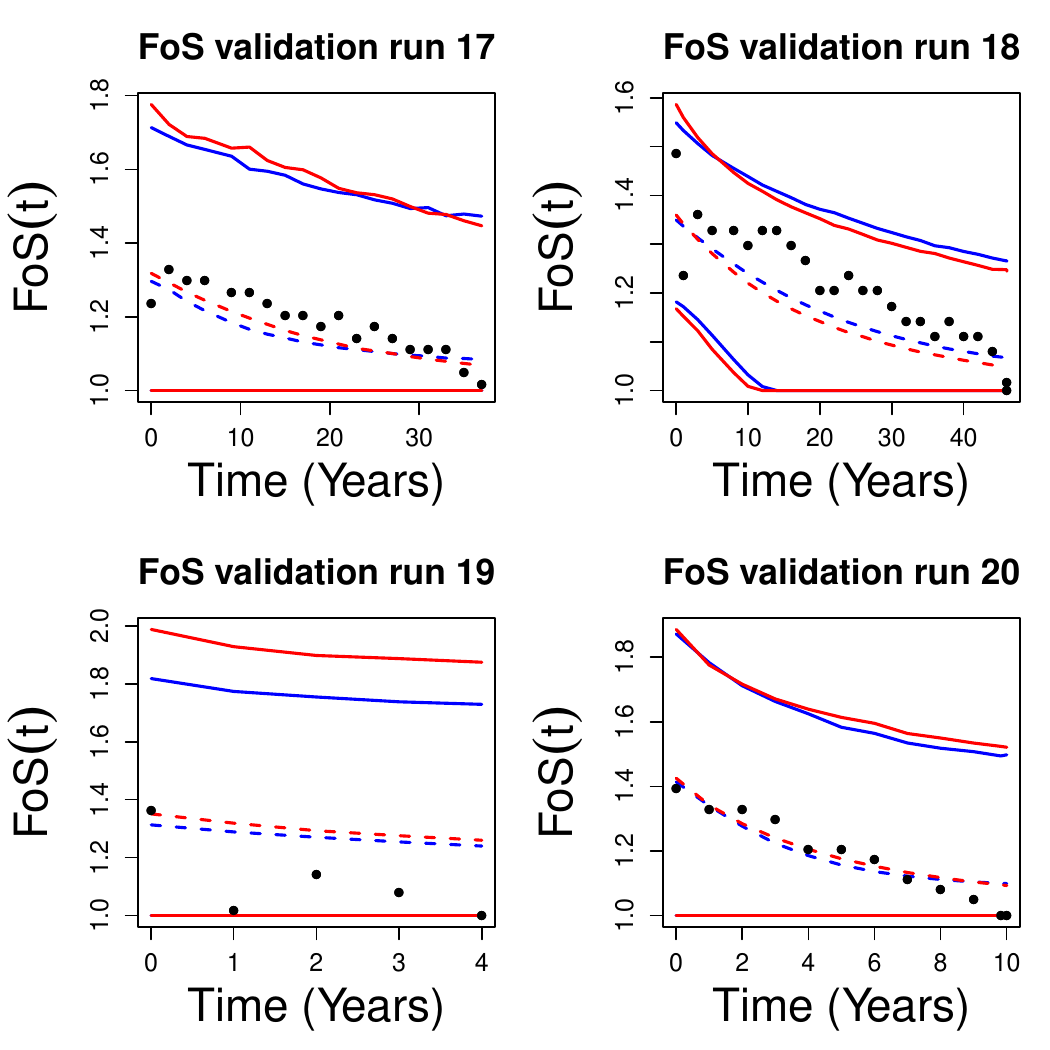}
	\caption*{Figure F2: Examples of posterior distributions of FoS using the quadratic model (blue) and the B-spline model (red) for four out-of sample computer runs. The dashed lines represent the posterior means and the solid lines represent the central posterior 95\% prediction intervals.}
	\label{FoS4}
\end{figure}
\end{appendices}

\end{document}